 \documentclass[10pt,preprint]{aastex}  
 \usepackage{natbib}
\def\ang{\AA}
\def\arcsec{\hbox{$^{\prime\prime}$}}

\def\gapprox{\lower.4ex\hbox{$\;\buildrel >\over{\scriptstyle\sim}\;$}}
\def\lapprox{\lower.4ex\hbox{$\;\buildrel <\over{\scriptstyle\sim}\;$}}

\shortauthors{ASCHWANDEN & XU 2014}
\shorttitle{Global Energetics of Solar Flares. I.}

\begin{document}

\title{         Global Energetics of Solar Flares: 
		I. Magnetic Energies }

\author{        Markus J. Aschwanden$^1$}

\affil{		$^1)$ Lockheed Martin, 
		Solar and Astrophysics Laboratory, 
                Org. A021S, Bldg.~252, 3251 Hanover St.,
                Palo Alto, CA 94304, USA;
                e-mail: aschwanden@lmsal.com }

\and

\author{        Yan Xu$^2$ and Ju Jing$^2$}

\affil{		$^2)$ Space Weather Research Laboratory, 
		Center for Solar-Terrestrial Research,
		New Jersey Institute of Technology,
		323 Martin Luther King Blvd., Newark, NJ 07102-1982, USA;
		e-mails: yan.xu@njit.edu, ju.jing@njit.edu }

\begin{abstract}
We present the first part of a project on the global energetics of
solar flares and coronal mass ejections (CMEs) that includes about
400 M- and X-class flares observed with AIA and HMI onboard SDO.
We calculate the potential ($E_p$), the nonpotential ($E_{np}$) 
or free energies ($E_{free} =E_{np}-E_p$), and the flare-dissipated 
magnetic energies ($E_{diss}$). We calculate these magnetic parameters  
using two different NLFFF codes: The COR-NLFFF code uses the line-of-sight 
magnetic field component $B_z$ from HMI to define the potential field, 
and the 2D coordinates of automatically detected coronal loops in 6 
coronal wavelengths from AIA to measure the helical twist of coronal 
loops caused by vertical currents, while the PHOT-NLFFF code 
extrapolates the photospheric 3D vector fields. We find agreement between
the two codes in the measurement of free energies and dissipated
energies within a factor of $\lapprox 3$. The size distributions of
magnetic parameters exhibit powerlaw slopes that are approximately
consistent with the fractal-diffusive self-organized criticality 
model. The magnetic parameters exhibit scaling laws for the
nonpotential energy, $E_{np} \propto E_p^{1.02}$, for the
free energy, $E_{free} \propto E_p^{1.7}$ and $E_{free} \propto
B_{\varphi}^{1.0} L^{1.5}$, for the dissipated
energy, $E_{diss} \propto E_p^{1.6}$ and $E_{diss} \propto E_{free}^{0.9}$,
and the energy dissipation volume, $V \propto E_{diss}^{1.2}$. 
The potential energies vary in the range of
$E_p = 1 \times 10^{31} - 4 \times 10^{33}$ erg, while the free energy
has a ratio of $E_{free}/E_p \approx 1\%-25\%$. The Poynting flux amounts to 
$F_{flare} \approx 5 \times 10^{8} - 10^{10}$ erg cm$^{-2}$ s$^{-1}$ 
during flares, 
which averages to $F_{AR} \approx 6 \times 10^6$ erg cm$^{-2}$ s$^{-1}$ 
during the entire observation period and is comparable with the 
coronal heating rate requirement in active regions.
\end{abstract}


\section{		INTRODUCTION				}

{\sl The whole is greater than the sum of its parts''}, mused Aristotle
to his disciples, written down in his {\sl Metaphysics} 350 B.C.E,
when he wondered whether there are additional substances besides
fire, earth, water, and air, that make up our universe.
His disciples may be intrigued about the mathematical paradoxon.
Exploring the global energetics of solar flares and associated
eruptive phenomena, we also wonder whether we can measure all
components of the energy output, partitioned by secondary processes, 
and whether the sum of their parts matches the whole of the energy input. 
In a nutshell, our concept or working hypothesis is that all primary
energy input is provided by dissipation of magnetic energies, 
which supply the energy output of secondary processes, such as
the thermal energies of the heated flare plasma, the nonthermal energies
of accelerated particles that produce hard X-rays, gamma-rays, or are
detected as {\sl solar energetic particles (SEP)}, and kinetic energies
of coronal mass ejections (CMEs). Regardless whether this hypothesis is true
or false, obtaining quantitative statistics of the different forms of
energies will be extremely useful for a host of reasons: (i) Do the known
forms of energy add up or do we miss important parts (similar to the
dark matter problem of the universe); (ii) do we have sufficient 
magnetic energy to supply all secondary processes, or are our
magnetic reconnection models insufficient; or (iii) what is the cause
and consequence, the efficiency and upper limits of various energy 
conversion processes? Essentially, every theoretical solar flare 
or CME model can be tested, evaluated, and disproved by energetic 
considerations.

The global energetics of solar flares and their energy partition
has been systematically addressed in some earlier studies, in two
papers on the energy partition of two solar flare/CME events
(Emslie et al.~2004, 2005), and in one on the global energetics of 38
large solar eruptive events (Emslie et al.~2012). Although these 
papers study various contents of energy, such as (i) the radiated
energy in soft X-rays detected by GOES, (ii) the total energy
radiated in soft X-rays, (iii) the peak energy in soft X-rays,
(iv) the bolometric radiated energy, (v) the non-thermal energy
in accelerated $>$20 keV electrons, (vi) in $>$1 MeV ions,
(vii) the kinetic energies in CMEs, and (viii) in solar energetic
particles (SEP) in interplanetary space, no measurement of the 
amount of available free magnetic energy that drives all these
energy conversion processes was attempted. Instead, the simple
(non-dissipative) magnetic potential field was calculated, and
an ad hoc value of 30\% was used to estimate the free magnetic
energy. In the meantime, various {\sl nonlinear force-free field
(NLFFF)} codes have been developed that are able to calculate the
free magnetic energy directly, either using information from
vector magnetic field measurements (e.g., Metcalf et al.~1995, 2005; 
Bobra et al.~2008; Jiao et al.~1997; Guo et al.~2008; Schrijver
et al.~2008; Thalmann et al.~2008, 2013), or employing tools
that take advantage of the geometry of coronal loops, which 
supposedly trace out the ``true'' coronal magnetic field
(Aschwanden 2013a,b,c; Aschwanden and Malanushenko 2013;
Malanushenko et al.~2011, 2012, 2014). Thus, the new capabilities
of calculating free magnetic energies as well as their decreases
during flares with high-quality HMI/SDO data, represent one 
important justification to investigate the global energetics of 
solar flares now. Another important reason is the unprecedented 
EUV and soft X-ray imaging capabilities of AIA/SDO,
which copiously display the twisted, sigmoid-like, and helical geometry
of coronal loops that define the non-potential magnetic field.
In addition, the EUV images provide spatial information (length scales, 
areas, and volumes of flares) which is a prerequiste to calculate
the volume-integrated thermal energies in flares. Moreover,
the EUV images from AIA/SDO and EUVI/STEREO yield also detailed
information on EUV dimming during the launch of a CME
(Aschwanden et al.~2009b), and thus allow us to determine masses, 
velocities, and kinetic energies of CMEs at any location on
the solar disk, while the traditional measurements of CME
masses using the polarized brightness of white-light images 
(e.g., Vourlidas et al. 2010) are only feasible near the 
solar limb.

\medskip
This first paper of a series on the global energetics of solar flares
is dealing with magnetic energies. Further studies will include thermal 
energies, non-thermal energies, kinetic energies of CMEs,
a comparative synthesis of the global energetics, and the application 
of the fractal-diffusive self-organized criticality model.
The organization of this first paper is as follows. Section 2 contains a brief
description of the two used magnetic calculation methods, the PHOT-NLFFF
and the COR-NLFFF codes, while more details about the COR-NLFFF code,
especially the recent improvements, are described in the Appendices
A (Automated tracing of coronal loops), B (Potential field parameterization),
C (Rotational invariance of magnetic field), and D (Forward-fitting
of non-potential fields). In Section 3 we present the observations and
statistical results, based on the data analysis of a comprehensive 
dataset that includes all 
M- and X-class flares observed with the SDO during the first 3.5 years of 
the mission. The results include measurements of the free energy and their
uncertainties, their time evolution, their timing, comparisons between
the two codes, scaling laws, geometric measurements, size distributions,
and the Poynting flux. In the Discussion in Section 4 we address aspects 
of measuring the coronal magnetic field, the illumination effects that
cause an apparent increase of free energies, a self-organized criticality
model, previous measurements of flare-dissipated energies, scaling laws
of magnetic energy dissipation, and aspects relevant to the coronal
heating problem. We summarize the Conclusions in Section 5.

\section{ 		DATA ANALYSIS METHODS		 	}

In this study we deal with two fundamentally different methods to
calculate magnetic energies of active regions during solar flare events,
which differ in their (photospheric versus coronal) origin of the 
observables. Photospheric codes calculate nonlinear force-free
fields (PHOT-NLFFF) based on an extrapolation of the photospheric 
3D magnetic field vectors ${\bf B}(x,y)=[B_x(x,y), B_y(x,y), B_z(x,y)]$ 
(Section 2.1), while the alternative method (COR-NLFFF) calculates 
coronal nonlinear force-free fields by fitting a force-free field
model to coronal loop geometries $[x(s), y(s)]$, which are obtained
by an automated feature detection method applied to multi-wavelength 
EUV images, as well as using the line-of-sight component $B_z(x,y)$ 
from a simultaneously observed magnetogram (Section 2.2).

\subsection{ The Photospheric Magnetic Field Extrapolation Method (PHOT-NLFFF)}

We use the photospheric vector magnetograms from the HMI onboard SDO 
(Scherrer et al.~2012; Hoeksema et al.~2014)
as the boundary condition for the photospheric non-linear 
forcefree field extrapolation method (PHOT-NLFFF). Since the photosphere
is not forcefree (Metcalf et al.~1995), 
while the corona is generally close to forcefree,
a pre-processing technique is used to make the photospheric boundary 
near-forcefree before extrapolation with the 3D NLFFF code 
(Wiegelmann et al.~2006, 2008; Wheatland and Regnier 2009).
For the pre-processing we use the weighted optimization method
of Wiegelmann (2004), which is an implementation of the original
work of Wheatland et al.~(2000). The extrapolations were performed
using non-uniformly rebinned magnetograms (approximately with a scale 
of 2 pixels $\times 0.5\arcsec = 1.0\arcsec$) within a computational domain
of $248 \times 248 \times 200$ (sometimes $248 \times 124 \times 200$)
uniform grid points, corresponding to $\approx (180 \times 180 \times
145)$ Mm$^3$, or $(180 \times 90 \times 145)$ Mm$^3$, respectively.
Since the Wiegelmann code requires a planar boundary perpendicular
to the line-of-sight, the vector magnetograms are de-rotated to 
the disk center and remapped using the Lambert (cylindrical) equal-area 
projection (see also Sun et al.~2012 and references therein).

The NLFFF extrapolation yields a 3D field 
${\bf B}({\bf x})=[B_x(x,y,z), B_y(x,y,z), B_z(x,y,z)]$ in each voxel
of the 3D cube. The free magnetic energy $E_{free}$ quantifies the
energy deviation of the coronal magnetic non-linear forcefree field 
${\bf B}_{np}$ from its potential field state ${\bf B}_p$, which is 
defined as
\begin{equation}
	E_{free}=E_{np}-E_p=\int \frac{(B_{np}^2-B_p^2)}{8\pi} dV \ ,
\end{equation}
where $V$ is the volume of the computational domain from the
photosphere to the corona, and the subscripts $np$ and $p$
represent the NLFFF and the potential field, respectively.
More details about the calculation of free magnetic energy
with the NLFFF code used here are given in Jing et al.~(2010). 

For the present study we calculated PHOT-NLFFF solutions at a
single time near the flare peak for 56 (out of the possible 172) 
flare events with $>$M1.0 GOES class, all at locations within 
$\le 45^\circ$ heliographic longitude difference to the central 
meridian. For the subset of the 11 largest events (of $>$X1.0 GOES 
class) we calculated a time series of free energies with a cadence 
of 12 minutes.  The calculation of a nonpotential field solution 
in a 3D computation box for a single time frame requires about 
$10-12$ hours computation time.

\subsection{    The Coronal Loop Fitting Method (COR-NLFFF)	}

An alternative method is the so-called {\sl Coronal Nonlinear Force-Free
Field (COR-NLFFF)} code, which we are using for the computation of 
nonpotential fields by employing coronal constraints. 
This novel method uses a line-of-sight magnetogram to define
a potential field solution, and applies forward-fitting of a
parameterized NLFFF model (in terms of vertical currents) to
the geometry of observed coronal loops, which supposedly trace out the true 
coronal magnetic field. The chief advantage of this alternative
model is that the NLFFF solution is not affected by the non-forcefreeness
of the photosphere and lower chromosphere (because it does not use 
the transverse photospheric magnetic field vectors in the extrapolation
of a NLFFF solution, as the standard PHOT-NLFFF codes do), and that the
obtained NLFFF solution matches closely the geometry of the observed 
coronal loops
(while standard PHOT-NLFFF codes have no capability to fit the
coronal loop geometry). Moreover, the COR-NLFFF code is orders of
magnitude faster than traditional PHOT-NLFFF codes, because
the COR-NLFFF model (in terms of vertical currents) represents 
an analytical NLFFF approximation (force-free to second order 
in the force-free $\alpha$-parameter) that can be forward-fitted 
fast and efficiently. The forward-fitting of a COR-NLFFF solution for
a single time frame is accomplished typically in $\approx 1-2$ min,
which is about 2-3 orders of magnitude faster than with a 
PHOT-NLFFF code.

Previous studies using this COR-NLFFF code
include the analytical derivation of a NLFFF solution in terms of 
vertical currents (Aschwanden 2013a), the numerical prototype code 
and tests with simulated data (Aschwanden and Malanushenko 2013),
calculations of the free energy (Aschwanden 2013b), potential-field
calculations of active regions (Aschwanden and Sandman 2010), 
nonpotential-field calculations of active regions using stereoscopic 
data (Aschwanden et al.~2012a), nonpotential forward-fitting 
with and without stereoscopic data (Aschwanden 2013c),
and comparisons of PHOT-NLFFF and COR-NLFFF solutions for the
2011 February 12-17 flares (Aschwanden, Sun, and Liu 2014a).
The COR-NLFFF code has been continuously developed over the 
last years, leading to substantial new improvements that are briefly
described in the Appendices A, B, C, and D. The COR-NLFFF code consists 
of three principal parts: (i) the automated tracing of coronal loops in 
EUV images (Appendix A), the potential-field parameterization 
(Appendix B), which is shown to be invariant to the solar rotation 
(Appendix C), and the forward-fitting of nonpotential fields (Appendix D).  

For the calculation of the free energy $E_{free}$ we use the
same definition as given for PHOT-NLFFF codes (Eq.~1), 
except that we correct for isotropic twist directions.
Since the analytical nonpotential field approximation includes only 
magnetically twisted field lines wound around vertical twist axes,
the obtained free energy $E_{\perp}^{free}$ is a lower limit
to the total free energy $E^{free}$. In order to obtain a first-order 
correction, we consider a current along a semi-circular loop or 
filament, which has a cosine-dependence along the loop and can be 
statistically included by introducing an {\sl isotropic twist 
correction factor} $q_{iso}=(\pi/2)^2 \approx 2.5$ (Aschwanden, 
Sun, and Liu 2014a),
\begin{equation}
         E^{free}_{\perp}   =  E^{free} \langle \cos(\theta)^{-2} \rangle
                            =  E^{free}  \left({2 \over \pi}\right)^2
                            =  E^{free} / q_{iso} \ .
\end{equation}
Thus we expect that the corrected free energy is about a factor of
$q_{iso}=(\pi/2)^2 \approx 2.5$ higher than the best-fit values of the
vertical-current free energies $E^{free}_{\perp}$.

\section{	OBSERVATIONS AND RESULTS			   } 

\subsection{AIA and HMI Observations}

The dataset we are analyzing for this project on the global energetics
of flares includes all M- and X-class flares observed with the
{\sl Solar Dynamics Observatory (SDO)} (Pesnell et al.~2011)
during the first 3.5 years of the mission (2010 June 1 to 2014 Jan 31),
which amounts to 399 flare events. The catalog of these flare events
is available online, see 
{\sl http://www.lmsal.com/$\sim$aschwand/RHESSI/flare$\_$energetics.html}.
Magnetic energies are determined
for events that have a heliographic longitude of $\lapprox 45^\circ$
(177 events), of which 5 events contained incomplete or corrupted AIA data, 
so that we are left with 172 events suitable for magnetic data analysis.
Using the COR-NLFFF code we calculate the evolution of
free (magnetic) energies for all of these 172 events, while a  
subset of 57 events is subjected to the (computationally 
more expensive) PHOT-NLFFF code also.

The analyzed SDO data set includes EUV images observed with the
{\sl Atmospheric Imaging Assembly (AIA)} (Lemen et al.~2012; 
Boerner et al.~2012), as well as magnetograms from the {\sl Helioseismic 
and Magnetic Imager (HMI)} (Scherrer et al.~2012).
The SDO started observations on 29 March 2010 and has produced 
essentially continuous data of the full Sun since then. 

AIA provides EUV images from four $4096 \times 4096$ detectors
with a pixel size of $0.6\arcsec$, corresponding to an effective
spatial resolution of $\approx 1.6\arcsec$. AIA contains ten different
wavelength channels, three in white light and UV, and seven EUV channels,
whereof six wavelengths (94, 131, 171, 193, 211, 335 \ang )
are centered on strong iron lines (Fe {\sc viii}, {\sc ix}, {\sc xii}, 
{\sc xiv}, {\sc xvi}, {\sc xviii}), covering the coronal range from 
$T\approx 0.6$ MK to $\gapprox 16$ MK.  AIA records a full set of 
near-simultaneous images in each temperature filter with a fixed cadence 
of 12 seconds. 

HMI provides full-disk magnetograms from measurements of the Doppler
shift using the 6173 \ang\ Fe {\sc i} absorption line. The HMI 
magnetograms are recorded with a $4096 \times 4096$ pixel camera
with a pixel size of $0.5\arcsec$, giving an overall cadence of
45 s for the Doppler velocity, intensity, and LOS magnetic field 
measurements, which have been processed from 135 s time intervals.

\subsection{    Example of a NLFFF Solution 		}

An example of a forward-fitting solution for one instant of time 
during the evolution of a solar flare is given 
in Fig.~1, which shows AR 11158 at heliographic position S21N12
observed on 2011 February 15, 01:14 UT, shortly before the 
GOES-class X2.2 flare. The theoretical field lines of the 
best forward-fit nonpotential field model (red curves in Fig.~1) 
are overlaid on the tracings
of observed loops (yellow curves in Fig.~1), calculated at the
intersection at the midpoints of each observed loop segment.
The observed magnetogram and the model with the decomposed magnetic
charges are shown in Fig.~20. The control parameters of the COR-NLFFF
code are listed
on the right side of Fig.~1, where also a histogram of the 
misalignment angles of all loops is shown, with a median of 
$\mu_2=6.8^{\circ}$. From the 6 AIA wavelength images, a total
of 508 loop structures were automatically detected, of which 
only 300 loops were used for forward-fitting, while a large
number (188) of structures with a ripple ratio of $q_{ripple} 
\ge 0.50$ have been discarded (which mostly contain 
``moss''-contaminted structures, see Appendix A).
The key results of this run are the median misalignment angle of
$\mu_2=6.8^{\circ}$ (which expresses the goodness-of-fit), 
the total potential energy of the active region,
$E_p=1.08 \times 10^{33}$ erg, the free energy,
$E_{free}= E_{np} - E_p = 1.15 \times 10^{32}$ erg, 
which yields a ratio of $q_{np}=E_{np}/E_p = 1.106$, 
or a free energy that is 10.6\% of the potential energy.

\subsection{    Uncertainties of Free Energy Measurements 	}

In Fig.~2 we show the time evolution of the magnetic free energy 
$E_{free}(t)$, the soft X-ray GOES 1-8 \ang\ flux profile
$F_{GOES}(t)$, the misalignment angles $\mu_2(t)$, and the number 
of detected $n_{det}(t)$ and fitted loops $n_{loop}(t)$. 
For all cases in our analysis we compute a time series with a length 
that covers the flare duration plus a half hour margin before and 
after the flare, with a cadence of 0.1 hours (6 min), yielding 
about 12-40 time frames per flare. For event \#12 shown in Figs.~1 and 2, 
we need 13 time intervals (with a step of $dt=0.1$ hr) to cover 
the time series.  

In order to quantify an error of the measurement of free energies
we vary most of the control parameters and find that the final
result of the free energy is somewhat sensitive to the loop 
selection criterion $q_{ripple}$ (Appendix A). Therefore we perform the 
forward-fitting with 3 different sets of loop selection parameters:
$q_{ripple}=0.25$ (RUN1; Fig.~2 left),
$q_{ripple}=0.50$ (RUN2; Fig.~2 middle), and 
$q_{ripple}=0.75$ (RUN3; Fig.~2 right).
This ``ripple criterion'' (Eq.~A2) discriminates between smooth
loop flux profiles ($q_{ripple} \gapprox 0$) and highly fluctuating
loop flux profiles ($q_{ripple} \lapprox 1$) that are most likely 
containing ``moss structures'' rather than loops, or a combination of
both (see Fig.~18 for examples of ``moss-like'' structures). 
A low value of the ripple criterion has the advantage of
selecting only ``good loop structures'', but has the disadvantage
that the number of selected loops is low, which may not be sufficient
in some cases to constrain the nonpotential energy near some magnetic
sources (see low ratio of fitted to detected loops in Fig.~2 
(bottom left panel). A high value, on the other side, 
contains more ``false (moss) structures'', but provides more 
statistics that enables ``good'' forward-fitting solutions
(see large ratio of fitted to detected loop structures in
Fig.~2: bottom right panel). Therefore, the optimum is
somewhere inbetween (say around $q_{ripple}\approx 0.5$). However, 
as the values shown in Fig.~2 demonstrate (see error bars in Fig.~2, 
top panel), the results of the
free energy $E_{free}(t)$ are fairly robust for different loop
selection parameters ($q_{ripple}=0.25-0.75$), and thus we use
the mean and standard deviation of these multiple 
(selection-dependent) solutions for error estimates of the free 
energy, $E_{free} \pm \sigma_{E,free}(t)$. 
We average the results from all the 3 trial 
runs (RUN1, RUN2, RUN3) and obtain a mean $E_{free}(t)$ and 
standard deviation $\sigma_{E,free}(t)$ of the free energy.
The uncertainty or error $e_{E,free}(t)$ of the mean value 
of the free energy (for each time frame $t$) is according to
standard statistics (Bevington and Robinson 1992)
\begin{equation}
	e_{E,free}(t) = { \sigma_{E,free}(t) \over \sqrt{N_{run}}} \ ,
\end{equation}
where $N_{run}=3$ for our 3 trial runs with different loop 
selection criteria. 

The evolutionary time profiles shown in Fig.~2 reveal a number
of interesting features of the COR-NLFFF code. The most instructive
property is that the free energy $E_{free}(t)$ systematically drops 
before the flare start time, for any chosen loop detection criterion 
(see error bars in Fig.~2, top, which are obtained from the scatter 
of the three runs), in contrast to the GOES flux that increases
slowly in the preflare phase and then grows rapidly during the
flare rise time (Fig.~2, second row). The drop in free energy is 
accompanied by a decrease of the nonlinear force-free parameter $\alpha_m$
for some of the strongest magnetic sources $m$, such as for the
magnetic charges $m=$0,1,2, and 3 (Fig.~2, third row).  The misalignment 
angle varies in the range of $\mu_2(t) \approx 5^\circ-10^\circ$ 
for the selection of smooth loops ($q_{ripple} \le 0.25$), while 
the misalignment is larger $\mu_2(t) \approx 8^\circ - 13^\circ$ 
for the selection of loops with a higher ripple ratio $q_{ripple}
\le 0.75$ (Fig.~2, fourth row), because of a higher contamination
of ``false loops'' or moss features. The largest misalignment occurs
around the flare peak time, when image saturation, pixel bleeding, and 
diffraction patterns occur and produce ``false loop structures'' 
that are difficult to completely remove (see details in Appendix A 
and Fig.~18). 

\subsection{ 	Time Evolution of the Coronal Free Energy		   }

Naively, we expect that the free energy $E_{free}(t)$ in a flaring 
active region has a near-constant value before a flare (because
build-up or storage of nonpotential magnetic energy is slow compared
with the rise time of a flare), which then 
decreases monotonically during the flare time interval, dropping 
to a lower level after the flare. We will see that this 
``single-step decrease'' behavior is sometimes observed in the free energy
when computed from photospheric field extrapolations (with PHOT-NLFFF
codes), but the measurements of the free energy based on coronal
loops (using COR-NLFFF codes) exhibit a more complex behavior 
that involves both apparent increases and decreases of the free 
energy before and during flare time intervals (Aschwanden, Sun, 
and Liu 2014a). The key aspect to understand this complex behavior 
is based on the fact that not all twisted
and current-carrying loop structures are illuminated before the
flare, and thus part of the free energy is invisible before the
flare. Once the flare starts, chromospheric evaporation gradually
fills up more and more helically twisted loops until all or most twisted
loop structures are ``illuminated'' and the full amount of free 
energy becomes detectable. At the same time, some free magnetic 
energy becomes dissipated during the flare, which is manifested
by decreases of the free energy. Thus, essentially we can interpret
the increases of free energy after flare start as temporary 
``coronal illumination'' effects, while decreases can be 
interpreted as episodes of ``magnetic energy dissipation''.  
In simple flares we expect then to observe one single increase
of the free energy at flare start, followed by a single decrease
during the flare rise time. In more complex flares, multiple phases
of illuminations and dissipations follow each other sequentially.
The simplest method to measure the total amount of dissipated
energy is then just to ignore the increases due to illumination
effects and to sum up all energy decreases during the flare time
interval. The principle of this method is illustrated in Fig.~3
for two subsequent illumination and dissipation phases. The
total amount of dissipated free energy is $\Delta E_{free}
=\Delta E_{3-2} + \Delta E_{5-4}$ in this example.

One might wonder whether the time intervals with decreases in the
free energy could be interpreted as an inverse illumination effect,
namely a disappearance of twisted flare loops by cooling, moving
a detected loop eventually out of the observed wavelength passband and
making it invisible again. This ambiguity, however, can entirely 
be ruled out during the risetime of soft X-ray or EUV light curves, 
because the rise time indicates a phase of increasing flux, emission 
measure, and electron density of the flaring structures, and thus
would contradict an interpretation in terms of cooling-related
flux decrease.

\subsection{ 	Measurement of Evolutionary Parameters 		   }

Based on the foregoing discussion we need to deconvolve the time
evolution of energy dissipation (which manifests as a temporary
decrease of free energy) from the illumination effects (which
is indicated by a temporary increase of free energy). In order
to achieve such a deconvolution we ignore the time steps with
increasing energy and derive a time profile $E_{neg}(t)$ that
includes only the negative energy decreases $dE_{free}(t)/dt < 0$,
which is shown in Fig.~4 (second panel), as derived from the
free energy time profile $E_{free}(t)$ of an observed flare (Fig.~4,
top panel), observed with SDO on 2012-Mar-07, 00:02 UT 
(event \# 147), being the largest flare of our analyzed dataset.
This monotonously decreasing time profile $E_{neg}(t)$ mimics
the free energy time profile that would be observed under ideal 
circumstances when all nonpotential loops would be illuminated 
in a coronal image (i.e., without illumination effects). 

The time profile of the dissipated energy $E_{diss}(t)$ 
(Fig.~4, third panel) can then be defined as a positive energy
by subtracting this monotonically decreasing energy time profile 
from its maximum value at the start time $t_0$,
\begin{equation}
	E_{diss}(t) = E_{neg}(t_0)-E_{neg}(t) \ ,
\end{equation}
or alternatively we can derive it directly from the free energy
profile $E_{free}(t)$ by adding the negative decreases as positive
increments, ignoring the positive increases,
\begin{equation}
	E_{diss}(t_i) = E_{diss}(t_{i-1}) + \left(
			[E_{free}(t_{i-1})-E_{free}(t_i)] > 0 \right) \ .
\end{equation}

In addition we define the magnetic energy dissipation rate
$F_{diss}(t)$, which is the time derivative of the dissipated 
energy $E_{diss}(t)$,
\begin{equation}
	F_{diss}(t_i) = {d E_{diss}(t_i) \over dt}
	= {E_{diss}(t_{i+1}) - E_{diss}(t_{i-1}) \over 2 \ dt_i} \ ,
\end{equation}
where $dt$ is the time step of the time profile, which is $dt=0.1$ hr
in our case. This energy dissipation rate is shown in Fig.~4 (fourth
panel, red-hatched curve) and coincides closely with the time 
derivative of the GOES 1-8 \ang\ time profile, $dF_{GOES}/dt$ 
(Fig.~4, fifth panel, blue-hatched curve), which is a good proxy for the
time profile of hard X-ray emission, particle acceleration, and the 
chromospheric thick-target heating rate in solar flares (Dennis and 
Zarro 1993; Brown 1972). Note that the two time profiles of energy
dissipation are determined from absolutely independent parameters:
the magnetic energy dissipation rate $F_{diss}(t)=dE_{diss}(t)/dt$ 
is entirely inferred from the geometry of untwisting coronal loops, 
while the GOES time derivative is derived from the soft X-ray
brightness of flare loops.

The example shown in Fig.~4 contains also error bars for each evolutionary
parameter at each time step, which we calculate based on the
uncertainties of the free energies $F_{free}(t)$ obtained from multiple
runs with different loop selections (Section 3.3), and using Monte-Carlo
simulations that propagate the errors using the definitions of the 
evolutionary time profiles given in Eqs.~4-6).

\subsection{	The Timing of Magnetic Energy Dissipation	   }

In the following we investigate the relative timing of the magnetic energy
dissipation rate $F_{diss}(t)=dE_{diss}/dt$ with respect to the flare 
peak time of the GOES soft X-ray flux. A representative subset of 
60 examples out of the 172 analyzed $>$M1.0 GOES-class flare events 
are shown in Figs.~5 to 8, in order of increasing complexity. 
For each case we show the temporal evolution of the (best-fit) 
magnetic free energy $F_{free}(t)$ and energy dissipation rate
$F_{diss}(t)$, along with the GOES 1-8 \ang\ light curve that
defines the flare start and end time (by NOAA convention), as well as 
its time derivative $dF_{GOES}(t)/dt$, which is a good proxy of 
the hard X-ray flux, the nonthermal emission, rate of particle acceleration,
and rate of chromospheric heating.

In a first group (Fig.~5) we show 12 examples with flare events where 
magnetic energy dissipation starts already before the GOES-defined
flare start. In some cases we see flare precursors in the GOES
time profile and its time derivative, which may indicate an early
trigger of magnetic energy dissipation.  Note, however, that the
free energy is generally not constant before the flare, but rather 
increases before flare start, probably thanks to an illumination 
effect of soft X-ray loops by gentle evaporation during the 
preflare phase.

In a second group (Fig.~6)
we show classic examples where the magnetic energy dissipation
coincides with the flare rise time of the soft X-ray (GOES) light
curve. This timing corresponds to our physical intuition that 
the major energy release phase of a flare occurs during the rise
time, when nonthermal particles are accelerated that heat the
chromosphere at the flare loop footpoints and drive chromospheric
evaporation according to the thick-target model (Brown 1972), 
which is manifested as a steady increase of the soft X-ray flux.
This scenario predicts a correlation between the soft X-ray
flux increase and the magnetic energy decrease, which is indeed
clearly fulfilled in the observed cases shown in Fig.~6.

In a third group (Fig.~7) we show examples where the peak of the
magnetic energy dissipation does occur slightly after the soft X-ray
flare peak, with continuing but decreasing energy dissipation in
the flare decay phase. This behavior could be explained by 
strongly driven chromospheric evaporation during the flare peak time,
which drives the illumination of twisted and soft X-ray emitting
flare loops with a higher emission increase rate than the magnetic
dissipation rate. Since the two effects of illumination and dissipation
are competing, it is no surprise that either one can be dominating
during particular time phases. 

In a fourth group (Fig.~8) we show cases with double flares
(although classified as a single event by NOAA), which
are clearly accompanied with two-step magnetic energy dissipation
phases. In these cases we can resolve two flare loop illumination
phases with two subsequent magnetic energy dissipation phases,
exactly as sketched in the cartoon of Fig.~3.

In a fifth group (Fig.~9) we show cases of complex flares, which
consequently have multi-step energy dissipation phases as a
consequence. These cases correspond to long-duration flares, which
typically last a few hours (say 1-4 hrs). These events were classified as
single flares according to the NOAA definition, but both the
soft X-ray time profile as well as the time derivative (i.e.,
the hard X-ray proxy) show multiple peaks that could possible be
considered as multiple flares. Nevertheless, because these events
last significantly longer and have many different soft X-ray and
hard X-ray emission peaks, it is natural that the free energy
dissipation rate reveals multiple phases also, which often 
exhibit a one-to-one correspondence. Note that the evolution of
the magnetic energy does not exhibit a monotonously dropping
staircase as a function of time, but rather an alternating
sequence of (illumination) increases and (dissipation) decreases.

\subsection{    Comparison of Photospheric versus Coronal NLFFF Results  }

For a subset of 57 flare events we calculated the potential,
nonpotential, and free energy with the PHOT-NLFFF code for one
single time frame. We compare these magnetic energies with the
COR-NLFFF code in form of scatterplots as shown in Fig.~10. We find
the following ratios: $q_{np}=E_{np}^{COR}/E_{np}^{PHOT}=0.998$
with a scatter by a factor of 1.4 for the nonpotential energy;
$q_{p}=E_p^{COR}/E_{p}^{PHOT}=1.080$ with a scatter by the same
factor of 1.4 for the potential energy, and
$q_{free}=E_{free}^{COR}/E_{free}^{PHOT}=0.343$ with a scatter 
by a factor of 2.2 for the free energy. Some differences can be
explained by slightly different field-of-views, but it is unknown
to what extent the pre-processing technique of the PHOT-NLFFF
code, or the deprojection of the magnetogram in the PHOT-NLFFF code
plays a role in the obtained absolute magnetic field strengths.
On the other side, a factor of $q_{iso} \approx 2.5$ has been
applied to the COR-NLFFF code to correct for isotropic twist
directions, which indeed improves the agreement between the two
codes, since the free energy would otherwise be a factor of
0.32/2.5=0.13 too low compared with the photospheric code.
We suspect that the slightly different spatial resolution
(2-pixel rebinning for the PHOT-NLFFF code versus 3-pixel
rebinning for the COR-NLFFF code) or unresolved twisted
magnetic structures could explain the slight underestimate
of the free energy.

For the subset of all (11) X-class flares we calculated the time
evolution of the free energy $E_{free}(t)$ with both the
PHOT-NLFFF and the COR-NLFFF codes, which are juxtaposed
in Figs.~11 and 12 and listed in Table 4. We find good agreement 
between the potential energies (within a factor $q_{E,pot} 
\lapprox 1.05\pm0.33$; Table 4). Differences in the potential
energy may partially be caused in the COR-NLFFF code by 
closely-spaced mixed magnetic polarities in the decomposition 
of the strongest fields in sunspots.

The agreement in the mean free energy is within a factor of 
$q_{E,free}=3.3\pm2.3$ (Table 4), which means that the PHOT-NLFFF
code detects about 3 times more free energy than the COR-NLFFF code. 
Systematic underestimates of the free energy with the COR-NLFFF code
may be be caused by (i) unresolved 
twisted structures, (ii) by an insufficient number of detectable 
coronal loops in magnetic field regions with high non-potentiality, 
or (iii) by the vertical-current approximation of our analytical 
NLFFF solution, which cannot model structures with horizontal 
twist axes, such as horizontal parts of helically twisted filaments.

The most important parameter is the decrease of the free energy
during the flares. The PHOT-NLFFF code does not detect a significant
decrease in one event (\#148). The COR-NLFFF code exhibits a highly 
significant decrease in all 11 cases (Figs.~11 and 12), while the
PHOT-NLFFF codes detects a factor of 
$q_{E,diss}=E_{diss}^{PHOT}/E_{diss}^{COR}=0.5\pm0.4$ less dissipated
energy, and thus the COR-NLFFF code appears to be more sensitive,
a pattern that was also found in previous work
(see Fig.~12 in Aschwanden, Sun, and Liu 2014a). 
Careful inspection of the time evolution of the free energy detected
with the PHOT-NLFFF code reveals sometimes increases shortly before
the flare (see event \#344 in Fig.~12, which may indicate new magnetic
flux emergence. We take such counter effects to energy dissipation
into account by using only the cumulative decreases of free energy
(thick blue curves in Figs.~11 and 12), the same way we do for
the COR-NLFFF code (thick red curves in Figs.~11 and 12). 

\subsection{ 	Scaling Laws of Magnetic Energies  }

In this study we present for the first time extensive statistics of
magnetic energies that are dissipated in M- and X-class flares.  The 
easiest magnetic quantity to measure is the total potential energy of 
an active region, for which we find a range of $E_{p}= 1 \times 10^{32}$ 
to $4 \times 10^{33}$ erg.  The other forms of magnetic energy are more
difficult to compute and hitherto could only be obtained with 
time-consuming runs of a PHOT-NLFFF code. In contrast, our COR-NLFFF 
code is much faster an can easily provide large statistics and 
useful scaling laws 
for the nonpotential magnetic parameters, which are listed in Table 3
and shown in Fig.~13.

The nonpotential energy of an active region is very closely correlated
to the potential energy, being a factor of $E_{np}/E_p=1.07 \pm 0.06$
or 7\% larger in the average. There is a slight nonlinearity between
the two parameters, which we determine with a linear regression fit
(Fig.~13 top left), 
\begin{equation}
	\left( {E_{np} \over 10^{30} {\rm erg}} \right) = 
	0.92 \left( {E_p \over 10^{30} {\rm erg}} \right)^{1.02} \ .
\end{equation}
This implies also that the free energy amounts to 7\% in the statistical
average, within a scatter by a factor of 2.3 (Fig.~13, top right). A linear 
regression fit reveals the following scaling law (fig.~13, top right panel),
\begin{equation}
	\left( {E_{free} \over 10^{30} {\rm erg}} \right)= 
	0.00034 \left( {E_p \over 10^{30}} {\rm erg}\right)^{1.73} \ ,
\end{equation}
so there is a strong nonlinearity of almost quadratic dependence. This
means that active regions with larger potential energy have an 
overproportional amount of free energy available for flaring. 
Note that a constant fraction of $E_{free}/E_p = 0.30$ was assumed in the
study of Emslie et al.~(2012), which matches our scaling law for the
very largest X-class flares only, but overestimates the free energy 
of M-class flares by about an order of magnitude.

The actually dissipated magnetic energy during a flare has a very
similar dependence on the potential energy, namely (Fig.~13, bottom left),
\begin{equation}
	\left( {E_{diss} \over 10^{30} {\rm erg}} \right)= 
	0.017 \left( {E_p \over 10^{30} {\rm erg}} \right)^{1.56} \ ,
\end{equation}
which implies that the magnetic energy $E_{diss}$ dissipated in a flare 
is almost identical to the available free energy $E_{free}$, within a scatter 
by a factor of 2.4 (Fig.~13, bottom right),
\begin{equation}
	\left( {E_{diss} \over 10^{30} {\rm erg}} \right)= 
	2.6 \left( {E_{free} \over 10^{30} {\rm erg}} \right)^{0.89} \ ,
\end{equation}
Note that the dissipated energy can exceed the free energy in our COR-NLFFF
code, because not all free energy is visible at the beginning of the flare.
In such cases, the dissipated free energy may still be accurate, but the
mean free energy averaged during the flare time interval is underestimated.
The ratio of free energies determined with the COR-NLFFF and PHOT-NLFFF code
differ indeed a factor of $E_{free}^{COR}/E_{free}^{PHOT} \approx 0.34$
(Fig.~10, bottom left panel). 

The scaling law of the free energy $E_{free}$ (Eq.~8) allows us to express the mean
twist angle $\varphi$ as a function of the potential field energy $E_{p}$ of an 
active region. The twist angle $\varphi$ is defined by the ratio of the
twisted azimuthal field component $B_{\varphi}$ to the radial potential field 
component $B_p$ by the relationship $\tan{(\varphi)}=B_\varphi/B_r$ 
(Aschwanden 2013a), and using the definition of the magnetic energies,
i.e., $E_p = B_p^2 / (8 \pi )$ and $E_{free} = B_{\varphi}^2 / (8 \pi)$
(Aschwanden 2013b) we obtain with Eq.~(8),
\begin{equation}
	\tan{(\varphi)} = \left( {B_{\varphi} \over B_p} \right) 
	= \left( {E_{free} \over E_p} \right)^{1/2}
	= 0.02 \left( {E_p \over 10^{30} {\rm erg}} \right)^{0.37} 
	\approx 1.2^{\circ} \left( {E_p \over 10^{30} {\rm erg}} \right)^{0.37} \ .
\end{equation}
According to this scaling law we expect mean twist angles of
$\varphi = 1.2^\circ, 2.8^\circ, 6.6^\circ, 15^\circ, 36^\circ$,
and $84^\circ$ for active regions with total potential
energies of $E_p = 10^{30}, 10^{31}, 10^{32}, 10^{33}, 10^{34}$,
and $10^{35}$ erg, respectively. The latter value 
corresponds to about the maximum possible twist and predicts a maximum
potential energy of $E_{p,max} \lapprox 10^{35}$ erg, which indeed 
represents a firm upper limit of all measured potential energies here,
as well as for the events studied in Emslie et al.~(2012).

\subsection{	Magnetic Energy Dissipation Areas 	   }

From the forward-fitting of the free energy $E_{free}(x,y,z)$ to the
geometry of coronal loops and their flare-related decreases 
$E_{diss}(x,y,z)$ we can obtain statistics on the spatial geometry 
of magnetic energy dissipation areas in flaring regions. An example of
the spatial distribution of the free energy before and at the peak of
the largest analyzed flare, observed on 2011-Feb-15, 01:40 UT,
is shown in Fig.~14 (top panels), with the evolution shown in Fig.~4 .
We show contours of constant free energies at levels of $E_{free}=(E_n-E_p)=
(B_{\varphi}^2/8\pi)$ corresponding to azimuthal magnetic field strengths
of $B_{\varphi}=5, 10, ..., 100$ G. In this example we witness an increase
of the energy dissipation rate by a factor of 13 with respect to half
an hour before the flare.

In order to characterize a geometric size $A_{diss}$ of the entire flare 
we determine a cumulative flare area $a_{cum}$ that contains all pixels
of the energy dissipation distribution $E_{diss}(x,y,t)$ that 
exceeded a threshold value $E_{thresh}$ at least ones during the
entire flare time interval. This procedure is identical to spatio-temporal
area definition of avalanche sizes in self-organized criticality models
applied in many other fields (e.g., Uritsky et al.~2002 on magnetospheric 
auroras).  The energy dissipation distribution
$E_{diss}(x,y,t)$ is defined by (negative) decreases of the
free energy during each time step $dt$,
\begin{equation}
	E_{diss}(x,y,t_i) = \left( E_{free}(x,y,t_{i-1}) - E_{free}(x,y,t_i)
	\right) > 0  \ .
\end{equation}
We define an energy dissipation area $a(t)$ above some threshold level 
$E_{thresh}=B_{\varphi}^2/(8\pi)$, with $B_{\varphi}=100$ G, unless the
maximum of the map $E_{free}(x,y,t)$ is below this threshold value, in
which case we take the full width at half maximum (FWHM) as a minimum 
width of the flare area. After counting all pixels above this threshold value
we obtain an instantaneous flare area map $a(t)$, from which we synthesize
a cumulative flare area $a_{cum}(t)$ that contains all partial flare areas
since flare start (with a margin of 0.5 hours earlier). The synthesized
cumulative area at the end of the flare (with a margin of 0.5 hours later)
represents then the total flare area,
\begin{equation}
	A = \sum a(t) = a_{cum}(t=t_{end}) \ .
\end{equation}
The cumulative flare area $a_{cum}(t)$ is a monotonously growing quantity,
$a_{cum}(t_i) \ge a_{cum}(t_{i-1})$. An example of this cumulative flare
area is shown for flare \#147 in Fig.~4 (bottom panel). The uncertainties
are calculated from the scatter between the three trial runs with different
loop selection parameters ($q_{ripple}=0.25, 0.5, 0.75$). 
We define also related geometric parameters by simple Euclidean
relationships, such as the flare length scale $L$,
\begin{equation}
	L = A^{1/2} \ ,
\end{equation}
and the flaring volume $V$,
\begin{equation}
	V = A^{3/2} \ .
\end{equation}

Since some active regions are located up to longitudes of $\le 45^\circ$ 
away from disk center of the visible hemisphere, we have to correct the 
projected areas (in the photosphere) with the cosine of the radial angle 
between Sun center and the heliographic position at longitude 
$(l)$ and latitude ($b$),
\begin{equation}
	A \approx {A_{proj} \over \cos{\sqrt{(l^2+b^2)}} } \ .
\end{equation}
	
In self-organized criticality models, the dissipated energy is often
assumed to scale with the size of an avalanche. Consequently, we expect
a correlation between the geometric flaring volume $V$ and the total
dissipated magnetic energy $E_{diss}$. We show a correlation plot
between these two measured parameters in Fig.~14 (bottom right) and
find almost proportionality. There is only a slight deviation from
proportionality that can be characterized by the scaling law (as
obtained from a linear regression fit between the logarithmic quantities),
\begin{equation}
	V \propto E_{diss}^{1.16} \ .
\end{equation}
For the scaling between the length scale $L$ and the dissipated energies
$E_{diss}$ we expect a powerlaw index that is 3 times smaller, i.e.,
$1.16/3=0.39$, which is indeed confirmed by a linear regression fit
(Fig.~14, bottom),
\begin{equation}
	L \propto E_{diss}^{0.39} \ .
\end{equation}
These scaling laws we quantified for magnetic dissipation area in solar flares
here for the first time, provide important information for physical
models of the energy release process (e.g., reconnection scaling law 
of Shibata and Yokoyama 1999).

\subsection{ 	Size Distributions of Magnetic Parameters  	   }

In Fig.~15 we plot the size distributions (or occurrence frequency 
distributions) of the various magnetic energy parameters measured here. 
Each size distribution follows a powerlaw at the upper end of the
distribution (and a roll-over at the lower end due to undersampling), as it is
typical for parameters of a nonlinear dissipative system that is governed
by self-organized criticality (Bak, Tang, and Wiesenfeld 1987), 
such as for solar flares (Lu and Hamilton 1991), in many wavelength
regimes (e.g., see recent review by Aschwanden et al.~2014b). Here
we measure the size distribution of magnetic parameters in solar 
flares for the first time and find the following powerlaw fits (Fig.~15):
The dissipated energies $E$, 
\begin{equation}
	N(E) dE \propto E^{-2.0\pm0.2} \ dE \ ,
\end{equation}
the peak energy dissipation rate $P$, 
\begin{equation}
	N(P) dP \propto P^{-2.3\pm0.2} \ dP \ ,
\end{equation}
the flare durations (measured from the GOES start and end times),
\begin{equation}
	N(T) dT \propto T^{-2.4\pm0.2} \ dT \ ,
\end{equation}
the flare length scale $L$,
\begin{equation}
	N(L) dL \propto L^{-3.8\pm0.3} \ dL \ ,
\end{equation}
the flare dissipation area $A$,
\begin{equation}
	N(A) dA \propto A^{-2.1\pm0.2} \ dA \ ,
\end{equation}
and the flaring volume $V$,
\begin{equation}
	N(V) dV \propto V^{-1.7\pm0.1} \ dV \ .
\end{equation}
The powerlaw slopes extend over 1-2 decades of the logarithmic values.
Our statistics is limited to $N=172$ events for which magnetic analysis
was suitable. From the geometric parameters, only the flare dissipation
area $A$ is directly measured, while the length $L$ and volume $V$ is
directly derived from the Euclidean relationships (Eqs.~14-15).

\subsection{ 		Poynting Flux  		 	   }

We provide also statistics on the azimuthal magnetic field
component $B_{\varphi}$, which is found to vary in the range of
$B_{\varphi}\approx 12-400$ G, and is strongly correlated with 
the dissipated flare energy $E$ (Fig~16, top left panel). Note,
that this azimuthal field component determines the free energy $E_{free}$
per voxel $dV$ of the computation grid,
\begin{equation}
	{dE_{free} \over dV} = {B_{\varphi}^2 \over 8 \pi} \ ,
\end{equation}
and is found to be nearly proportional to the total (volume-integrated)
dissipated energy,
$E_{free} = \int (B_{\varphi}^2 / 8\pi) dV$, as the scatterplot in Fig.~16
(top right panel) demonstrates. This implies that most of the magnetic
energy is contained in a compact core (that is of similar size in
different flares) around the location with the maximum azimuthal 
magnetic field strength $B_{\varphi}$, and does not scale with the
overall flare volume. 

Finally we calculate also the Poynting flux $F$, 
\begin{equation}
	F = {E \over A T} = {E \over L^2 T} \ ,
\end{equation}
which specifies the energy flux per unit area $A$ and time $T$,
where $E$ represents the total dissipated energy per flare.
The scatterplot in Fig.~16 (bottom left panel) shows that the
Poynting flux $F$ is somewhat correlated with the dissipated energy $E$
and has a range of $F \approx 5 \times 10^{8} - 10^{10}$ erg cm$^{-2}$
s$^{-1}$. 
A theoretical estimate of the Poynting flux into a reconnection region,
i.e., $F = v_{inflow} B^2/(4\pi)$, with $v_{inflow} \approx 0.1 v_A$,
$B \approx 100$ G, and $v_A \approx 1000$ km s$^{-1}$,
yields a similar value, $F \approx 8 \times 10^9$ erg cm$^{-2}$ s$^{-1}$.
Thus, the average Poynting flux 
during flaring time intervals exceeds that of the steady-state heating of 
the corona in active regions ($F \lapprox 10^7$ erg) by several orders 
of magnitude. The total duration of the 172 analyzed
flares is $T_{flare}=\sum_{i=1}^{172} T_i = 75.3$ hrs = $2.6 \times 10^5$ s,
which corresponds to an average flare duration of $<T_i>=0.43$ hr.

We can estimate the time-averaged Poynting flux in active regions by
dividing the total sum $E_{tot}$ of all flare-dissipated energies
by the average active region area $A_{AR}$ and the total time span $T_{tot}$ 
of observations, for which we obtain
\begin{equation}
	F_{AR} = {E_{tot} \over A_{AR} T_{tot}} 
	= 5.8 \times 10^6 \ \left( {L_{AR} \over 0.1 R_{\odot}} \right)^{-2} 
	\quad [{\rm erg}\ {\rm cm}^{-2} {\rm s}^{-1}] \ ,
\end{equation}
where the total energy $E_{tot}=\sum E_i = 3.1 \times 10^{34}$ erg is obtained
from summing all dissipated energies of each of the 172 flares (Table 3),
the total observing time is $T_{obs}= \sum T_i = 3.5$ years $= 1.1 \times 10^8$ s,
and the active region size $A_{AR}=L^2$ is normalized to the 
length scale of $L=0.1 R_{\odot} \approx 70,000$ km.
Interestingly, this average Poynting flux in active regions is close
to the average coronal heating requirement of $F_{heat}
\lapprox 10^7$ erg cm$^{-2}$ s$^{-1}$ (Withbroe et al.~1977),
which we will discuss in the context of the coronal heating problem
in Section 4.6.

\section{		DISCUSSION				   }

\subsection{ 	Measuring the Coronal Magnetic Field  		   }

It has often been stated that we have no direct method to measure the
coronal magnetic field, except for some special methods that can infer the
magnetic field at particular locations only, such as in some layers
above sunspots by means of gyroresonance emission 
(Alissandrakis et al.~1980), in the core of active regions by means of
polarized bremsstrahlung (Brosius and Holman 1988), 
both measured in radio wavelengths, by spectropolarimetry of forbidden
coronal lines in infrared (Lin, Kuhn, and Coulter 2004; Judge et al.~2001), 
or by coronal
seismology applied to oscillating loops (Roberts, Edwin, and Benz 1984).
With the two NLFFF methods used in this study, however, we have new
tools that are able to measure the 3D magnetic field ${\bf B}({\bf x})$
in a space-filling coronal volume that encompasses entire active 
regions. We introduced two methods in Section 2, the PHOT-NLFFF
method that uses the 3D magnetic field at the photospheric boundary as
input, and the COR-NLFFF method that uses a LOS-magnetogram $B_z(x,y)$
and the projected 2D coordinates of coronal loops. With these two methods,
the coronal magnetic field can be measured in principle in the entire corona.
However, challenges for the PHOT-NLFFF codes are the non-forcefreeness 
of the photosphere (DeRosa et al.~2009) and the heavy computational demands
(ca. 10-12 hrs per run), while the COR-NLFFF code does not have these 
problems, but may partially suffer from sparseness of suitable loop 
structures (uncontaminated by ``moss'') in the immediat proximity 
of sunspots, where the highest field strengths and thus the largest 
amounts of free energies are measured. Nevertheless, we improved the 
COR-NLFFF code substantially in recent times and obtained reasonable 
results of the measured magnetic energies in all analyzed flares. 
The computational efficiency of the COR-NLFFF code makes it possible 
to obtain these results fast for a large number of flares and many time 
steps (in the order of minutes per time step and active region.
While the present version of the COR-NLFFF code uses
an approximative NLFFF solution in terms of vertical currents, more
accurate NLFFF solutions have been carried out elsewhere (Malanushenko
et al. 2014) with similar results (though with manual rather than 
automated tracing of coronal loops, and with significantly longer
computation times). 

When we talk about magnetic energies in the solar corona, we have to be
aware that there are at least three different quantities that can be
measured, which are, in order of increasing difficulty: 
(i) the potential energy $E_p$,
(ii) the free energy $E_{free}$, or the non-potential energy 
which is the sum of the potential and free energy, $E_{np}=E_P+E_{free}$,
and (iii) the dissipated energy $E_{diss}=E_{free}(t_2)-E_{free}(t_1)$
during a certain time interval $\Delta t=t_2-t_1$. All these energies
are volume-integrated quantities, $E = \int (dE/dV) dV$, while the
energy density $(dE/dV)$ is directly related to the magnetic field 
by $(dE/dV)=B_{\varphi}^2/(8\pi)$. The potential energy density $(dE/dV)_p$ is
related to the potential field $B_p(x,y,z)$, while the free energy
density $(dE/dV)_{free}$ is related to an azimuthal magnetic field
component $B_{\varphi}(x,y,x)$ that is perpendicular to the potential field
component $B_p(x,y,z)$, because the definition of the nonpotential field,
$E_{np}=E_p + E_{free} = B_p^2 +B_{\varphi}^2$ implies that $B_p$ and
$B_{\varphi}$ are perpendicular, according to the Pythagoras' theorem
(Aschwanden 2013b). Consequently, the non-potentiality of a magnetic field
can easily be inferred from the misalignment angle 
$\mu = \arctan{(B_{\varphi}/B_p)}$ between the potential $B_p(x,y,z)$ and 
non-potential field $B_{np}(x,y,z)$. This misalignment angle is
constant along a uniformly helically twisted field line, which corresponds
to a constant $\alpha$-value of a nonlinear force-free field, and thus
constitutes a nonlinear force-free field (NLFFF) solution. Since the
automated tracing of coronal loops yields a direct measurement of the
projected misalignment angle $\mu_2$ in a 2D image observed in soft
X-rays or EUV, the COR-NLFFF method is particularly sensitive to
deviations of the nonpotential field from the potential field.
Previous assessments of the non-potentiality of active regions were
mostly based on visual inspection of EUV images and overlaid
potential field lines (e.g., Schrijver et al.~2005). 

What is the accuracy of our modeling of the coronal magnetic field ?
While the potential field $B_{r}$ (Eqs.~B1 and D2) is a zero-order 
approximation of the coronal magnetic field, the azimuthal field 
$B_{\varphi}$ (Eq.~D3) due to helical twist is a first-order approximation,
the associated free energy $E_{free} \propto B_{\varphi}^2$ is a 
second-order effect, and the decreases of free energy during a flare,
which we call the dissipated energy, is a third-order effect.
In Table 3 we provide uncertainties for all these measured magnetic
quantities. For the potential field energy, we find a scatter of 
$\sigma_B/B_p=0.05 \pm 0.12$ ($\approx 5\%$), which includes the 
variation of the potential energy during a flare time interval 
as well as the uncertainty in the forward-fitting of a NLFFF solution.
Comparing the COR-NLFFF with the PHOT-NLFFF code (Table 4), we find
a similar degree of accuracy, namely $E_p^{PHOT}/E_p^{COR}=1.05\pm0.33$
($\approx 5\%$). For the free energy, which is a second-order effect,
we find an agreement of $E_{free}^{PHOT}/E_{free}^{COR}=2.8 \pm 2.0$
(or a factor of $\lapprox 3$) (Table 4), which includes methodical
differences between both codes, such as uncertainties of the transverse
field component of the vector magnetograms and spatial averaging 
effects due to pre-processing in the PHOT-NLFFF method, as well as
sparseness of suitable loops free of moss contamination in the
proximity of sunspots and separation problems of closely-spaced
mixed magnetic polarities in the COR-NLFFF method. Even more important, 
for the
dissipated energy, which is a third-order effect, we find an agreement of
$E_{diss}^{PHOT}/E_{diss}^{COR}=0.5 \pm 0.4$ (or a factor of 2) between
the two methods. We note that the COR-NLFFF 
code is more sensitive than the PHOT-NLFFF code, and detects a 
significant amount of dissipated energy in all of the 172 analyzed
flares, with a significance ratio of $\sigma_{E,diss}/E_{diss}
=10 \pm 8$. As a caveat, we have to add that these results were derived
under the assumption that all decreases in the free energy during a
flare time interval are due to energy dissipation, and that all
energy increases during a flare time interval are due to illumination
effects (such as by chromospheric evporation), an assumption that we 
will discuss further in the following section.

\subsection{ Coronal Illumination Effects of Magnetic Structures   }

As the 60 examples of measurements with the COR-NLFFF code shown in 
Figs.~5-9 demonstrate, we almost never observe the naively expected 
scenario of a constantly elevated level of free energy before a flare, 
followed by a single-step decrease during the impulsive flare phase, 
with a constant depleted value afterward. On the other side, the
PHOT-NLFFF code shows in about half of the cases such a single-step
decrease behavior (Figs.~11-12), but detects a significantly smaller
decrease of free energy in the other half of the cases (Figs.~11-12),
which raises some questions about the sensitivity of the PHOT-NLFFF
code. It may be hampered due to the averaging effects of the
pre-processing technique (which tries to suppress the 
non-forcefreeness of the photosphere). So, what can explain this
different behaviour in the measurement of free energies of the COR-NLFFF 
method?

Let us discuss first the positive increases of free energy during
flaring time intervals. There are essentially two possibilities:
(i) incremental storage of free energy, either by continued twisting
of the magnetic field, or by new flux emergence with vertical currents,
or (ii) progressive illumination of nonpotential field structures, such as
twisted loops, sigmoids, or twisted filaments, manifested as brightening 
EUV structures, as it can be produced by chromospheric evaporation
in the thick-target scenario (e.g., Antonucci et al.~1982; Brown 1972).
The first argument can be largely eliminated by the argument of time
scales. A statistical study of the nonpotentiality of 95 active regions
has lead to the conclusion that the electric currents associated with
the nonpotentiality have a characteristic growth and decay time scale
of 10-30 hrs. Here we analyze the preflare time interval of 172 flares
over a much shorter time margin of 0.5 hrs, which is a factor of
20-60 times shorter than the characteristic growth and decay time of 
nonpotentiality,
and thus it can readily be neglected. Hence, the only obvious alternative
explanation of the observed increases of free energy is due to
chromospheric evaporation, which can illuminate twisted loop structures
in the preflare phase (Fig.~5), during the rise time of the impulsive
flare phase (Fig.~6), as well as during the decay time of the impulsive
flare phase (Fig.~7). Therefore, we ignore the time intervals with
positive increases in the calculations of the dissipated energy
(in terms of cumulative decreases of free energy). 
A sceptic may even raise the argument that positive
increases of the free energy could be caused by uncertainties in the 
forward-fitted NLFFF model. In order to convince ourselves that this
is not the case, we repeated each forward-fit with three substantially
different sets of loops (see Fig.~2) and obtain error bars that reflect
the uncertainty of the forward-fits due to loop selections, but we find
that these error bars are in most cases significantly smaller than
the cumulative positive energy increases during the flare time interval.  
This means that the energy increases are due to a systematic effect
that is significantly above the random noise of the NLFFF solutions.

What about the measured energy decreases of free energy. Are they all
due to energy dissipation, as we assume in our data analysis technique
(Fig.~4)? In principle, additional contributions to negative energy steps 
could arise from (i) decay of nonpotentiality, 
(ii) cooling of the flare plasma that renders twisted structures
(such as sigmoids, helical loops, or twisted filaments) invisible, or 
(iii) from random fluctuations in the forward-fitting method. Again, 
we can argue in terms of time scales. Statistical studies of
transient magnetic features associated with significant currents
in active regions decay on time scales of $\approx 27$ hrs
(Pevtsov et al.~1994),  $\approx 20$ hrs (Schrijver et al.~2005),
or 1-2 days (Welsch et al.~2011),
which is much shorter than the time interval of 0.5 hrs we analyze
after flares in our study. The second option of plasma cooling, 
can also largely
be ruled out by the argument of flare decay time scales observed
in EUV and soft SXR. Although the theoretical time scales of 
radiative and conductive cooling for a single loop structure can
be in the order of $\approx 0.2$ hrs (Rosner et al.~1978; 
Antiochos 1980, Culhane et al.~1994), the overall cooling time that
is observed in a postflare loop system amounts to $T = 0.4\pm 0.5$
hrs (as averaged from the flare durations listed in Table 3), which
is generally longer than the rapid decay times of $\Delta t_{free}
\lapprox 0.1$ hr that are seen for the decreases of free energy
(Figs.~5-10). This time scale ratio can be inspected in the 60 
time profiles that we show for the evolution of the free energy
$E_{free}(t)$ and the GOES light curve $F_{GOES}(t)$ in Figs.~5-10.
And the third argument can also be eliminated by the fact that the
error bars in the free energy solutions (Figs.~5-10 and Table 3)
are generally much lower than the negative energy jumps, which is
found to have a significance of $\sigma_{E,diss}/E_{diss}
=10 \pm 8$ (as averaged from Table 3).

Based on these arguments we justify the assumption made in our 
data analysis that the impulsive increases of free energy are largely
due to ``coronal illumination effects'', and the rapid energy increases
represent the energy dissipation of magnetic energies during flares
(as depicted in Fig.~3), which are caused by untwisting and relaxing 
of field lines after a magnetic reconnection process, according to
our model of free energy produced by vertical currents. 

\subsection{ 	Previous Estimates of Dissipated Flare Energies    }

Estimates of the free magnetic energy that is partially dissipated
in a solar flare have initially been made with the virial theorem, 
which yields an upper limit of twice the potential energy for a 
simple dipole field (Metcalf et al.~1995, 2005; Emslie et al.~2012). 
Other methods include flux-rope modeling (Bobra et al.~2008), which 
yielded a misalignment angle of $\mu_2\approx 10^\circ$ between the 
helical flux rope and the potential field, which translates into a 
free energy ratio of $E_{free}/E_p = (B_{\varphi}/B_r)^2 \approx
\tan^2{(10^0)} \approx 3\%$. Calculations with nonlinear force-free
field (NLFFF) codes using photospheric vector magnetograph data
have been carried out in a number of studies, yielding free energy 
ratios in active regions of 
$q_{free}=E_{free}/E_p \approx 30\%$ (Metcalf et al.~1995), 
$q_{free}=10\%$ (Jiao et al.~1997), $q_{free}=2\%$ (Guo et al.~2008),
a scatter of $q_{free}\approx -12\%$ to $+32\%$ from a test comparison
between 14 different NLFFF codes (Schrijver et al.~2008), 
$q_{free}=0.6\%-6.3\%$ (Thalmann et al.~2008), 
$q_{free}=9\%-36\%$ (Thalmann et al.~2013), 
$q_{free}=4\%-32\%$ (Malanushenko et al.~2014), 
$q_{free}=14\%$ (Sun et al.~2012). In summary, we can say that the
ratio of the free energy to the potential energy is found in a
range of $q_{free}\approx 0.4\%-25\%$. This fits well with our 
statistical result of 172 flares, where the ratio of free energy
is found in a range of $q_{free}\approx 0.6\%-36\%$ (Fig.~13, top right
panel), for potential energies in the range of 
$E_p \approx 1 \times 10^{31} - 4 \times 10^{33}$ erg.
Moreover, we find a scaling law of $E_{free} \propto E_p^{1.73}$ that
implies a near-quadratic dependence between the two quantities.

Now, the next question is what fraction of the free energy is dissipated
in solar flares, which requires to measure the evolution of the free
energy $E_{free}(t)$ during an entire flare event. This is a computationally
more challenging task and has been computed only for few cases.
Schrijver et al.~(2008) compared the free energies calculated by 14
different NLFFF codes before and after a flare, where only two codes
yielded a negative decrease of the free energy during the flare, in
the amount of $7\%-13\%$. Guo et al.~(2008) measure a decrease of
$\approx 2\%$ during an X3.4 flare. Thalmann et al.~(2008) measure
an energy decrease that corresponds to $q_{free}=2.3\%$ of the
potential energy, which translates into $E_{diss}/E_{free}
\approx 38\%$, so about a third of the free energy becomes dissipated
during the flare. Malanushenko et al.~(2012) obtains free energies
in the range of $q_{free} \approx 4\%-32\%$, but the reference
potential field changes during the flare, so that it is not trivial
to estimate the dissipated energy during the flare. Sun et al.~(2012)
provide a detailed study of the evolution of the free energy in
active region 11158 over 5 days and find the free energy decreases
from $q_{free}=29\%$ before the flare to $q_{free}=25\%$ after the
flare, so a fraction of $E_{diss}/E_p=4\%\pm1\%$ is dissipated,
which is about 14\% of the available free energy. 
Thus, these previous studies find that the actually dissipated energy
in flares amounts to $E_{diss}/E_{free} \approx 7\% - 38\%$.
In a statistical study of 38 eruptive flare events, the dissipated
energy in flares was assumed (ad hoc) to a fraction $E_{diss}/E_p \approx 30\%$
(Emslie et al.~2012). In a follow-on study (Aschwanden, Sun, and Liu 2014a)
we applied both a PHOT-NLFFF and a COR-NLFFF code, and found that the 
PHOT-NLFFF code underestimates the dissipated flare energy by a factor 
of $\approx 3-8$ from C-class to X-class flares, compared with the 
PHOT-NLFFF code, which is also consistent with the new findings of 
$E_{diss}^{PHOT}/E_{diss}^{COR} =0.5\pm0.4$ (Table 4). 
In our statistical study of 172 M and X-class flares here
we find that the amount of dissipated energy scales with the potential 
energy of the active region and follows a scaling of 
$E_{diss} \propto E_p^{1.56}$ (Eq.~9), which yields a ratio from 
$E_{diss}/E_{free} \approx 20\%$ at $E_p=10^{32}$ erg to   
$E_{diss}/E_{free} \approx 80\%$ at $E_p=10^{33}$ erg
(see also Fig.~13, bottom left).    
Note that the dissipated energy can exceed 100\% of the available 
free energy using our COR-NLFFF code, because parts of the free
energy is hidden in invisible loops that become illuminated around
the peak time of the flare only.

\subsection{ 	   Self-Organized Criticality Models		}

In this study we obtained for the first time statistical data on the
primary form of energy that is dissipated in solar flares. The dissipated
magnetic energy is believed to constitute the primary source of energy 
that supplies both flares as well as coronal mass ejection (CME) phenomena, 
while the conversion into thermal energies, non-thermal energies,
and CME motion represent secondary energy conversion
processes, each one consuming a partial amount of the primary energy.
Statistics of secondary energy processes in solar flares, such as hard X-ray 
emission, have been interpreted early on as a manifestation of nonlinear
energy dissipation processes that are governed by self-organized criticality 
(SOC) (Lu and Hamilton 1991), a concept that was originally developed to explain
the powerlaws of earthquake magnitude distributions, originally modeled with
cellular automaton models and sandpile avalanches (Bak, Tang, and 
Wiesenfeld 1987). A recent review on the application of this SOC concept
in solar and astrophysics summarizes the developments over the last 25
years (Aschwanden et al.~2014b).

We present the occurrence frequency distributions (also known as size 
distributions, which are generally plotted in log(N)-log(S) format)
of dissipated magnetic energies $(E)$, peak dissipation rates $(P)$,
durations $(T)$, length scales $(L)$, flare areas $(A$), and volumes $(V)$
in Fig.~15 and present the retrieved powerlaw scalings in Section 3.10.
Let us compare these results with the theoretical expectations.
The {\sl fractal-diffusive self-organized criticality model (FD-SOC)}
(Aschwanden 2012)
considers the spatial scale $(L$) as the most fundamental quantity of
SOC systems, which has a scale-free probability distribution of
$N(L) \propto L^{-3}$ in Euclidean space dimension $d=3$, which is
not too far from the observed value of $\alpha_L=3.75 \pm 0.26$
(Fig.~15, top right panel), given the small-number statistics. 
The associated size distributions for
areas are predicted to have values of $\alpha_A=1+(d-1)/2=2.0$
and $\alpha_V=1+(d-1)/d=5/3$ (Aschwanden 2012), which are consistent
with our measurements $\alpha_A=2.08 \pm 0.17$ and $\alpha_V=1.72\pm0.11$
(Fig.~15, right middle and bottom panel). For the dissipated energy $E$
and peak dissipation rate $P$, the FD-SOC model predicts powerlaw slopes
of $\alpha_E=3/2$ and $\alpha_P=5/3$, for Euclidean dimension $d=3$
and classical diffusiont $(\beta=1)$. Our observed values are
$\alpha_E=2.00\pm0.21$ and $\alpha_P=2.30\pm0.15$, which is somewhat 
steeper. The simplest version of the FD-SOC model assumes a
proportionality between the volume $V$ and total dissipated energy $E$,
which is not exactly true, because the scatterplot in Fig.~14 (bottom right
panel) indicates a slight nonlinearity of $V \propto E^{1.16}$, or inversely,
$E \propto V^{0.86}$, which can explain the differences to the values
predicted by the simplest version of the FD-SOC model. This result 
constrains physical scaling laws of the
energy release process, as we will discuss in the next section.

\subsection{	    Scaling Law of Magnetic Energy Dissipation	  }

Statistics of magnetic parameters can reveal physical scaling laws
of the magnetic energy release process, such as a particular type
of magnetic reconnection. An exhaustive list of magnetic scaling
laws for different types of coronal heating models has been compiled
in Mandrini et al.~(2000), which includes stressing models with a 
heating rate produced by stochastic build-up,
$E_H \propto B^2 L^{-2} V^2 \tau$ (Sturrock and Uchida 1981),
or for stressing models with a critical angle,
$E_H \propto B^2 L^{-1} V \tan(\varphi)$ (Parker 1988, Berger 1993),
where $B$ is the magnetic field, $L$ the length of a magnetic field
line, $V$ the volume, $\tau$ the time scale, and $\varphi$ the critical
shearing angle. Our NLFFF model involves a helically twisted field
line with vertical currents, for which the scaling law of the dissipated 
energy is defined as $E_{diss} \propto (B_\varphi)^2 V \propto 
B^2 \tan^2{(\varphi)} V$ (Aschwanden 2013b). However, we have to be
aware that this represents a microscopic scaling law that applies
to one single (helically twisted) magnetic field line, while a
macroscopic scaling law represents the integral over the entire
volume of an inhomogeneous active region. The microscopic parameters
in an inhomogeneous medium average out in such a way that 
macroscopic parameters, such as the average (azimuthal) magnetic
field strength $<B_{\varphi}>$ 
or the average length scale $<L>$ of loops produce a different
scaling law. The best we can hope is that the 
volume integration still preserves some scaling law between the
dissipated energy $E_{diss}$ and the maximum field strength $B_{\varphi,max}$
and length scale $L_{AR}$ of an active region. We test such a
hypothetical scaling law by defining two a priori unknown exponents
$\beta$ and $\lambda$,
\begin{equation}
	E_{diss}= \int \left({B_{\varphi} \over 8\pi}\right)^2 \ dV 
		\propto \ B_{\varphi,max}^\beta L_{AR}^\lambda \ .
\end{equation}
Since we measured the maximum field strength $B_{\varphi,max}$ and length
scale $L_{AR}$ of flare areas in active regions, we can
perform a linear regression fit for this scaling law, which yields
a slope $\gamma$ for an arbitrary choice of $\beta$ and $\lambda$, i.e.,
$E_{diss} \propto (B_{\varphi,\max}^\beta L_{AR}^\lambda)^\gamma$.
For the best fit of a hypothetical scaling law we demand a slope
of $\gamma=1$ and a minimum uncertainty $\sigma_{\gamma}$ of the 
fitted slope, 
which we can express with the goodness-of-fit criterion $\chi$
that needs to be minimized, 
\begin{equation}
	\chi = (\gamma -1) + \sigma_{\gamma} \ .
\end{equation}
We calculate a goodness-of-fit map $\chi(\beta, \lambda)$ in the
range of $-3 \le \beta, \lambda \le +3$ (Fig.~17, top), where we 
find a minimum value at $\beta=1.0$ and $\lambda=1.5$, for a slope of 
$\gamma=1.00\pm0.16$, with $\chi_{min}=0.16$. We show the 
corresponding best fit in Fig.~17 (bottom panel), which reveals
a scaling law of
\begin{equation}
	E_{diss} \propto \ B_{\varphi,max}^{1.0} L_{AR}^{1.5} \ .
\end{equation}
This scaling law applies also approximately to the dissipated
energy in a flare, since we found $E_{diss} \propto E_{free}^{0.9}$
(Eq.~10).

It is interesting to note that this macroscopic scaling law does not 
preserve the same exponents as the theoretical (microscopic) scaling 
law predicts for one single
twisted field line, i.e., $E_B \propto B^2 L^3$, but
only about the half value of the exponents, which is a consequence
of the averaging effects over a highly inhomogeneous active region
volume. In comparison, a scaling law of $F_H \propto B L^{-1}$ was
found for the heating flux $F_H$ from hydrostatic modeling of
a multi-loop corona in Schrijver et al.~(2004). The heating flux
$F_H$ (in units or erg cm$^{-2}$ s$^{-1}$) corresponds to a
volumetric heating rate of $H \propto F_H/L \propto B L^{-2}$
(in units of erg cm$^{-3}$ s$^{-1}$), or to a volume-integrated heating 
energy flux of $E_{heat} = F_H V = F_H L^3 = B L$ (in units of
erg s$^{-1}$). The time-integrated heating energy would then be
$E_{heat} = F_{heat} T \propto B L T$ (in units of erg).
We find that the flaring time scale $T$ is not significantly
correlated with any other parameter, and thus the empirical
scaling law $E_{heat} \propto B L$ of Schrijver et al.~(2004)
is similar to our scaling law of magnetically dissipated energies,
$E_{diss} \propto B L^{1.5}$.

\subsection{ 	      The Coronal Heating Problem  	 	}

We calculated the Poynting flux during the analyzed energy
dissipation episodes (or flares) and found values of $F \approx 
5 \times 10^{8} - 10^{10}$ erg cm$^{-2}$ s$^{-1}$ that occur temporarily,
averaged over the flare duration. If we average these Poynting
fluxes over the entire time span of observations (3.5 years) and
an average active region area with a length scale of 
$L_{AR}=0.1 R_{\odot}$, we find an average Poynting flux of
$<F_{AR}> \approx 5.8 \times 10^6$ erg cm$^{-2}$ s$^{-1}$. This
average Poynting flux meets the average coronal heating requirement
of active regions, which is commonly quoted as 
$<F_{heat}> \lapprox 10^7$ erg cm$^{-2}$ s$^{-1}$, which is needed to
balance the observed conductive and radiative losses from the
corona (Withbroe et al.~1977).

We have to be aware that this value of the Poynting flux with a
total dissipated magnetic energy of $\sum E_{diss} \approx 3 \times
10^{34}$ erg during 3.5 years represents the energy content of
172 flares with a magnitude of $\ge$ M1.0 GOES class only. 
As the size distribution of dissipated energies in Fig.~15 (top
left panel) shows, the dissipated energies exhibit a distribution
with a powerlaw slope of $\alpha_E \approx 2.0 \pm 0.2$ in the range
of $10^{32}$ and $10^{33}$ erg. If we extend this powerlaw
distribution down to the range of nanoflares with energies of
$E \gapprox 10^{24}$ erg (Parker 1988), the total dissipated
energy increases only by a factor of
\begin{equation}
	{\int_{E_{nano}}^{E_X} E \ N(E) \ dE \over 
	 \int_{E_M}^{E_X} E \ N(E) \ dE }
	= {\ln{(E_X/E_{nano})} \over \ln{(E_X/E_M)}} = 9 \ ,
\end{equation}
based on a size distribution of $N(E) \propto E^{-\alpha_E}$ with
a powerlaw slope of $\alpha_E \approx 2$, with lower energy limits
of $E_{nano}\approx 10^{24}$ erg for nanoflares, $E_{M} \approx
10^{32}$ erg for M-class flares, and an upper limit of
$E_X\approx 10^{33}$ erg for X-class flares (see Fig.~15 top left
panel). Thus, extrapolating the observed energy distribution
in the energy range of $[E_M, E_X]$ to the microflare and nanoflare
range $[E_{nano},E_X]$, we estimate a total dissipated magnetic 
energy that is about
a factor of 9 higher, i.e., corresponding to a Poynting flux of 
$<F_{AR}> \approx 4 \times 10^7$ erg cm$^{-2}$ s$^{-1}$.
It might be somewhat higher during more active
solar cycles, since the observed period of 2010-2014 belongs to
a relatively weak solar cycle. All previous estimates of the
energy budget for coronal heating were based on thermal energies
or non-thermal energies (e.g., Crosby et al.~1993; Shimizu 1995), 
which appear to be only a lower limit 
to the magnetic energy budget (Emslie et al.~2008). This explains
that some of those global energy estimates were found to be
slightly below the coronal heating requirement (see discussion
in Section 9.8.3 of Aschwanden 2004). 

In conclusion we find that the magnetic energy dissipated during
solar flares is sufficient to explain the coronal heating problem
in active regions. Since most parts of the Quiet Sun (essentially
all parts of the closed-field corona) are magnetically connected 
with active regions, the heating of the Quiet Sun can equally
be explained as a by-product of plasma heating in active regions.
Only coronal hole regions, which are not magnetically connected with
active regions, require a different mechanism to explain a (low)
coronal temperature ($T_e \lapprox 0.8$ MK) in such open-field regions.
We conclude that the dissipated magnetic energies in solar flares,
which are measured here with unprecedented statistics,
represent the most relevant constituent to identify the energy
source of coronal heating. Our results support the view that
the solar corona is largely heated by impulsive magnetic energy
dissipation processes that reduce the helical twist of the
stressed coronal magnetic field during solar flares, 
most likely faciliated by a magnetic reconnection process. 

\section{               CONCLUSIONS                             }

We started a project on the global energetics of solar flares,
using the most recent data from the SDO mission, which contains
about 400 GOES M and X-class flares during the first 3.5 years of
the mission. In this first study we measure the magnetic energy
that is dissipated during solar flares, for 172 events that are
located within a longitude of $\le 45^\circ$ from disk center. 
The major results and conclusions can be summarized as follows.

\begin{enumerate}
\item{We are using two complementary nonlinear forcefree field
(NLFFF) codes to measure the dissipated energies during flares.
The PHOT-NLFFF code uses the vector magnetic field (from HMI/SDO)
measured at the photospheric boundary and extrapolates the forcefree
field after pre-processing in order to improve the forcefreeness
condition. The COR-NLFFF code uses the line-of-sight (LOS)
magnetic field component $B_z$ from magnetograms (from HMI/SDO)
and the geometry of coronal loops as measured in EUV images
(from AIA/SDO) in six coronal wavelengths. The numerical procedure
of the COR-NLFFF code consists of 3 major steps: (i) the
decomposition of the magnetogram into buried magnetic charges that
define the potential field, (ii) automated tracing of coronal
loops to obtain the projected 2D coordinates of coronal loops, and
(iii) forward-fitting of an analytical NLFFF approximation, which 
is based on vertical currents that twist coronal loops, by varying
the nonlinear forcefree $\alpha$-parameters of the nonpotential field
until the misalignment between the model field lines and the
observed loop directions is minimized. The average misalignment
angle of all forward-fits is $\mu_2=8.6^\circ \pm 2.1^\circ$.}

\item{We measure the evolution of the potential energy $E_p(t)$,
the free energy $E_{free}(t)$, and the dissipated energy
$E_{diss}(t)$ during 172 flare events. While the PHOT-NLFFF code
mostly detects a step-wise decrease of the free energy during
most flares, the COR-NLFFF code detects both increases and
decreases of the free energy during flares. We interpret the
episodes of increasing free energy as ``coronal illumination 
effects" of twisted loop structures during the impulsive flare 
phase (such as by chromospheric evaporation), while the episodes
with decreasing free energies indicate the dissipation of 
magnetic energies, which can occur before flare start, during
the impulsive flare phase, and sometimes are even detected
during the flare decay phase.}

\item{Comparing the COR-NLFFF with the PHOT-NLFFF code, which
could be done only for 11 X-class flares due to computational 
time limitations, we find that the potential and nonpotential 
energies agree within a few percents for the average of all
cases, but vary by a factor of $\lapprox 1.4$ for individual
flares, which corresponds to a factor of $\lapprox 1.2$ in the
magnetic field. The agreement of the free energies varies
by a factor of $q_{free} = 3.3\pm2.3$, which
could be due to model assumptions of the COR-NLFFF code
(vertical currents with helical twist cannot reproduce
horizontally twisted structures), or numerical procedures
(pre-processing and heliographic deprojection) of the
PHOT-NLFFF code. The dissipated energies, which is a third-order
effect of the magnetic field model, agree within a factor of
$q_{diss} = 0.5\pm0.4$ between the two codes, where the COR-NLFFF code
is more sensitive and detects decreases of the free energy in
all 172 analyzed flares.}

\item{From the statistics of 172 events analyzed with the COR-NLFFF
code we find the following empirical scaling laws between the
magnetic potential energies ($E_p$), nonpotential energies $(E_{np})$,
the free energies ($E_{free}$), and the dissipated energies $(E_{diss})$:
$E_{np} \propto E_p^{1.02}$, $E_{free} \propto E_p^{1.73}$,
$E_{diss} \propto E_p^{1.56}$, and $E_{diss} \propto E_{free}^{0.89}$.
The mean twist angle in a flaring active region is related to the
potential energy of the active region by the relationship:
$\tan{(\varphi)} \approx (E_{free}/E_p)^{1/2} \approx 1.2^\circ
(E_p/10^{30}$ erg$)^{0.37}$. This relationship allows us to predict
the magnitude of the largest flare to occur in an active region
based on the average twist angle (or misalignment angle to the
potential field). Furthermore we found a semi-empirical scaling
law between the dissipated energy $E_{free}$, the maximum (azimuthal)
magnetic 
field strength $B_{\varphi,max}$, and the length scale $L_{AR}$ of the 
active region: $E_{free} \propto B_{\varphi}^{1.0} L^{1.5}$, which is similar
to a scaling law found by Schrijver et al.~(2004) for coronal 
heating, i.e., $E_{heat} \propto B L$.}

\item{The size distributions, which we derive here for the first time
for magnetic parameters, are found
to have the following powerlaw slopes: $\alpha_E=2.0\pm0.2$ for
dissipated energies, $\alpha_P=2.3\pm0.2$ for the peak energy
dissipation rate, $\alpha_T=2.4\pm0.2$ for the flare duration,
$\alpha_L=3.8\pm0.3$ for flare length scales, 
$\alpha_A=2.1\pm0.2$ for flare areas, and
$\alpha_V=1.7\pm0.1$ for flare volumes.
The flare volume $V = L^3$ and the dissipated flare energy $E_{diss}$
are found to scale as $V \propto E_{diss}^{1.16}$. These results are
approximately consistent with the predictions of the fractal-diffusive 
self-organized criticality (FD-SOC) model. Since SOC models describe
the statistics of nonlinear energy dissipation processes, the 
measurement of primary energy parameters, such as the dissipated 
magnetic energy measured here, are more important
then secondary energy parameters, such as thermal or nonthermal
energies in solar flares that have been subjected to SOC models
previously.}

\item{The Poynting fluxes of dissipated magnetic energies are found
to have values in the range of 
$F \approx 5 \times 10^{8} - 10^{10}$ erg cm$^{-2}$ s$^{-1}$ 
during flare time intervals. The sum of all magnetic energies
dissipated in solar flares is $E_{tot} \approx 3 \times 10^{34}$ erg
during the 3.5 years of observations, yields a temporally
and spatially averaged flux of $P \approx 6 \times 10^6$ 
erg cm$^{-2}$ s$^{-1}$ for a mean active region size of $L=0.1\ R_{\odot}$,
and of $P \approx 4 \times 10^7$ erg cm$^{-2}$ s$^{-1}$ when extrapolated
down to the nanoflares. 
This amount of dissipated magnetic energies is sufficient to explain
coronal heating in active regions (and quiet-Sun regions). 
Previous estimates of the global energy budget of the solar corona
were based on thermal and nonthermal energies, which represent lower
limits to the dissipated magnetic energy only and thus underestimate
coronal heating rate. Our results support the view that
the solar corona is largely heated by impulsive magnetic energy
dissipation processes that reduce the helical twist of the
stressed coronal magnetic field during solar flares.}
\end{enumerate}

The comparison between two completely different NLFFF codes has 
demonstrated that both codes yield commensurable results (within
a factor of $\approx 3$), which gives us more confidence in either code.
In future studies we will calculate other forms of energies obtained
during flares, such as thermal energies of the heated flare plasma,
non-thermal energies of accelerated hard X-ray producing particles,
and kinetic energies of CMEs. We will investigate whether those
secondary energy products add up to the total dissipated magnetic 
energies inferred here, and what the relative energy partition 
in the various flare processes is. The unprecedented statistics of
flare energies may reveal the underlying physical scaling laws that
govern flares and CME processes.

\bigskip
\acknowledgements
We thank the referee for insightful comments and we appreciate 
helpful discussions with Bart De Pontieu, Mark DeRosa, Brian
Dennis, Gordon Emslie, Allen Gary, Anna Malanushenko, 
Aidan O'Flannagain, Karel Schrijver, Daniel Ryan, Manuela Temmer, 
Astrid Veronig, and Brian Welsch. 
Part of the work was supported by NASA contract NNG 04EA00C of the 
SDO/AIA instrument and the NASA STEREO mission under 
NRL contract N00173-02-C-2035.
YX and JJ are supported by  NSF AGS-1345513, 1153424, NASA NNX11AQ55G 
and NNX13AG13.

\section*{ APPENDIX A: Automated Tracing of Coronal Loops	}

The key input of the COR-NLFFF code is the geometry of coronal loops,
which can be measured in 2D images in form of cartesian 
coordinates $[x(s), y(s)]$ as a function of a loop length coordinate
$s$ from highpass-filtered EUV images, 
or in form of 3D coordinates $[x(s), y(s), z(s)]$
from stereoscopic reconstruction. In principle, 3D coordinates would
be preferable because they provide stronger and more unique constraints
for any type of loop modeling (Aschwanden 2009, 2011; 
Aschwanden et al.~2008a, 2008b, 2009a, 2012a, 2012b), but 
the inferior spatial resolution of the EUVI/STEREO
imagers (W\"ulser et al.~2004), compared with AIA/SDO
(Lemen et al.~2012), and the restricted time range suitable 
for small-angle stereoscopy (Aschwanden et al.~2012) make it 
impractical. 

There exists a (Grad-Rubin method) COR-NLFFF code that calculates 
a NLFFF solution by fitting the geometry of coronal loops 
(Malanushenko et al.~2009, 2011, 2012, 2014), but its application
is restricted to manually traced loops and was applied to 
very few flares only.  Therefore, automated loop tracing is a 
prerequisite for efficient and objective NLFFF forward-fitting 
codes. The pioneering phase of automated
coronal loop tracing started with an initial comparison of the
performance of five different methods (Aschwanden et al.~2008c).
One of these codes, the {\sl Oriented Coronal CUrved Loop Tracing
(OCCULT)} code, was further developed by specializing the automated
pattern recognition of curvi-linear features to the geometric
property of large curvature radii, which achieved a performance
close to visual perception (Aschwanden 2010).
The guiding criterion of the oriented-directivity method for
curvi-linear tracing was then further refined by including
second-order terms (OCCULT-2; Aschwanden, DePontieu, and Katrukha 2013).
In a recent study with AIA/SDO data, the automated loop tracing
was extended to all available 7 coronal wavelengths (94, 131, 171,
193, 211, 304, 335 \ang; Lemen et al.~2012), and the effect of
loop selection in different filters on the NLFFF solution was
investigated (Aschwanden et al.~2014). 

The basic steps of the OCCULT-2 automated loop tracing code are: 
(a) read EUV images in 6 coronal wavelengths and apply a 
highpass-filter (with a typical highpass boxcar of $nsm_1=3$
and lowpass boxcar of $nsm_2=5$; (b) evaluation of a flux
threshold based on the flux mean and standard deviations 
in 10$\times$10 macropixels; (c) automated tracing using the
control parameters of: minimum curvature $r_{min}=25$ pixels,
minimum structure length $l_{min}=25$ pixels; maximum gap
along coherent structure $n_{gap}=3$ pixels, and threshold 
of $q_{thresh,2}=3$ times the median flux of the background;
(d) coordinate transformation of pixel units into units of
solar radii relative to Sun center; and (e) rejection of
unwanted loop structures.

From previous experience we learned that the convergence of a 
forward-fit of a NLFFF solution can be substantially degraded if 
there is a significant amount of false loop structures. It is
therefore imperative to remove as many false loop structures as
possible, when using an automated pattern recognition code. 
In the present study we improved the loop selection
criteria further by automated feature detection of 6 types of false
loop structures that are visible in EUV images. 

(1) Curvi-linear structures that have no footpoint directly 
connected to a magnetic source (sunspot or magnetic flux 
concentration in the magnetogram) are likely to be false loop
structures, because most coronal loops are best visible at their 
footpoints, where usually the maximum of the electron density, 
emission measure, and EUV brightness occurs, due to the 
hydrostatic stratification. We detect such unwanted loop structures
by the following magnetic proximity requirement, 
$$
	\left[ (x_{foot,i}-x_{mag,j})^2 + (y_{foot,i}-y_{mag,j})^2 
	\right]^{1/2} \le d_{foot} = 0.015 \ R_{\odot} \ ,
        \eqno(A1)
$$
where $(x_{foot,i},y_{foot,i})$ is the starting point ($i=1$) or
end point ($i=n_s$) of a curvi-linear structure, and $(x_{mag,j}, 
y_{max,j}), j=0,...,n_m$ is the image position of the next buried 
magnetic charge (decomposed from the line-of-sight magnetograms)
of any of the $n_m$ magnetic charges. About 3.7\% of automatically
detected curvi-linear structures do not meet the magnetic proximity
condition (Table 1). 

(2) Parts of some AIA/SDO images contain saturated pixels
at the datanumber limit (i.e., $> 2^{14}$ DN/s in 2-byte encoded 
images), which can produce curvi-linear features along the boundaries 
of saturated areas in the CCD images (often occurring during the
peak time of flares), which is found in about 0.3\% of automatically
detected structures (see Fig.~18; red curves). 

(3) Saturated images display also ``bleeding pixels'', which 
are manifested in form of vertical streaks in the CCD readout, 
which we found to produce about 0.1\% false loop structures (Table 1).

(4) EUV images of active regions display often ``moss structure''
(Berger et al.~1999), which is a reticulated spongy fine structure
that indicates the footpoint transition regions of hot coronal loops. 
Often, cooler coronal loops overlay fields with moss structures, which
produces a ripple of the EUV flux profile $F(s)$ along the loop.
Moreover, chains of moss dots often form a curved structure by chance
coincidence and lead to false loop detections. We eliminate such false
moss structure by a ripple criterion,.
$$
	q_f = {1 \over (n_s-1)} \sum_{i=0}^{n_s} 
	{|F(s_{i})-F(s_{i+1})| \over max[F_{s_i}, F_{s_{i+1}} ]}  
	\le q_{ripple} \ ,
	\eqno(A2)
$$
which essentially quantifies the average degree of fluctuations,
modulation depth, or smoothness of a flux profile $F(s)$ along the
loop coordinate $0 < s < L$. Flux profiles that are absolutely smooth
have a ripple ratio of $q_f \approx 0$, while strongly fluctuating
flux profiles have a maximum ripple ratio of $q_f \lapprox 1.0$.
We perform three types of runs (RUN1, RUN2, RUN3) with different 
ripple ratio limits of $q_f <$ 0.25, 0.50, or 0.75, respectively. The ripple
ratio limit has the biggest influence on the number of selected loops,
ranging form 22\% for smooth loops with $q_f \le 0.25$ (RUN1) to
80\% for very inhomogeneous loops with $q_f \ge 0.75$ (RUN3) (Table 1).  
We apply the ripple criterion only to short structures ($L \le 2 l_{min}
=50$ pixels $\approx 0.03 R_{\odot}$), because longer structures
are much less likely to form a regularly curved loop structure by 
chance coincidence (Fig.~18; green curves). 

(5) A particular instrumental effect is the diffraction pattern
that occurs from the EUV entrance mesh filter at high brightness
levels during flares, which is detected from a clustering
of directivity angles either in parallel or perpendicular direction 
in a directivity histogram. This applies to about 1.9\% of the 
automatically detected loops (Table 1).

(6) After calculating a potential field ${\bf B}_p(s)$, we can 
measure the 2D misalignment angles between the potential field 
(projected in the plane-of-sky) and the automatically traced loops.
Structures that have a large 2D misalignment angle to the potential
field, say $\mu_2 > 45^\circ$, are unlikely to fit a non-potential
field, which we discard also in the forward-fitting of our NLFFF model.
This is the case in about 1.4\% of the automatically detected loops
(Table 1).

Some examples of such automatically detected structures are shown 
in Fig.~18, including coronal loop structures (blue curves),
rippled (moss) structures (green curves), and boundaries of saturated 
image areas (red curves).

\section*{ APPENDIX B: Potential Field Parameterization	}		

In contrast to standard potential field codes, which
generally extrapolate a potential field using the eigenfunction
(spherical harmonic) expansion (Green's function) method, originally 
derived by Altschuler \& Newkirk (1969) and Sakurai (1982), the
COR-NLFFF code deconvolves a line-of-sight magnetogram into a finite
number of buried unipolar magnetic charges (Aschwanden \& Sandman 2010).
The chief advantage of the magnetic charge decomposition method is that
it automatically provides also a suitable parameterization for
NLFFF solutions with vertical currents, which can be defined 
for each unipolar magnetic charge and can be forward-fitted efficiently. 

The decomposition of a potential field into uni-polar magnetic charges 
is defined in terms of $m=1,...,n_m$ sub-photospheric locations 
($x_m, y_m, z_m$) and a vertical field strength $B_m$ at the
photospheric surface, vertically above the buried magnetic charge.
The field strength $B(r)$ of each unipolar magnetic source decreases 
with the square of the radial distance $r$. A arbitrary large number 
$n_m$ of magnetic charges can be superimposed, which yield the
resulting potential field ${\bf B}_p$,
$$
        {\bf B}_p({\bf x}) = \sum_{m=1}^{N_{\rm m}} {\bf B}_m({\bf x})
        = \sum_{m=1}^{N_{\rm m}}  B_m
        \left({d_m \over r_m}\right)^2 {{\bf r_m} \over r_m} \ ,
	\eqno(B1)
$$
where $r_m=[(x-x_m)^2+(y-y_m)^2+(z-z_m)^2]^{1/2}$ is the distance of an
arbitrary coronal location ${\bf x}=(x,y,z)$ to the subphotospheric charge
location $(x_m, y_m, z_m)$, while $d_m=1-[x_m^2+y_m^2+z_m^2]^{1/2}$ is the
depth of the buried charge, and $B_m$ is the magnetic field strength
at the solar surface in vertical direction above the buried charge.
The square-dependence of the radial field component $B(r) \propto r^{-2}$
warrants that each magnetic charge fulfills Maxwell's divergence-free
condition,
$$
        \nabla \cdot {\bf B} = 0 \ ,
	\eqno(B2)
$$
which it is also true for the summed magnetic field according to Eq.~(B1),
because the linear superposition of divergence-free fields is
divergence-free too, i.e., $\nabla \cdot {\bf B} = \nabla \cdot
(\sum_m {\bf B_m}) = \sum_m (\nabla \cdot {\bf B_m}) = 0$.

The decomposition of a LOS magnetogram $B_z(x,y)$ into a finite
number $n_m$ of magnetic charges is carried out by iterative
decomposition of local maxima of the magnetic field into individual 
magnetic charges, each one yielding four model parameters,
$(B_m, x_m, y_m, z_m), m=1,..., n_m$. The numerical procedure is 
demonstrated in Aschwanden and Sandman (2010), and an analytical 
treatment is derived in Appendix A of Aschwanden et al.~(2012a).  

We start with the absolute peak in the magnetogram, which is measured 
at the location $(x_p, y_p)$ and has the value $B_z$ for the LOS 
component of the magnetic field vector. We extract then a local magnetogram
map around this peak that has an extension of $(w \times w)$, where
$w$ corresponds to the numerically determined full width at a level
of 25\% of the peak flux. From the observables $(B_z, x_p, y_p)$ and the 
variable $d_m$ for the depth of the buried magnetic charge we can
calculate the projected disk center distance $\rho_p$, the LOS 
coordinate $z_p$ at the photospheric height, the angle $\alpha$
between the LOS and solar surface vertical, the angle $\beta_p$ 
between the solar surface vertical and the LOS field component
$B_z$, which yield then the field strength $B_m$ and the coordinates 
($x_m, y_m, z_m, r_m, \rho_m$) of the buried magnetic charge $m$
(see Fig.~19 and Eqs.~A1-A11 in Aschwanden et al.~(2012a),
$$
        \begin{array}{ll}
	\rho_p  &=\sqrt{(x_p^2+y_p^2)} \\
	z_p	&=\sqrt{1 - \rho_p^2} \\
        \alpha  &\approx \arctan({\rho_p / z_p}) \\
        \gamma  &=\arctan({y_p / x_p}) \\
        \beta   &=\arctan{\left[ \left( \sqrt{9 + 8 \tan^2 \alpha}-3 \right)
                  / 4\ \tan{\alpha} \right]} \\
        B_m     &={ B_z / [\cos^2{\beta} \ \cos{(\alpha-\beta)}]} \\
        r_m     &=(1-d_m)       \\
        \rho_m  &=\rho_p - d_m {\sin{(\alpha-\beta)} /
                \cos{\beta} } \\
        z_m     &=\sqrt{r_m^2-\rho_m^2} \\
        x_m     &=\rho_m \ \cos{\gamma} \\
        y_m     &=\rho_m \ \sin{\gamma} \\
	\end{array}
        \eqno(B3)
$$
While the width $w$ was obtained as a direct observable in Aschwanden et 
al.~(2012a), we found a more robust procedure here by varying the depth
parameter $d_m$ until the spatially integrated unsigned magnetic flux 
of the local peak map yields the best match between the model and 
the observed local map $B_z(x,y)$,
$$
	\Phi = \int |B_z^{obs}(x,y)|\ dx\ dy 
	= \int |B_z^{model}(x,y; w)|\ dx\ dy \ .
	\eqno(B4)
$$
In this way we obtain an inversion of the observables $(B_z, x_p, y_p)$
and by varying $d_m$ to find the model parameters $(B_m, x_m, y_m, z_m)$.
After the deconvolution of the global maximum in the magnetogram,
which yields the first 4 parameers of the model map, we subtract the
model distribution $B_z(x,y)$ of the first magnetic source and
continue in the same way by iterating additional magnetic source components.
Since the magnetogram has positive and negative magnetic field values,
the iteration is performed at the unsigned magnetogram, while the
correct sign of the magnetic polarity is applied to each deconvolved
component. Typically, a number of $n_m \approx 100$ magnetic
sources is sufficient to obtain a realistic potential field model of
a solar active region. In the end we renormalize the total unsigned
magnetic flux of the model magnetogram to that of the observed
magnetogram, in order to compensate for numerical residuals, which is
typically in the order of a few percents. An example of a unipolar
magnetic charge decomposition is shown in Fig.~20, for the same
observation as shown in Fig.~18 (first panel). Note the negligible difference 
(Fig.~20 top right) between the observed (Fig.~20, top left) and
the model magnetogram (Fig.~20, bottom left), which is also
visualized with a 1D-scan across the sunspot with maximum magnetic 
field strength (Fig.~20, bottom right).

\section*{ APPENDIX C:  Rotational Invariance of Magnetic Fields	}

Most of the existing PHOT-NLFFF codes require a cartesian coordinate
system with a planar boundary at the bottom of the computation box,
oriented in perpendicular direction to the line-of-sight of the 
observed magnetogram. Active regions with a heliographic position 
that are some distance away from Sun center are therefore de-rotated
to the disk center and remapped using the Lambert (cylindrical) 
equal-area projection (see also Sun et al.~2012 and references 
therein). Since the accuracy of the LOS component $B_z$ is much
higher than that of the transverse components $(B_x, B_y)$, a
de-rotation of the magnetogram implies also a variable weighting 
in the accuracy of the horizontal and vertical magnetic field 
components. In the extreme case of an active region near the
solar limb, the horizontal field component $B_x$ is measured
with the highest accuracy, while the other horizontal component
$B_y$ and the vertical component $B_z$ are measured much less
accurately. The accuracy of measuring vertical currents thus
varies considerably from center to limb.
To our knowledge, no validation test has been done
to demonstrate whether a NLFFF solution is invariant to the
heliographic position, or whether there is a center-to-limb
dependency.

In contrast, the COR-NLFFF code takes the full sphericity of the
solar surface into account and no de-rotation of magnetograms is
required. Since the COR-NLFFF code uses only the observed LOS
component of the magnetogram to infer the potential field
solution, which is measured with highest accuracy, and does not
require the knowledge of the transverse components, which are
measured with much less accuracy, the inferred potential field 
solution should be invariant to solar rotation to first order,
as long as small-scale magnetic sources are neglected that suffer
from degraded spatial resolution due to projection effects near
the limb. We perform a validation test of this rotational invariance
hypothesis in Fig.~21. A bipolar active region is simulated at
various longitudes from $0^\circ$ to $80^{\circ}$ (Fig.~21 top
panels) to mimic observed magnetograms. Then we decompose the 
simulated LOS magnetogram $B_z^{obs}(x,y)$ into two magnetic 
sources, and calculate a model map $B_z^{model}(x,y)$, from
which we show the profiles $B_z(x)$ along the East-West direction $x$
(Fig.~21, middle panels). Then we calculate the ratio of the
magnetic energies from the observed and the model map,
$$
	q_E = {\int B_{z,model}^2(x,y)\ dx \ dy \over
	       \int B_{z,obs}^2(x,y)\ dx \ dy } 
	    = {\int B_{model}^2(x,y)\ dx \ dy \over
	       \int B_{obs}^2(x,y)\ dx \ dy } 
		\eqno(C1)
$$
and find a mean ratio of $q_E=1.000 \pm 0.024$ when averaged over
the different longitudes (Fig.~21, bottom panel). This result proves
that the magnetic potential field energy is conserved and invariant
to the solar rotation or center-limb-distance, as computed with our
COR-NLFFF code. The accuracy starts to degrade at longitudes
$\gapprox 80^\circ$, which corresponds to a projected distance of
$\sin(80^\circ) \approx 0.98$ solar radii. Thus, our code is able
to calculate potential field solutions for a fraction of
$\approx 90\%$ of active regions that are observed on the solar disk.

\section*{  APPENDIX D:  Forward-Fitting of Non-Potential Fields  }

The COR-NLFFF code is designed to forward-fit an approximate
NLFFF solution in terms of vertical currents to the geometry
of coronal loops. We use the same
parameterization of the potential field solution ${\bf B}_p$
as described above (Eq.~B1), i.e., $(B_m, x_m, y_m, z_m)$
for $m=1,...,n_m$, but add a nonpotential parameter, the so-called
force-free $\alpha$-parameter, so that we have 5 variables for each
magnetic source, i.e., $(B_m, x_m, y_m, z_m, \alpha_m)$.
This force-free $\alpha$-parameter represents a helical twist
of the non-potential field lines about a vertical axis, for each
magnetic charge. Requiring a force-free solution that fulfills
Maxwell's equation,
$$
       {\bf j}/c = {1 \over 4\pi} (\nabla \times {\bf B}) = 0 \ , 
	\eqno(D1)
$$
we calculated an analytical approximation in spherical coordinates
$(r, \varphi, \theta)$ (Aschwanden 2013a),
$$
        B_r(r, \theta) = B_0 \left({d^2 \over r^2}\right)
        {1 \over (1 + b^2 r^2 \sin^2{\theta})} \ ,
	\eqno(D2)
$$
$$
        B_\varphi(r, \theta) =
        B_0 \left({d^2 \over r^2}\right)
        {b r \sin{\theta} \over (1 + b^2 r^2 \sin^2{\theta})} \ ,
	\eqno(D3)
$$
$$
        B_\theta(r, \theta) \approx 0
        \ ,
	\eqno(D4)
$$
$$
        \alpha(r, \theta) \approx {2 b \cos{\theta} \over
        (1 + b^2 r^2 \sin^2{\theta})}  \ .
	\eqno(D5)
$$
$$
        b = {2 \pi N_{twist} \over L} ,
	\eqno(D6)
$$
that is accurate to second-order in the parameter $\alpha$ or 
$r \sin(\theta)$. While these equations are expressed in a
spherical coordinate system that is aligned along the axis $r$
with the solar vertical, the sphericity of the Sun is taken into
full account by transforming the coordinates from each magnetic
charge system into a common cartesian coordinate system that
has the $z$-axis aligned with the observer's line-of-sight.
The resulting nonpotential field ${\bf B}_{np}$ is then summed from all
magnetic charges, 
$$
        {\bf B}_{np}({\bf x}) = \sum_{m=1}^{N_{\rm m}} {\bf B}_m({\bf x}) \ ,
	\eqno(D7)
$$
which is also accurate to second-order in the parameter $\alpha$.
In the limit of $\alpha_m=0$, this solution degenerates to a 
potential field solution (Eq.~B1),

The numerical fitting technique of the nonpotential field 
${\bf B}_n ({\bf r})$ to observed loop coordinates $(x_s, y_s)$
has been initially described in Aschwanden \& Malanushenko (2013b)
and was gradually improved over time. Essentially, the nonpotential
model parameters $\alpha_m, m=1,...,n_m$ have to be optimized until
they match the observed loop geometries, while the potential
model parameters $(B_m, x_m, y_m, z_m), m=1,...,n_m$ are left
unchanged. The convergence criterion of the forward-fitting method 
is a minimum value of the median misalignment angle $\mu({\bf x})$
between theoretical field lines and observed loop geometries
(Sandman et al.~2009; Aschwanden and Sandman 2010; Sandman and
Aschwanden 2011) at a number of locations ${\bf x}$,
$$
        \mu({\bf x}) =
        cos^{-1} \left({ {\bf B}^{theo}({\bf x}) \cdot
        {\bf B}^{obs}({\bf x}) \over
        |{\bf B}^{theo}({\bf x})|\ |{\bf B}^{obs}({\bf x})| }\right) \ .
	\eqno(D8)
$$
where ${\bf x}$ refers to a number of loop positions, for which we
choose $n_{seg}=9$ loop segments. The misalignment angle can be defined
in 2D ($\mu_2$), or in 3D ($\mu_3$), but we will use only the 2D values
$\mu_2$ here, since we are not using any observational information from 
the third (line-of-sight) coordinate.

\medskip
The current version of the COR-NLFFF forward-fitting code, for which
we list the settings of the standard control parameters in Table 2,
contains the following major steps:

\begin{enumerate}
\item{\underbar{Initialization of force-free parameter $\alpha_m$:}
	The initial guess of the variables start with the potential-field
	value $\alpha_m=0$, if no near-simultaneous NLFFF solution exists, 
	while previous solutions of $\alpha_m \neq 0$ are used for 
	time series with $n_t \ge 2$ time frames. 
	This strategy warrants more continuouity of the NLFFF solution
	for sequential calculations with small time steps (say with
	time steps of $\Delta t \approx 0.1$ hour). Most of the
	magnetic energy is contained in the strongest sources, typically
	the 10 strongest magnetic sources contain about 90\% of the 
	magnetic energy. Therefore we need to vary only 
	a subset of values $\alpha_m$ that correspond to the strongest 
	magnetic sources, say $n_{nlfff}\approx 10$ for 
	$B_m \gapprox 0.1 B_{max}$, which represent $E_B \gapprox 
	0.01 E_{max}$, and thus about 99\% of the magnetic energy. 
	Moreover we apply also a minimum distance criterion 
	($d_{foot}$) between nonpotential magnetic
	charges, in order to avoid magnetic flux cancellation of 
        spatially overlapping magnetic sources with opposite magnetic
	polarity. A typical separation distance requirement of
	$d_{foot} \ge 0.015\ R_{\odot}$ reduces the maximum number of magnetic
	charges ($n_m$), used in the decomposition of the magnetogram,
	to $n_{nlfff} \approx 20-50$ values of $\alpha_m$, which 
	yields also a more unique solution and speeds up the convergence 
	of the forward-fitting code.}
\item{\underbar{Optimization of 3D loop geometry:}
	Since the loop tracing method provides 2D coordinates $[x(s), y(s)]$
	only, while 3D coordinates $[x(s), y(s), z(s)]$ are required to
	enable forward-fitting with a 3D NLFFF model, we have to estimate
	the third coordinate $z(s)$ for each loop position in every
	forward-fitting iteration cycle. We estimate the 3D geometry of
	loops by using a circular geometry for each fitted loop segment,
	parameterized by a loop curvature radius $r_{loop}$, a loop apex
	altitude $h_{loop}$, and two footpoint locations at either end 
	of the traced loop segments (Fig.~22). We perform for each loop a global
	search of the minimum 2D misalignment angle $\mu_2$ within a 
	physically plausible range (e.g., $h_{loop} \le h_{max} 
	= 0.2\ R_{\odot}$, $L \le r_{loop} \le h_{max}$, where $L$ is 
	the detected projected loop length and $h_{max}$ is the altitude
	of the computation box). The reconstruction of the best-fit
	3D loop geometry is updated with every optimization cycle of
	$\alpha$-values. Alternative methods to parameterize the 3D
        coordinates of coronal loops include cubic B\'ezier curves
	(Gary et al. 2014a,b).}
\item{\underbar{Optimization of force-free parameters $\alpha_m$:}
	Each value $\alpha_m$ of the $n_{nlfff}$ (strongest and
	well-separated) magnetic source components
	is optimized by minimizing the 2D misalignment angle $\mu_2({\bf x})$
	between the nonpotential field ${\bf B}_{np}({\bf x})$ and the
	automatically traced loop segments $B_{obs}[x(s), y(s)]$,
	where ${\bf x}$ refers to different loop positions
	(typically $n_{seg}=9$ segments) and different loops
	synthesized from all 6 coronal AIA filters (typically
	$n_{loop} \approx 200-500$ for one time frame).
	The median value of this misalignment angle $<\mu>$ is 
	minimized by at least $n_{iter,min}=25$ iteration 
	cycles of all $\alpha$ values and all selected loops ($n_{loop})$.
	The $\alpha_m$ minimization is accomplished with a {\sl direction
	set (Powell's) method in multi-dimensions} (Press et al.~1986, p.294),
	which calculates in each iteration cycle all gradients
	$(\partial \mu/\partial \alpha_m)$ produced by each magnetic 
	source, and improves the next iteration value by
	$\alpha_m^{new} = \alpha_m^{old} - \Delta \alpha_0  
	(\partial \mu/\partial \alpha_m)/max[(\partial \mu/\partial \alpha_m)]$,
	which optimizes the misalignment angles by 	
	$\mu^{new} = \mu^{old} + \Delta \alpha_0 (\partial \mu/\partial \alpha_m)$.}
\end{enumerate}
 
The final result of a NLFFF solution is contained in a set of 
coefficients $(B_m, x_m, y_m, z_m, \alpha_m), m=1,...,n_{nlfff}$,
from which a volume-filling NLFFF solution 
${\bf B}_{np}=[B_x(x,y,z), B_y(x,y,z), B_z(x,y,z)]$ can be computed
in the entire computation box. Individual field lines can be
calculated from any starting point $(x,y,z)$ by sequential
extrapolation of the local B-field vectors in both directions,
until the field line hits a boundary of the computation box. 


\clearpage

\begin{deluxetable}{lrrrrr}
\tabletypesize{\normalsize}
\tablecaption{Statistics of automated feature detection in analyzed data.
Three different runs (RUN1, RUN2, RUN3) have been executed with different
thresholds in the flux profile smoothness criterion ($q_{ripple} \le$ 0.25,
0.50, and 0.75).}
\tablewidth{0pt}
\tablehead{
\colhead{Feature type}&
\colhead{Number}&
\colhead{Percentage}&
\colhead{Percentage}&
\colhead{Percentage}&
\colhead{Percentage}\\
\colhead{}&
\colhead{All}&
\colhead{All}&
\colhead{RUN1}&
\colhead{RUN2}&
\colhead{RUN3}}
\startdata
Ripple criterion $q_{ripple}$	&	     &  & $\le0.25$ & $\le0.50$ & $\le0.75$ \\
Number of runs			&     	  6  &	       &         &         &        \\
Number of wavelengths   	&         6  &         &         &         &        \\
Number of flares		&       172  &	       &         &         &        \\
Number of time frames 		&     2,584  &         &         &         &        \\
Number of detected loops	& 6,900,000  & 100.0\% & 100.0\% & 100.0\% & 100.0\%\\
Number of fitted loops          & 1,500,000  &  22.4\% &  22.4\% &  46.2\% &  79.8\%\\
Number of eliminated loops:     &            &         &         &         &        \\
- magnetic proximity            &   260,000  &   3.7\% &   3.7\% &   6.0\% &  10.2\%\\
- saturated pixels              &    19,000  &   0.3\% &   0.3\% &   0.5\% &   0.7\%\\
- pixel bleeding                &    10,000  &   0.1\% &   0.1\% &   0.3\% &   0.5\%\\
- rippled flux profile (moss)   & 4,800,000  &  70.2\% &  70.2\% &  42.8\% &   1.0\%\\
- diffraction pattern           &   130,000  &   1.9\% &   1.9\% &   1.9\% &   3.1\%\\
- misalignment ($>45^\circ$)    &   930,000  &   1.4\% &   1.4\% &   2.3\% &   4.7\%\\
\enddata
\end{deluxetable}

\begin{deluxetable}{ll}
\tabletypesize{\normalsize}
\tablecaption{Standard control parameters of COR-NLFFF code used in this study.}
\tablewidth{0pt}
\tablehead{
\colhead{Symbol}&
\colhead{Parameter}}
\startdata
			& \underbar{Instruments :}			\\
HMI/SDO			& \qquad  Magnetograms				\\
AIA/SDO			& \qquad  EUV images					\\
$\Delta_{HMI}=0.0005 R_{\odot}$   & \qquad Spatial resolution HMI magnetogram            \\
$\Delta_{EUV}=0.0015 R_{\odot}$   & \qquad Spatial resolution AIA image         \\
$\lambda=94, 131, 171, 193, 211, 335$ \ang & \qquad Wavelengths of EUV images	\\
FOV$=0.35\ R_{\odot}$   & \qquad Field-of-view				\\ 
			& \underbar{Automated Loop Tracing :}		\\
$r_{min}=25$ EUV pixels & \qquad  Minimum loop curvature radius		\\
$l_{min}=25$ EUV pixels & \qquad  Minimum loop segment length			\\
$n_{gap}=3$ EUV pixels  & \qquad  Maximum gap with zero flux along loop 	\\
$n_{thresh}=3$		& \qquad  Significance level of noise threshold	\\
$n_{sig}=4$  		& \qquad  Significance level of diffraction pattern	\\
$n_{point}=200$		& \qquad  Maximum number of points per loop		\\
$n_{loop}=200$		& \qquad  Maximum number of loops per wavelengths	\\
$n_{high}=3$		& \qquad  Highpass filter boxcar			\\
$n_{low}=5$		& \qquad  Lowpass filter boxcar			\\
$\Delta s=0.002 R_{\odot}$ & \qquad Spatial resolution of loop length coordinate\\
$q_{ripple}=0.25,0.50,0.75$ & \qquad Loop flux profile ripple criterion         \\
$\mu=45^\circ$		& \qquad  Maximum tolerated misalignment angle        \\
			& \underbar{Magnetogram Decomposition :}	\\
$n_m=100$		& \qquad  Maximum number of magnetic charges		\\
$\Delta_m=3 \Delta_{HMI}$ & \qquad Spatial resolution of decomposed magnetogram\\
$d_{foot}=0.015	R_{\odot}$ & \qquad Footpoint separation of magnetic charges	\\
$d_{prox}=0.015	R_{\odot}$ & \qquad Magnetic proximity requirement of loop footpoints\\
	 	        & \underbar{Forward-Fitting Algorithm :}	\\
$n_{iter,min}=25$	& \qquad  Minimum number of iteration cycles		\\
$n_{iter,max}=100$	& \qquad  Maximum number of iteration cycles		\\
$n_{seg}=9$		& \qquad  Number of loop segments 			\\
$h_{min}=\Delta s/2$	& \qquad  Minimum altitude				\\
$h_{max}=0.2 R_{\odot}$ & \qquad  Maximum altitude of computation box		\\
$n_{dim}=2$		& \qquad  Dimension of misalignment angle             \\
$\Delta \alpha_0=1/R_{\odot}$ 
			& \qquad  Increment of force-free parameter $\alpha$  \\
			& \underbar{Time Series Forward-Fitting :}	\\
$dt=0.1$ hr		& \qquad  Time step of EUV multi-wavelength dataset	\\
$t_{margin}=0.5$ hr	& \qquad  Margin of preflare and postflare time window  \\
$lon_{max}=45^\circ$	& \qquad  Maximum longitude difference to central meridian\\
\enddata
\end{deluxetable}

\begin{deluxetable}{rrrrrrrrrrrl}
\tabletypesize{\footnotesize}
\tablecaption{Magnetic energy parameters calculated with the COR-NLFFF 
code for 172 M and X-class flares with a longitude difference of 
$<45^\circ$ to the central meridian.}
\tablewidth{0pt}
\tablehead{
\colhead{\#}&
\colhead{Flare}&
\colhead{GOES}&
\colhead{Helio-}&
\colhead{Mis-}&
\colhead{Potential}&
\colhead{Free}&
\colhead{Energy}&
\colhead{Dissipated}&
\colhead{Peak}&
\colhead{Length}&
\colhead{Duration}\\
\colhead{}&
\colhead{start time}&
\colhead{class}&
\colhead{position}&
\colhead{align}&
\colhead{energy}&
\colhead{energy}&
\colhead{ratio}&
\colhead{energy}&
\colhead{dissipation}&
\colhead{scale}&
\colhead{}\\
\colhead{}&
\colhead{}&
\colhead{}&
\colhead{graphic}&
\colhead{angle}&
\colhead{$E_p$}&
\colhead{$E_{free}$}&
\colhead{$E_{free}$}&
\colhead{$E_{diss}$}&
\colhead{rate $P_{diss}$}&
\colhead{}&
\colhead{}\\
\colhead{}&
\colhead{}&
\colhead{}&
\colhead{}&
\colhead{(deg)}&
\colhead{$(10^{30}$ erg)}&
\colhead{$(10^{30}$ erg)}&
\colhead{$/E_p$}&
\colhead{$(10^{30}$ erg)}&
\colhead{$(10^{30}$ erg}&
\colhead{$L$ (Mm)}&
\colhead{$T$ (hrs)}\\
\colhead{}&
\colhead{}&
\colhead{}&
\colhead{}&
\colhead{}&
\colhead{}&
\colhead{}&
\colhead{}&
\colhead{}&
\colhead{$/0.2$ hr)}&
\colhead{}&
\colhead{} }
\startdata
           3 & 2010-08-07 17:55 & M1.0 & N13E34 &  4.0$^\circ$ & $ 311\pm   8$ & $   6\pm   3$ &  0.050 & $  15\pm   1$ & $   9\pm   1$ & $  31\pm   1$ &   0.87 \\
           4 & 2010-10-16 19:07 & M2.9 & S18W26 &  4.0$^\circ$ & $ 202\pm   6$ & $   9\pm   4$ &  0.138 & $  27\pm   6$ & $  15\pm   4$ & $  20\pm   1$ &   0.13 \\
          10 & 2011-02-13 17:28 & M6.6 & S21E04 &  3.4$^\circ$ & $ 842\pm  19$ & $  23\pm   8$ &  0.101 & $  85\pm  20$ & $  47\pm  15$ & $  25\pm   4$ &   0.32 \\
          11 & 2011-02-14 17:20 & M2.2 & S20W07 &  3.5$^\circ$ & $1154\pm  14$ & $  96\pm  12$ &  0.069 & $  79\pm  11$ & $  47\pm  11$ & $  30\pm   2$ &   0.20 \\
          12 & 2011-02-15 01:44 & X2.2 & S21W12 &  4.4$^\circ$ & $1065\pm  14$ & $  52\pm  20$ &  0.113 & $ 120\pm  10$ & $  53\pm  11$ & $  49\pm   3$ &   0.37 \\
          13 & 2011-02-16 01:32 & M1.0 & S22W27 &  4.2$^\circ$ & $ 823\pm  18$ & $ 128\pm  17$ &  0.136 & $ 111\pm   7$ & $  45\pm   8$ & $  38\pm   4$ &   0.23 \\
          14 & 2011-02-16 07:35 & M1.1 & S23W30 &  4.6$^\circ$ & $ 930\pm  19$ & $ 172\pm  26$ &  0.230 & $ 213\pm  44$ & $  82\pm  19$ & $  65\pm   6$ &   0.33 \\
          15 & 2011-02-16 14:19 & M1.6 & S23W33 &  4.4$^\circ$ & $ 855\pm  22$ & $ 140\pm  53$ &  0.217 & $ 185\pm  11$ & $  93\pm  12$ & $  50\pm   4$ &   0.17 \\
          16 & 2011-02-18 09:55 & M6.6 & N15E05 &  3.4$^\circ$ & $ 875\pm  16$ & $   8\pm   3$ &  0.016 & $  13\pm   2$ & $  11\pm   3$ & $  33\pm   6$ &   0.33 \\
          17 & 2011-02-18 10:23 & M1.0 & N17E07 &  3.3$^\circ$ & $ 936\pm  19$ & $  12\pm   5$ &  0.022 & $  20\pm   3$ & $  14\pm   4$ & $  40\pm   2$ &   0.23 \\
          19 & 2011-02-18 14:00 & M1.0 & N17E04 &  3.5$^\circ$ & $1016\pm  39$ & $  16\pm   6$ &  0.026 & $  26\pm   3$ & $  24\pm   7$ & $  40\pm   2$ &   0.25 \\
          20 & 2011-02-18 20:56 & M1.3 & N15E00 &  3.4$^\circ$ & $ 956\pm  18$ & $  11\pm   4$ &  0.016 & $  15\pm   1$ & $  14\pm   4$ & $  37\pm   5$ &   0.30 \\
          22 & 2011-02-28 12:38 & M1.1 & N22E35 &  5.5$^\circ$ & $ 428\pm  17$ & $  15\pm  12$ &  0.070 & $  29\pm   7$ & $  29\pm  10$ & $  21\pm   2$ &   0.42 \\
          27 & 2011-03-07 13:45 & M1.9 & N11E21 &  4.0$^\circ$ & $1135\pm  24$ & $  11\pm   7$ &  0.044 & $  50\pm   8$ & $  27\pm   9$ & $  36\pm   6$ &   1.18 \\
          37 & 2011-03-09 23:13 & X1.5 & N10W11 &  4.3$^\circ$ & $1790\pm  23$ & $ 149\pm  38$ &  0.150 & $ 268\pm  26$ & $ 142\pm   3$ & $  41\pm   2$ &   0.27 \\
          38 & 2011-03-10 22:34 & M1.1 & N10W25 &  4.0$^\circ$ & $1618\pm  26$ & $ 151\pm  55$ &  0.173 & $ 280\pm  12$ & $ 138\pm  34$ & $  38\pm   4$ &   0.25 \\
          39 & 2011-03-12 04:33 & M1.3 & N07W41 &  4.6$^\circ$ & $1172\pm  20$ & $  34\pm  16$ &  0.096 & $ 112\pm  20$ & $  52\pm  12$ & $  41\pm   2$ &   0.25 \\
          44 & 2011-03-25 23:08 & M1.0 & S18E34 &  3.8$^\circ$ & $1200\pm  48$ & $ 130\pm  21$ &  0.109 & $ 131\pm   7$ & $  52\pm   8$ & $  34\pm   3$ &   0.37 \\
          45 & 2011-04-15 17:02 & M1.3 & N13W24 &  3.6$^\circ$ & $1427\pm  23$ & $  31\pm  10$ &  0.022 & $  31\pm   7$ & $  17\pm   6$ & $  22\pm   1$ &   0.43 \\
          46 & 2011-04-22 04:35 & M1.8 & S19E40 &  3.8$^\circ$ & $ 713\pm  29$ & $  21\pm   7$ &  0.064 & $  45\pm   9$ & $  20\pm   7$ & $  19\pm   2$ &   0.65 \\
          47 & 2011-04-22 15:47 & M1.2 & S19E34 &  4.6$^\circ$ & $1005\pm  46$ & $  74\pm  41$ &  0.229 & $ 229\pm  16$ & $ 152\pm  43$ & $  53\pm   6$ &   0.40 \\
          52 & 2011-07-27 15:48 & M1.1 & N20E41 &  4.0$^\circ$ & $ 313\pm   9$ & $   7\pm   5$ &  0.104 & $  32\pm   4$ & $  17\pm   2$ & $  22\pm   2$ &   0.57 \\
          53 & 2011-07-30 02:04 & M9.3 & N16E35 &  5.0$^\circ$ & $ 575\pm  21$ & $  33\pm  12$ &  0.163 & $  93\pm  22$ & $  55\pm  20$ & $  32\pm   2$ &   0.13 \\
          54 & 2011-08-02 05:19 & M1.4 & N16W11 &  5.8$^\circ$ & $1018\pm  28$ & $  64\pm  36$ &  0.112 & $ 114\pm  21$ & $  96\pm  13$ & $  29\pm   5$ &   1.48 \\
          55 & 2011-08-03 03:08 & M1.1 & N15W23 &  5.7$^\circ$ & $ 793\pm  15$ & $  25\pm  18$ &  0.027 & $  21\pm   1$ & $  12\pm   3$ & $  32\pm   1$ &   0.72 \\
          56 & 2011-08-03 04:29 & M1.7 & N16E10 &  4.0$^\circ$ & $1855\pm  17$ & $ 290\pm  41$ &  0.124 & $ 230\pm  26$ & $ 106\pm  29$ & $  79\pm   9$ &   0.10 \\
          57 & 2011-08-03 13:17 & M6.0 & N17W30 &  5.7$^\circ$ & $ 741\pm  17$ & $  68\pm  28$ &  0.113 & $  84\pm   4$ & $  44\pm  13$ & $  33\pm   5$ &   0.88 \\
          58 & 2011-08-04 03:41 & M9.3 & N18W36 &  4.8$^\circ$ & $ 600\pm  16$ & $  54\pm  28$ &  0.191 & $ 114\pm  12$ & $  56\pm   8$ & $  31\pm   3$ &   0.38 \\
          65 & 2011-09-06 01:35 & M5.3 & N15W03 &  4.6$^\circ$ & $1213\pm  33$ & $  33\pm  19$ &  0.093 & $ 112\pm  14$ & $  64\pm  20$ & $  38\pm   3$ &   0.50 \\
          66 & 2011-09-06 22:12 & X2.1 & N16W15 &  4.7$^\circ$ & $ 922\pm  24$ & $  49\pm  34$ &  0.203 & $ 187\pm   9$ & $  97\pm  21$ & $  30\pm   5$ &   0.20 \\
          67 & 2011-09-07 22:32 & X1.8 & N16W30 &  4.7$^\circ$ & $ 582\pm  17$ & $  89\pm  16$ &  0.161 & $  93\pm   7$ & $  50\pm  16$ & $  19\pm   4$ &   0.20 \\
          68 & 2011-09-08 15:32 & M6.7 & N17W39 &  4.5$^\circ$ & $ 508\pm  18$ & $  89\pm  28$ &  0.265 & $ 134\pm   4$ & $  75\pm   7$ & $  31\pm   1$ &   0.33 \\
          89 & 2011-09-25 08:46 & M3.1 & N14E43 &  4.8$^\circ$ & $1477\pm 467$ & $ 136\pm  98$ &  0.379 & $ 560\pm  70$ & $ 288\pm  49$ & $  63\pm   4$ &   0.10 \\
          91 & 2011-09-25 15:26 & M3.7 & N15E39 &  4.7$^\circ$ & $1620\pm  55$ & $ 245\pm  68$ &  0.283 & $ 458\pm  85$ & $ 265\pm  67$ & $  47\pm   4$ &   0.20 \\
          92 & 2011-09-25 16:51 & M2.2 & N16E38 &  4.2$^\circ$ & $1788\pm  74$ & $ 265\pm  95$ &  0.113 & $ 202\pm  33$ & $ 131\pm  21$ & $  44\pm   6$ &   0.30 \\
          93 & 2011-09-26 05:06 & M4.0 & N15E35 &  4.5$^\circ$ & $2207\pm  34$ & $ 350\pm  85$ &  0.322 & $ 710\pm 169$ & $ 355\pm 102$ & $  74\pm  12$ &   0.12 \\
          94 & 2011-09-26 14:37 & M2.6 & N16E25 &  4.7$^\circ$ & $2164\pm  86$ & $ 302\pm  98$ &  0.110 & $ 237\pm  49$ & $ 157\pm  50$ & $  47\pm  10$ &   0.42 \\
          96 & 2011-09-30 18:55 & M1.0 & N11E08 &  4.2$^\circ$ & $ 686\pm  21$ & $   9\pm   7$ &  0.037 & $  25\pm   5$ & $  17\pm   3$ & $  30\pm   1$ &   0.33 \\
          97 & 2011-10-01 08:56 & M1.2 & N10W03 &  4.5$^\circ$ & $ 660\pm 255$ & $  62\pm  34$ &  0.110 & $  72\pm   6$ & $  51\pm  19$ & $  31\pm   0$ &   1.35 \\
          98 & 2011-10-02 00:37 & M3.9 & N10W13 &  5.4$^\circ$ & $ 771\pm  14$ & $  78\pm  17$ &  0.081 & $  62\pm   6$ & $  30\pm   8$ & $  31\pm   3$ &   0.37 \\
         111 & 2011-11-05 11:10 & M1.1 & N22E43 &  2.7$^\circ$ & $2742\pm 100$ & $ 207\pm  62$ &  0.107 & $ 294\pm  47$ & $ 133\pm  31$ & $  37\pm   4$ &   0.53 \\
         112 & 2011-11-05 20:31 & M1.8 & N21E37 &  2.9$^\circ$ & $3356\pm  91$ & $  83\pm  33$ &  0.049 & $ 163\pm  19$ & $  80\pm  13$ & $  31\pm   3$ &   0.38 \\
         113 & 2011-11-06 00:46 & M1.2 & N22E34 &  3.0$^\circ$ & $3572\pm  80$ & $ 125\pm  50$ &  0.067 & $ 239\pm  36$ & $ 177\pm  51$ & $  38\pm   3$ &   0.63 \\
         114 & 2011-11-06 06:14 & M1.4 & N21E31 &  3.3$^\circ$ & $3656\pm  77$ & $ 102\pm  71$ &  0.040 & $ 146\pm  41$ & $ 126\pm  26$ & $  53\pm   8$ &   0.45 \\
         115 & 2011-11-09 13:04 & M1.1 & N20E36 &  5.0$^\circ$ & $ 989\pm  32$ & $  24\pm  11$ &  0.014 & $  14\pm   1$ & $  10\pm   1$ & $  36\pm   5$ &   1.13 \\
         117 & 2011-11-15 12:30 & M1.9 & S19E36 &  4.1$^\circ$ & $ 333\pm   8$ & $   7\pm   6$ &  0.101 & $  33\pm   4$ & $  23\pm   6$ & $  26\pm   1$ &   0.33 \\
         119 & 2011-12-25 18:11 & M4.0 & S20W26 &  4.9$^\circ$ & $ 235\pm   9$ & $   5\pm   2$ &  0.048 & $  11\pm   2$ & $   4\pm   1$ & $  13\pm   0$ &   0.15 \\
         120 & 2011-12-26 02:13 & M1.5 & S18W34 &  4.5$^\circ$ & $ 210\pm  11$ & $   3\pm   0$ &  0.045 & $   9\pm   0$ & $   3\pm   0$ & $  20\pm   1$ &   0.38 \\
         121 & 2011-12-26 20:12 & M2.3 & S18W44 &  4.9$^\circ$ & $ 320\pm  90$ & $   7\pm   8$ &  0.079 & $  25\pm   4$ & $  22\pm   8$ & $  23\pm   1$ &   0.40 \\
         126 & 2011-12-31 16:16 & M1.5 & S22E42 &  3.9$^\circ$ & $ 926\pm  31$ & $  59\pm  20$ &  0.177 & $ 163\pm  29$ & $ 102\pm  29$ & $  33\pm   2$ &   0.30 \\
         130 & 2012-01-19 13:44 & M3.2 & N32E24 &  4.7$^\circ$ & $1204\pm  42$ & $ 251\pm  66$ &  0.354 & $ 426\pm  30$ & $ 179\pm  14$ & $  50\pm   2$ &   4.10 \\
         131 & 2012-01-23 03:38 & M8.7 & N30W21 &  5.0$^\circ$ & $ 828\pm  20$ & $  50\pm  31$ &  0.059 & $  48\pm   9$ & $  28\pm   4$ & $  24\pm   3$ &   0.93 \\
         138 & 2012-03-05 19:27 & M1.8 & N16E45 &  5.9$^\circ$ & $1358\pm 673$ & $ 242\pm 281$ &  0.691 & $ 938\pm 150$ & $ 777\pm 111$ & $  28\pm   0$ &   0.08 \\
         139 & 2012-03-05 22:26 & M1.3 & N16E43 &  6.1$^\circ$ & $1473\pm 532$ & $ 237\pm 197$ &  0.882 & $1299\pm  70$ & $ 785\pm 103$ & $ 160\pm  16$ &   0.27 \\
         140 & 2012-03-06 00:22 & M1.3 & N18E42 &  6.0$^\circ$ & $1775\pm 541$ & $ 159\pm 152$ &  0.478 & $ 848\pm  11$ & $ 555\pm  67$ & $ 236\pm   6$ &   0.15 \\
         141 & 2012-03-06 01:36 & M1.2 & N18E41 &  5.6$^\circ$ & $1545\pm  38$ & $  59\pm  32$ &  0.125 & $ 193\pm  33$ & $ 119\pm  37$ & $  33\pm   3$ &   0.23 \\
         142 & 2012-03-06 04:01 & M1.0 & N18E39 &  4.4$^\circ$ & $1609\pm  48$ & $  92\pm  45$ &  0.170 & $ 273\pm  32$ & $ 210\pm  32$ & $  55\pm  11$ &   0.12 \\
         143 & 2012-03-06 07:52 & M1.0 & N18E40 &  4.6$^\circ$ & $1634\pm  37$ & $  95\pm  75$ &  0.287 & $ 469\pm  74$ & $ 256\pm  84$ & $  77\pm   6$ &   0.13 \\
         144 & 2012-03-06 12:23 & M2.1 & N21E40 &  5.6$^\circ$ & $1781\pm 556$ & $ 304\pm 265$ &  0.868 & $1546\pm 110$ & $1021\pm  69$ & $  84\pm  13$ &   0.52 \\
         145 & 2012-03-06 21:04 & M1.3 & N18E32 &  6.0$^\circ$ & $1795\pm  20$ & $ 253\pm  54$ &  0.186 & $ 333\pm  24$ & $ 140\pm  28$ & $  71\pm   4$ &   0.17 \\
         146 & 2012-03-06 22:49 & M1.0 & N18E32 &  6.0$^\circ$ & $1720\pm  39$ & $ 212\pm  75$ &  0.139 & $ 239\pm  70$ & $ 148\pm  40$ & $  49\pm   9$ &   0.37 \\
         147 & 2012-03-07 00:02 & X5.4 & N18E31 &  7.5$^\circ$ & $1740\pm  32$ & $ 198\pm  66$ &  0.158 & $ 275\pm  46$ & $ 164\pm  42$ & $  55\pm   4$ &   0.63 \\
         148 & 2012-03-07 01:05 & X1.3 & N18E29 &  7.9$^\circ$ & $1780\pm  34$ & $ 168\pm  55$ &  0.113 & $ 200\pm   9$ & $ 146\pm   3$ & $  39\pm   1$ &   0.30 \\
         149 & 2012-03-09 03:22 & M6.3 & N17W00 &  5.2$^\circ$ & $1815\pm  55$ & $ 211\pm  65$ &  0.080 & $ 144\pm  29$ & $  73\pm  18$ & $  31\pm   3$ &   0.93 \\
         150 & 2012-03-10 17:15 & M8.4 & N16W21 &  4.5$^\circ$ & $1459\pm  30$ & $ 110\pm  23$ &  0.039 & $  57\pm  13$ & $  63\pm  15$ & $  25\pm   2$ &   1.25 \\
         152 & 2012-03-14 15:08 & M2.8 & N14E07 &  3.8$^\circ$ & $ 355\pm   9$ & $   5\pm   2$ &  0.039 & $  13\pm   2$ & $   8\pm   2$ & $  22\pm   2$ &   0.47 \\
         153 & 2012-03-15 07:23 & M1.8 & N16W04 &  1.8$^\circ$ & $ 131\pm 156$ & $   2\pm   3$ &  0.000 & $   0\pm   0$ & $   0\pm   0$ & $   0\pm   0$ &   0.75 \\
         154 & 2012-03-17 20:32 & M1.3 & S25W28 &  3.7$^\circ$ & $ 319\pm  10$ & $  11\pm   4$ &  0.089 & $  28\pm   1$ & $  14\pm   2$ & $  20\pm   1$ &   0.17 \\
         157 & 2012-04-27 08:15 & M1.0 & N13W26 &  3.8$^\circ$ & $ 324\pm   7$ & $   2\pm   0$ &  0.014 & $   4\pm   1$ & $   2\pm   0$ & $  24\pm   2$ &   0.23 \\
         164 & 2012-05-09 12:21 & M4.7 & N13E29 &  4.1$^\circ$ & $2522\pm  35$ & $  77\pm  66$ &  0.117 & $ 295\pm  54$ & $ 169\pm  24$ & $  50\pm   4$ &   0.25 \\
         165 & 2012-05-09 14:02 & M1.8 & N12E29 &  3.8$^\circ$ & $2640\pm  41$ & $  97\pm  46$ &  0.091 & $ 239\pm  38$ & $ 129\pm  43$ & $  46\pm   5$ &   0.20 \\
         166 & 2012-05-09 21:01 & M4.1 & N13E24 &  4.1$^\circ$ & $3039\pm  51$ & $ 114\pm  98$ &  0.115 & $ 349\pm  22$ & $ 216\pm  61$ & $  56\pm   3$ &   0.13 \\
         167 & 2012-05-10 04:11 & M5.7 & N12E19 &  3.8$^\circ$ & $3189\pm  65$ & $  42\pm  30$ &  0.045 & $ 145\pm  27$ & $ 121\pm  43$ & $  39\pm   7$ &   0.20 \\
         168 & 2012-05-10 20:20 & M1.7 & N12E10 &  3.9$^\circ$ & $3180\pm  44$ & $  34\pm  10$ &  0.040 & $ 128\pm  16$ & $  45\pm  10$ & $  27\pm   1$ &   0.17 \\
         170 & 2012-06-03 17:48 & M3.3 & N15E33 &  3.6$^\circ$ & $ 731\pm  14$ & $  23\pm  11$ &  0.062 & $  45\pm   4$ & $  23\pm   6$ & $  38\pm   6$ &   0.15 \\
         171 & 2012-06-06 19:54 & M2.1 & S18W04 &  4.1$^\circ$ & $ 562\pm  16$ & $  46\pm   8$ &  0.096 & $  53\pm   6$ & $  26\pm   4$ & $  36\pm   5$ &   0.32 \\
         175 & 2012-06-13 11:29 & M1.2 & S18E21 &  4.0$^\circ$ & $1755\pm  73$ & $ 407\pm 110$ &  0.173 & $ 303\pm  41$ & $ 141\pm  42$ & $  62\pm   2$ &   3.03 \\
         176 & 2012-06-14 12:52 & M1.9 & S19E06 &  3.7$^\circ$ & $2534\pm  85$ & $ 634\pm 109$ &  0.122 & $ 310\pm  77$ & $ 221\pm  55$ & $  76\pm   4$ &   3.07 \\
         178 & 2012-06-29 09:13 & M2.2 & N15E37 &  5.3$^\circ$ & $ 337\pm   9$ & $  12\pm  22$ &  0.289 & $  97\pm   4$ & $  79\pm   4$ & $  35\pm   1$ &   0.15 \\
         179 & 2012-06-30 12:48 & M1.0 & N15E21 &  5.5$^\circ$ & $ 370\pm  14$ & $   9\pm   6$ &  0.109 & $  40\pm   5$ & $  23\pm   5$ & $  25\pm   4$ &   0.10 \\
         180 & 2012-06-30 18:26 & M1.6 & N14E18 &  5.1$^\circ$ & $ 422\pm  10$ & $  11\pm   5$ &  0.095 & $  40\pm   3$ & $  23\pm   3$ & $  25\pm   1$ &   0.13 \\
         181 & 2012-07-01 19:11 & M2.8 & N15E04 &  5.2$^\circ$ & $ 642\pm  20$ & $  25\pm  13$ &  0.104 & $  66\pm   9$ & $  45\pm  10$ & $  31\pm   4$ &   0.17 \\
         182 & 2012-07-02 00:26 & M1.1 & N15E01 &  6.0$^\circ$ & $ 668\pm  24$ & $  48\pm  30$ &  0.098 & $  65\pm   9$ & $  42\pm   4$ & $  33\pm   3$ &   0.23 \\
         183 & 2012-07-02 10:43 & M5.6 & S17E06 &  3.9$^\circ$ & $1641\pm  34$ & $ 155\pm  56$ &  0.039 & $  63\pm  10$ & $ 118\pm  25$ & $  35\pm   8$ &   0.23 \\
         184 & 2012-07-02 19:59 & M3.8 & S17E00 &  4.8$^\circ$ & $1713\pm  50$ & $ 148\pm  56$ &  0.072 & $ 123\pm  20$ & $  70\pm  21$ & $  29\pm   1$ &   0.23 \\
         185 & 2012-07-02 23:49 & M2.0 & S16W09 &  4.4$^\circ$ & $ 884\pm  15$ & $  75\pm  21$ &  0.120 & $ 106\pm   8$ & $  51\pm  16$ & $  28\pm   3$ &   0.23 \\
         186 & 2012-07-04 04:28 & M2.3 & S18W18 &  3.8$^\circ$ & $1872\pm  29$ & $ 221\pm  65$ &  0.064 & $ 119\pm  30$ & $  93\pm  38$ & $  32\pm   3$ &   0.28 \\
         187 & 2012-07-04 09:47 & M5.3 & S17W18 &  4.2$^\circ$ & $1993\pm  24$ & $ 186\pm  41$ &  0.083 & $ 166\pm  46$ & $ 117\pm  38$ & $  28\pm   6$ &   0.17 \\
         188 & 2012-07-04 12:07 & M2.3 & S17W19 &  4.1$^\circ$ & $2011\pm  39$ & $ 164\pm  52$ &  0.073 & $ 147\pm  33$ & $  82\pm  28$ & $  45\pm   5$ &   0.42 \\
         189 & 2012-07-04 14:35 & M1.3 & S18W20 &  3.6$^\circ$ & $2052\pm  50$ & $ 143\pm  67$ &  0.026 & $  53\pm   6$ & $  46\pm  16$ & $  24\pm   3$ &   0.12 \\
         190 & 2012-07-04 16:33 & M1.8 & N14W33 &  4.5$^\circ$ & $ 498\pm  28$ & $   7\pm   3$ &  0.057 & $  28\pm   5$ & $  11\pm   2$ & $  27\pm   4$ &   0.25 \\
         191 & 2012-07-04 22:03 & M4.6 & S16W28 &  4.0$^\circ$ & $2095\pm  48$ & $ 160\pm  38$ &  0.058 & $ 121\pm  16$ & $  66\pm  18$ & $  29\pm   2$ &   0.20 \\
         192 & 2012-07-04 23:47 & M1.2 & S19W28 &  4.0$^\circ$ & $1935\pm  42$ & $ 164\pm  63$ &  0.035 & $  68\pm  14$ & $  37\pm  14$ & $  22\pm   3$ &   0.25 \\
         193 & 2012-07-05 01:05 & M2.4 & S19W29 &  4.2$^\circ$ & $1887\pm  29$ & $ 120\pm  58$ &  0.033 & $  62\pm  16$ & $  47\pm  23$ & $  26\pm   4$ &   0.17 \\
         194 & 2012-07-05 02:35 & M2.2 & S18W26 &  3.9$^\circ$ & $1892\pm  21$ & $ 186\pm  75$ &  0.071 & $ 133\pm  14$ & $ 113\pm  10$ & $  44\pm   6$ &   0.20 \\
         195 & 2012-07-05 03:25 & M4.7 & S18W29 &  4.4$^\circ$ & $2007\pm  26$ & $ 194\pm  76$ &  0.083 & $ 166\pm  49$ & $  75\pm  34$ & $  25\pm   5$ &   0.23 \\
         196 & 2012-07-05 06:49 & M1.1 & S17W29 &  4.5$^\circ$ & $2029\pm  32$ & $ 166\pm  67$ &  0.063 & $ 127\pm  15$ & $  60\pm  13$ & $  25\pm   1$ &   0.27 \\
         197 & 2012-07-05 07:40 & M1.3 & S18W30 &  4.2$^\circ$ & $2029\pm  34$ & $ 141\pm  50$ &  0.048 & $  96\pm  17$ & $  80\pm  19$ & $  27\pm   2$ &   0.13 \\
         198 & 2012-07-05 10:44 & M1.8 & S18W30 &  3.9$^\circ$ & $1917\pm  39$ & $  63\pm  44$ &  0.030 & $  56\pm  11$ & $  33\pm  13$ & $  42\pm   5$ &   0.10 \\
         199 & 2012-07-05 11:39 & M6.1 & S18W32 &  4.1$^\circ$ & $1946\pm  52$ & $  86\pm  43$ &  0.067 & $ 129\pm  24$ & $  95\pm  23$ & $  33\pm   2$ &   0.17 \\
         200 & 2012-07-05 13:05 & M1.2 & S18W36 &  4.1$^\circ$ & $1866\pm  32$ & $ 167\pm  81$ &  0.063 & $ 118\pm   3$ & $  83\pm  19$ & $  37\pm   3$ &   0.45 \\
         201 & 2012-07-05 20:09 & M1.6 & S18W39 &  3.8$^\circ$ & $1755\pm  28$ & $  99\pm  43$ &  0.068 & $ 118\pm  19$ & $  89\pm  28$ & $  33\pm   5$ &   0.32 \\
         202 & 2012-07-05 21:37 & M1.6 & S18W41 &  4.0$^\circ$ & $1707\pm  31$ & $  67\pm  34$ &  0.101 & $ 172\pm  21$ & $ 114\pm  24$ & $  39\pm   9$ &   0.23 \\
         203 & 2012-07-06 01:37 & M2.9 & S18W43 &  3.8$^\circ$ & $1502\pm 507$ & $  34\pm  23$ &  0.048 & $  72\pm   8$ & $  36\pm   8$ & $  48\pm   2$ &   0.08 \\
         206 & 2012-07-06 10:24 & M1.8 & S17W44 &  3.2$^\circ$ & $1555\pm  36$ & $  44\pm  29$ &  0.026 & $  40\pm   7$ & $  35\pm   8$ & $  39\pm   1$ &   0.13 \\
         217 & 2012-07-09 23:03 & M1.1 & S17E38 &  6.1$^\circ$ & $3278\pm  56$ & $ 635\pm 167$ &  0.258 & $ 847\pm 120$ & $ 383\pm  89$ & $  93\pm   2$ &   0.13 \\
         218 & 2012-07-10 04:58 & M1.7 & S16E35 &  6.0$^\circ$ & $3166\pm  66$ & $ 661\pm 214$ &  0.153 & $ 484\pm  86$ & $ 283\pm  85$ & $  96\pm  10$ &   0.55 \\
         219 & 2012-07-10 06:05 & M2.0 & S16E30 &  5.5$^\circ$ & $3523\pm  57$ & $ 778\pm 115$ &  0.238 & $ 839\pm 163$ & $ 461\pm 109$ & $  96\pm   4$ &   0.70 \\
         220 & 2012-07-12 15:37 & X1.4 & S15W03 &  4.8$^\circ$ & $3915\pm 766$ & $ 951\pm 324$ &  0.357 & $1399\pm  89$ & $ 721\pm 222$ & $ 103\pm   3$ &   1.88 \\
         221 & 2012-07-14 04:51 & M1.0 & S20W23 &  5.7$^\circ$ & $3485\pm 106$ & $ 863\pm 275$ &  0.240 & $ 835\pm 141$ & $ 324\pm  80$ & $  91\pm   4$ &   0.23 \\
         227 & 2012-07-30 15:39 & M1.1 & S21E28 &  3.8$^\circ$ & $1358\pm  42$ & $  87\pm  33$ &  0.076 & $ 102\pm   5$ & $  52\pm   7$ & $  40\pm   2$ &   0.23 \\
         228 & 2012-08-06 04:33 & M1.6 & N16W12 &  5.5$^\circ$ & $ 396\pm  10$ & $   3\pm   1$ &  0.033 & $  12\pm   2$ & $   6\pm   1$ & $  27\pm   3$ &   0.13 \\
         229 & 2012-08-11 11:55 & M1.0 & S25W41 &  3.7$^\circ$ & $ 142\pm   4$ & $   7\pm   2$ &  0.115 & $  16\pm   0$ & $  10\pm   1$ & $  22\pm   0$ &   1.03 \\
         239 & 2012-09-08 17:35 & M1.4 & S14W40 &  4.4$^\circ$ & $1004\pm  28$ & $  66\pm  19$ &  0.153 & $ 154\pm  38$ & $  56\pm  12$ & $  28\pm   2$ &   0.75 \\
         253 & 2012-11-13 05:42 & M2.5 & S26E44 &  4.0$^\circ$ & $ 275\pm  19$ & $  65\pm  23$ &  0.397 & $ 109\pm  11$ & $  47\pm   4$ & $  27\pm   1$ &   0.20 \\
         254 & 2012-11-13 20:50 & M2.8 & S23E31 &  4.1$^\circ$ & $ 275\pm  88$ & $  25\pm  10$ &  0.128 & $  35\pm   3$ & $  17\pm   2$ & $  21\pm   2$ &   0.12 \\
         255 & 2012-11-14 03:59 & M1.1 & S23E27 &  3.7$^\circ$ & $ 308\pm  13$ & $  30\pm   6$ &  0.137 & $  42\pm   5$ & $  17\pm   3$ & $  23\pm   1$ &   0.13 \\
         256 & 2012-11-20 12:36 & M1.7 & N10E22 &  4.2$^\circ$ & $ 534\pm  13$ & $   6\pm   3$ &  0.040 & $  21\pm   3$ & $  11\pm   2$ & $  26\pm   2$ &   0.17 \\
         257 & 2012-11-20 19:21 & M1.6 & N10E19 &  3.8$^\circ$ & $ 680\pm  23$ & $  12\pm   5$ &  0.047 & $  31\pm   4$ & $  17\pm   4$ & $  24\pm   0$ &   0.18 \\
         258 & 2012-11-21 06:45 & M1.4 & N10E12 &  3.9$^\circ$ & $ 806\pm 233$ & $  34\pm  14$ &  0.072 & $  58\pm   4$ & $  37\pm  10$ & $  28\pm   1$ &   0.38 \\
         259 & 2012-11-21 15:10 & M3.5 & N10E08 &  3.4$^\circ$ & $1075\pm  23$ & $  48\pm  12$ &  0.072 & $  77\pm   7$ & $  50\pm  17$ & $  35\pm   2$ &   0.47 \\
         261 & 2012-11-27 21:05 & M1.0 & S13W42 &  3.9$^\circ$ & $ 788\pm  17$ & $  18\pm  11$ &  0.054 & $  42\pm   9$ & $  15\pm   3$ & $  27\pm   3$ &   0.42 \\
         264 & 2013-01-11 08:43 & M1.2 & N05E42 &  3.7$^\circ$ & $1804\pm  66$ & $  62\pm  23$ &  0.114 & $ 206\pm  30$ & $ 119\pm  27$ & $  37\pm   1$ &   0.57 \\
         265 & 2013-01-11 14:51 & M1.0 & N06E42 &  3.0$^\circ$ & $1970\pm  51$ & $  47\pm  24$ &  0.037 & $  73\pm   7$ & $  32\pm   5$ & $  32\pm   4$ &   0.55 \\
         266 & 2013-01-13 00:45 & M1.0 & N18W15 &  4.6$^\circ$ & $ 811\pm  16$ & $  25\pm  14$ &  0.024 & $  19\pm   2$ & $  15\pm   6$ & $  37\pm   6$ &   0.12 \\
         267 & 2013-01-13 08:35 & M1.7 & N17W18 &  3.9$^\circ$ & $ 834\pm   7$ & $  43\pm  18$ &  0.059 & $  49\pm  11$ & $  37\pm  10$ & $  29\pm   1$ &   0.08 \\
         268 & 2013-02-17 15:45 & M1.9 & N12E23 &  4.2$^\circ$ & $ 219\pm   4$ & $   3\pm   2$ &  0.083 & $  18\pm   2$ & $   7\pm   1$ & $  19\pm   1$ &   0.12 \\
         273 & 2013-04-11 06:55 & M6.5 & N11E13 &  5.2$^\circ$ & $ 805\pm  15$ & $  18\pm  15$ &  0.062 & $  50\pm   4$ & $  29\pm   6$ & $  34\pm   4$ &   0.57 \\
         275 & 2013-04-22 10:22 & M1.0 & N13W27 &  3.5$^\circ$ & $1197\pm  47$ & $ 104\pm  45$ &  0.223 & $ 267\pm  14$ & $ 157\pm   7$ & $  36\pm   2$ &   0.15 \\
         276 & 2013-05-02 04:58 & M1.1 & N10W19 &  5.4$^\circ$ & $1115\pm  21$ & $  27\pm  21$ &  0.056 & $  62\pm  15$ & $  38\pm  21$ & $  36\pm   1$ &   0.35 \\
         277 & 2013-05-03 16:39 & M1.3 & N11W38 &  4.7$^\circ$ & $ 548\pm  16$ & $  21\pm   7$ &  0.027 & $  14\pm   4$ & $  15\pm   5$ & $  30\pm   3$ &   0.72 \\
         289 & 2013-05-16 21:36 & M1.3 & N11E40 &  4.2$^\circ$ & $ 391\pm  15$ & $  13\pm   6$ &  0.056 & $  22\pm   2$ & $  10\pm   2$ & $  27\pm   3$ &   0.45 \\
         290 & 2013-05-17 08:43 & M3.2 & N11E36 &  4.6$^\circ$ & $ 389\pm   9$ & $  12\pm   8$ &  0.041 & $  16\pm   2$ & $  17\pm   7$ & $  24\pm   3$ &   0.60 \\
         293 & 2013-05-31 19:52 & M1.0 & N12E42 &  4.1$^\circ$ & $  85\pm   3$ & $   1\pm   0$ &  0.018 & $   1\pm   0$ & $   1\pm   0$ & $  15\pm   1$ &   0.23 \\
         299 & 2013-08-12 10:21 & M1.5 & S21E17 &  5.1$^\circ$ & $ 388\pm   9$ & $  15\pm   6$ &  0.043 & $  16\pm   0$ & $  13\pm   1$ & $  22\pm   3$ &   0.43 \\
         300 & 2013-08-17 18:16 & M3.3 & S04W30 &  5.2$^\circ$ & $ 595\pm  23$ & $  97\pm  52$ &  0.120 & $  71\pm   5$ & $  35\pm   5$ & $  33\pm   1$ &   0.32 \\
         301 & 2013-08-17 18:49 & M1.4 & S04W30 &  4.6$^\circ$ & $ 600\pm  28$ & $ 124\pm  47$ &  0.360 & $ 216\pm  35$ & $ 124\pm  19$ & $  53\pm   3$ &   1.08 \\
         304 & 2013-10-13 00:12 & M1.7 & S22E17 &  4.9$^\circ$ & $ 501\pm  10$ & $  31\pm  11$ &  0.162 & $  81\pm  10$ & $  39\pm  10$ & $  23\pm   4$ &   0.88 \\
         305 & 2013-10-15 08:26 & M1.8 & S21W14 &  4.7$^\circ$ & $ 353\pm   8$ & $  10\pm   3$ &  0.068 & $  24\pm   1$ & $  15\pm   4$ & $  18\pm   3$ &   0.37 \\
         306 & 2013-10-15 23:31 & M1.3 & S21W22 &  4.7$^\circ$ & $ 294\pm   5$ & $  16\pm   4$ &  0.134 & $  39\pm  10$ & $  14\pm   6$ & $  22\pm   2$ &   0.17 \\
         308 & 2013-10-22 00:14 & M1.0 & N08E20 &  3.4$^\circ$ & $1226\pm  20$ & $  43\pm  24$ &  0.121 & $ 148\pm  10$ & $  78\pm   7$ & $  34\pm   5$ &   0.23 \\
         309 & 2013-10-22 14:49 & M1.0 & N08E11 &  3.5$^\circ$ & $1959\pm  47$ & $  43\pm  17$ &  0.059 & $ 115\pm  21$ & $  61\pm  16$ & $  28\pm   4$ &   0.65 \\
         310 & 2013-10-22 21:15 & M4.2 & N07E03 &  3.5$^\circ$ & $2090\pm  44$ & $   7\pm   4$ &  0.013 & $  26\pm   2$ & $  14\pm   3$ & $  37\pm   3$ &   0.12 \\
         311 & 2013-10-23 20:41 & M2.7 & N08W06 &  3.5$^\circ$ & $3015\pm  64$ & $ 231\pm  57$ &  0.055 & $ 166\pm  21$ & $  79\pm  15$ & $  43\pm   1$ &   0.30 \\
         312 & 2013-10-23 23:33 & M1.4 & N09W08 &  4.2$^\circ$ & $2392\pm 788$ & $ 216\pm 105$ &  0.147 & $ 350\pm  19$ & $ 294\pm  27$ & $  71\pm   3$ &   0.23 \\
         313 & 2013-10-23 23:58 & M3.1 & N09W09 &  3.8$^\circ$ & $2923\pm  43$ & $ 254\pm  87$ &  0.051 & $ 149\pm   7$ & $ 130\pm  37$ & $  45\pm   3$ &   0.30 \\
         314 & 2013-10-24 00:21 & M9.3 & S09E12 &  4.2$^\circ$ & $1468\pm  42$ & $  81\pm  40$ &  0.163 & $ 239\pm  24$ & $ 125\pm  26$ & $  86\pm   3$ &   0.23 \\
         315 & 2013-10-24 09:59 & M2.5 & N09W14 &  3.6$^\circ$ & $2812\pm  35$ & $ 230\pm  55$ &  0.077 & $ 217\pm  12$ & $ 156\pm  34$ & $  46\pm   3$ &   0.30 \\
         316 & 2013-10-24 10:30 & M3.5 & N09W14 &  3.7$^\circ$ & $2848\pm  48$ & $ 311\pm  77$ &  0.053 & $ 150\pm  40$ & $ 115\pm  47$ & $  49\pm   9$ &   0.12 \\
         334 & 2013-10-28 14:46 & M2.7 & S08E27 &  4.8$^\circ$ & $ 682\pm  24$ & $  16\pm  11$ &  0.076 & $  52\pm  11$ & $  42\pm   5$ & $  31\pm   3$ &   0.30 \\
         335 & 2013-10-28 15:07 & M4.4 & S06E28 &  4.2$^\circ$ & $ 697\pm  18$ & $  37\pm  19$ &  0.144 & $ 100\pm   3$ & $  53\pm  10$ & $  34\pm   2$ &   0.23 \\
         339 & 2013-11-01 19:46 & M6.3 & S12E01 &  3.7$^\circ$ & $1523\pm  23$ & $  34\pm  12$ &  0.054 & $  82\pm   7$ & $  47\pm  10$ & $  29\pm   0$ &   0.20 \\
         340 & 2013-11-02 22:13 & M1.6 & S12W12 &  4.6$^\circ$ & $1060\pm  27$ & $  40\pm  21$ &  0.082 & $  87\pm   3$ & $  40\pm   5$ & $  29\pm   1$ &   0.20 \\
         341 & 2013-11-03 05:16 & M5.0 & S10W17 &  4.2$^\circ$ & $ 968\pm  22$ & $  14\pm   7$ &  0.090 & $  87\pm   8$ & $  42\pm  13$ & $  32\pm   2$ &   0.17 \\
         344 & 2013-11-05 22:07 & X3.3 & S08E44 &  3.8$^\circ$ & $2066\pm  65$ & $  72\pm  30$ &  0.132 & $ 273\pm  66$ & $ 141\pm  41$ & $  50\pm   5$ &   0.13 \\
         345 & 2013-11-06 13:39 & M3.8 & S09E35 &  3.9$^\circ$ & $2256\pm  60$ & $  71\pm  22$ &  0.080 & $ 180\pm   5$ & $  67\pm  10$ & $  40\pm   3$ &   0.23 \\
         347 & 2013-11-07 03:34 & M2.3 & S08E26 &  4.6$^\circ$ & $2812\pm  36$ & $ 119\pm  58$ &  0.132 & $ 370\pm  63$ & $ 167\pm  19$ & $  47\pm   4$ &   0.15 \\
         348 & 2013-11-07 14:15 & M2.4 & S08E18 &  4.0$^\circ$ & $2915\pm  49$ & $  98\pm  58$ &  0.097 & $ 283\pm  23$ & $ 117\pm  36$ & $  48\pm   3$ &   0.27 \\
         349 & 2013-11-08 04:20 & X1.1 & S11E11 &  4.5$^\circ$ & $3290\pm  56$ & $ 187\pm  58$ &  0.077 & $ 252\pm  29$ & $ 186\pm  31$ & $  41\pm   2$ &   0.15 \\
         350 & 2013-11-08 09:22 & M2.3 & S17W29 &  4.2$^\circ$ & $ 236\pm   9$ & $   8\pm   2$ &  0.102 & $  24\pm   2$ & $   8\pm   1$ & $  23\pm   3$ &   0.15 \\
         351 & 2013-11-10 05:08 & X1.1 & S13W13 &  4.2$^\circ$ & $2010\pm  22$ & $ 109\pm  49$ &  0.127 & $ 254\pm  51$ & $ 154\pm  31$ & $  46\pm   4$ &   0.17 \\
         355 & 2013-11-16 04:47 & M1.2 & S14W29 &  3.7$^\circ$ & $ 767\pm  36$ & $  65\pm  61$ &  0.054 & $  41\pm  12$ & $  26\pm  14$ & $  26\pm   2$ &   0.17 \\
         356 & 2013-11-16 07:45 & M1.6 & S17W30 &  3.7$^\circ$ & $ 806\pm  36$ & $  54\pm  30$ &  0.062 & $  49\pm   8$ & $  55\pm  18$ & $  29\pm   1$ &   0.13 \\
         357 & 2013-11-17 05:06 & M1.0 & S19W41 &  3.6$^\circ$ & $ 680\pm  34$ & $ 135\pm  28$ &  0.180 & $ 122\pm  16$ & $  70\pm  20$ & $  44\pm   1$ &   0.12 \\
         367 & 2013-12-22 14:24 & M1.6 & S16E44 &  4.5$^\circ$ & $ 553\pm  10$ & $  49\pm  19$ &  0.118 & $  65\pm   6$ & $  35\pm   9$ & $  23\pm   4$ &   0.40 \\
         372 & 2013-12-29 07:49 & M3.1 & S16E03 &  3.3$^\circ$ & $ 936\pm  12$ & $  18\pm   8$ &  0.042 & $  39\pm   5$ & $  22\pm   6$ & $  24\pm   2$ &   0.18 \\
         373 & 2013-12-31 21:45 & M6.4 & S19W36 &  4.2$^\circ$ & $ 848\pm  29$ & $  29\pm  25$ &  0.194 & $ 164\pm  18$ & $ 105\pm  14$ & $  36\pm   0$ &   0.58 \\
         374 & 2014-01-01 18:40 & M9.9 & S19W45 &  5.1$^\circ$ & $ 624\pm  11$ & $  41\pm  16$ &  0.191 & $ 119\pm  25$ & $  59\pm  16$ & $  31\pm   2$ &   0.38 \\
         380 & 2014-01-04 18:47 & M4.0 & S15E30 &  4.5$^\circ$ & $3436\pm 109$ & $ 737\pm 170$ &  0.328 & $1125\pm 224$ & $ 465\pm 126$ & $ 101\pm   7$ &   1.60 \\
         382 & 2014-01-07 03:49 & M1.0 & N07E07 &  2.5$^\circ$ & $ 835\pm  18$ & $  61\pm  16$ &  0.089 & $  74\pm   7$ & $  23\pm   4$ & $  21\pm   1$ &   0.12 \\
         383 & 2014-01-07 10:07 & M7.2 & S13E13 &  4.4$^\circ$ & $3872\pm  48$ & $ 443\pm 137$ &  0.137 & $ 529\pm  61$ & $ 263\pm  70$ & $  73\pm   3$ &   0.50 \\
         384 & 2014-01-07 18:04 & X1.2 & S12E08 &  4.3$^\circ$ & $3949\pm  67$ & $ 478\pm 148$ &  0.074 & $ 292\pm  74$ & $ 184\pm  59$ & $  60\pm   3$ &   0.90 \\
         399 & 2014-01-31 15:32 & M1.1 & N07E34 &  3.4$^\circ$ & $ 717\pm  20$ & $  27\pm  11$ &  0.128 & $  91\pm   9$ & $  32\pm   6$ & $  31\pm   4$ &   0.35 \\
\enddata
\end{deluxetable}
\clearpage

\begin{deluxetable}{rlrcrcrc}
\tabletypesize{\footnotesize}
\tablecaption{Comparison of magnetic energies calculated with the COR-NLFFF and PHOT-NLFFF
code for 11 M and X-class flares with a longitude difference of $<45^\circ$ to the
central meridian.}
\tablewidth{0pt}
\tablehead{
\colhead{\#}&
\colhead{GOES}&
\colhead{Potential}&
\colhead{Potential}&
\colhead{Free}&
\colhead{Free}&
\colhead{Dissipated}&
\colhead{Dissipated}\\
\colhead{}&
\colhead{class}&
\colhead{energy}&
\colhead{energy ratio}&
\colhead{energy}&
\colhead{energy ratio}&
\colhead{energy}&
\colhead{energy ratio}\\
\colhead{}&
\colhead{}&
\colhead{$E_p^{COR}$}&
\colhead{$E_p^{PHOT}/E_p^{COR}$}&
\colhead{$E_{free}^{COR}$}&
\colhead{$E_{free}^{PHOT}/E_{free}^{COR}$}&
\colhead{$E_{diss}^{COR}$}&
\colhead{$E_{diss}^{PHOT}/E_{diss}^{COR}$}}
\startdata
          12 & X2.2 & $1065\pm  15$ & 0.94 & $  52\pm  21$ & 5.16 & $ 120\pm  11$ & 0.88 \\
          37 & X1.5 & $1791\pm  24$ & 0.91 & $ 150\pm  38$ & 1.21 & $ 269\pm  27$ & 0.28 \\
          66 & X2.1 & $ 922\pm  24$ & 0.64 & $  50\pm  35$ & 3.22 & $ 188\pm  10$ & 0.37 \\
          67 & X1.8 & $ 582\pm  17$ & 1.04 & $  89\pm  17$ & 0.85 & $  94\pm   8$ & 1.10 \\
         147 & X5.4 & $1741\pm  32$ & 1.22 & $ 198\pm  66$ & 5.61 & $ 275\pm  47$ & 1.27 \\
         148 & X5.4 & $1781\pm  35$ & 1.19 & $ 168\pm  56$ & 6.64 & $ 200\pm  10$ & .... \\
         220 & X1.4 & $3916\pm 766$ & 1.01 & $ 951\pm 324$ & 1.12 & $1399\pm  89$ & 0.09 \\
         344 & X3.3 & $2067\pm  65$ & 1.61 & $  72\pm  31$ & 6.64 & $ 273\pm  67$ & 0.32 \\
         349 & X1.1 & $3290\pm  57$ & 0.62 & $ 188\pm  58$ & 1.37 & $ 253\pm  29$ & 0.29 \\
         351 & X1.1 & $2011\pm  23$ & 0.85 & $ 109\pm  49$ & 2.25 & $ 255\pm  52$ & 0.22 \\
         384 & X1.2 & $3950\pm  68$ & 1.57 & $ 479\pm 149$ & 1.83 & $ 293\pm  75$ & 0.56 \\
	     &      &               &      &               &      &               &      \\
Average	     &      &               &$1.05\pm0.33$&        &$3.3\pm2.3$&          &$0.5\pm0.4$\\
\enddata
\end{deluxetable}
\clearpage


\begin{figure}
\plotone{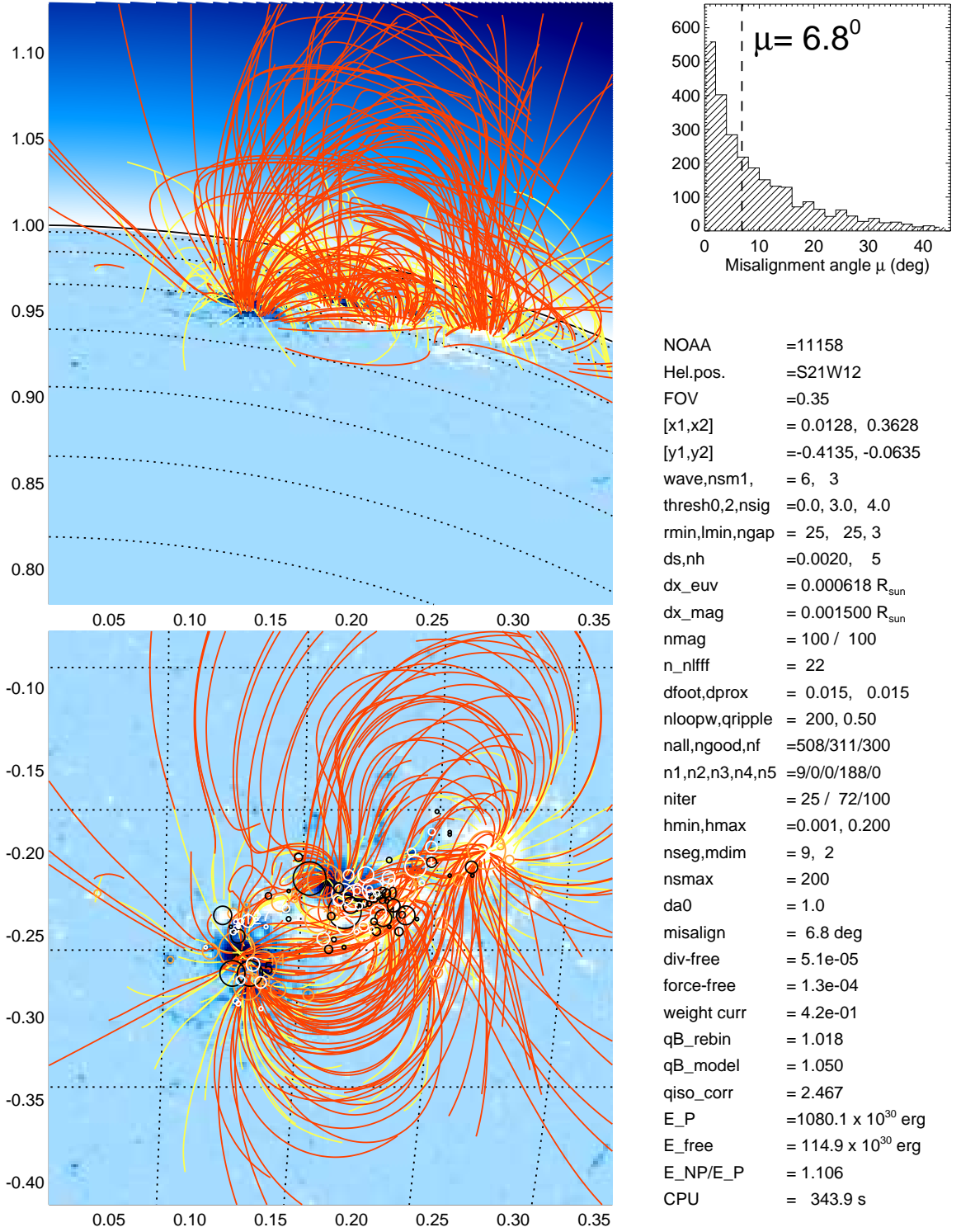}
\caption{Example of a forward-fitting run (RUN2) for flare event \#12
(2011 February 15, 01:14 UT), for time frame $i_t=0$ (out of $n_t=13$ 
flare time intervals).
RUN2 selects loop structures with a ripple ratio of 
$q_{ripple} \le 0.5$ in a field-of-view of $FOV=0.35$ solar radii.
The magnetogram (blue) and observed loops (yellow curves) and best-fit
magnetic field lines (red) are shown in the image plane $(x,y)$ (bottom 
panel) and rotated by $90^\circ$ to the north (top panel), which corresponds to a
projection into the $(x,z)$ plane. The magnetic charges are indicated 
with (white/black) circles according to their (positive/negative) 
magnetic polarity and depth (radius or circles). The distribution of 2D 
misalignment angles is shown in a histogram (top right panel), 
measured in $n_{seg}=9$ loop segment positions in $n_{loop}=300$ loops.}
\end{figure}
\clearpage

\begin{figure}
\plotone{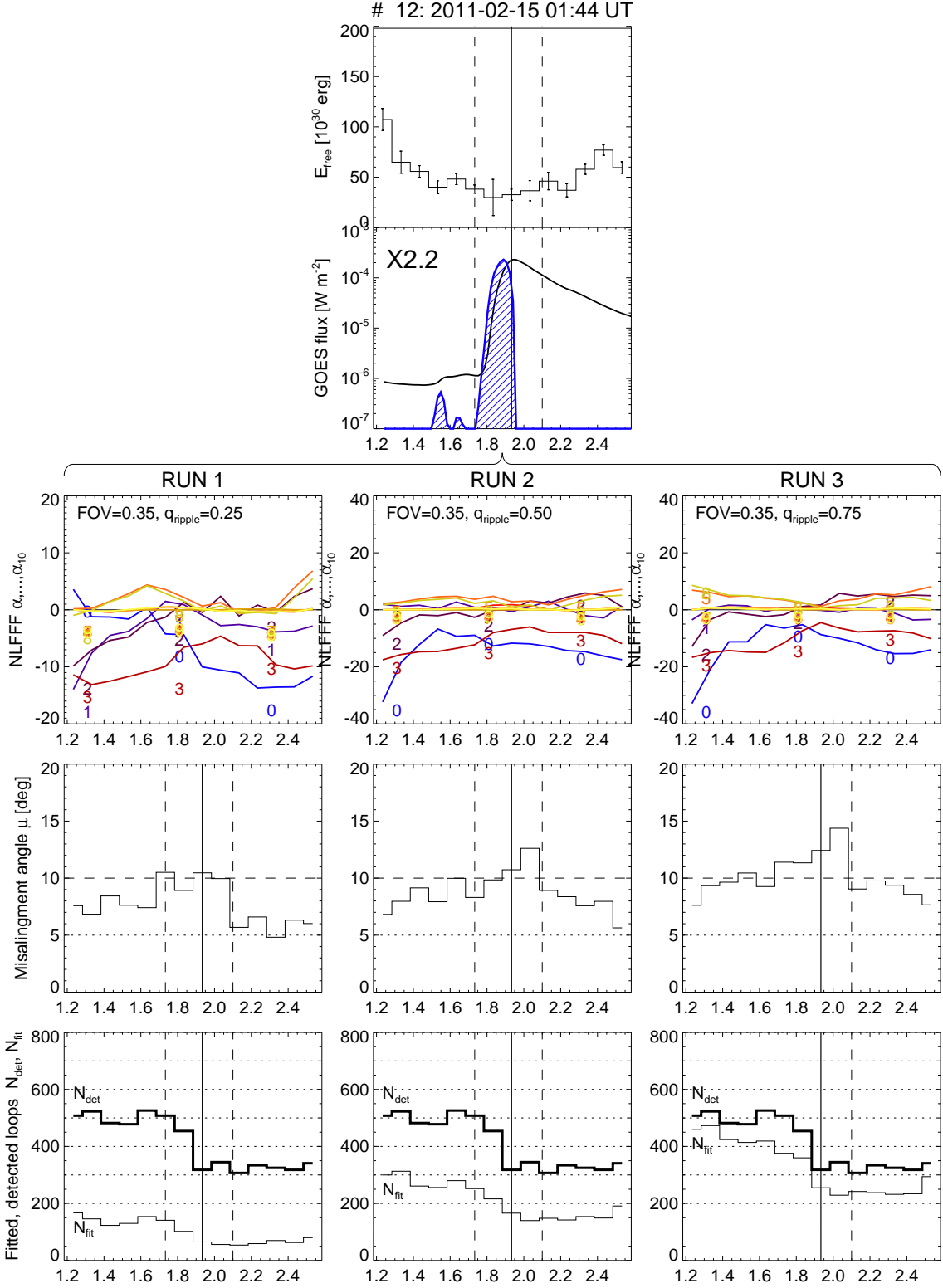}
\caption{The time evolution of the free energy $E_{free}(t)$ (top
middle), the GOES flux $F_{GOES}(t)$ (second row from top, black curve) 
and its time derivative (second row; blue hatched curve), averaged from
three runs with different loop selection parameters (3 columns:
$q_{ripple}=0.25$ (RUN1, left), 
$q_{ripple}=0.50$ (RUN2, middle), 
$q_{ripple}=0.75$ (RUN3, right),
yielding error bars for the free energy measurements.
The time evolution of the force-free parameter
$\alpha(t)$ for the 10 strongest magnetic sources (third row),
the median misalignment angle $\mu_2(t)$ (fourth row), the number of 
detected $N_{det}(t)$ (bottom row; thick linestyle) and number 
of fitted loops $N_{fit}(t)$ (bottom row; thin linestyle), are shown
for each of the 3 runs separately. 
The vertical lines indicate the flare peak time (solid linestyle),
the flare start and end times (dashed linestyle), based on the GOES flux.}
\end{figure}
\clearpage

\begin{figure}
\plotone{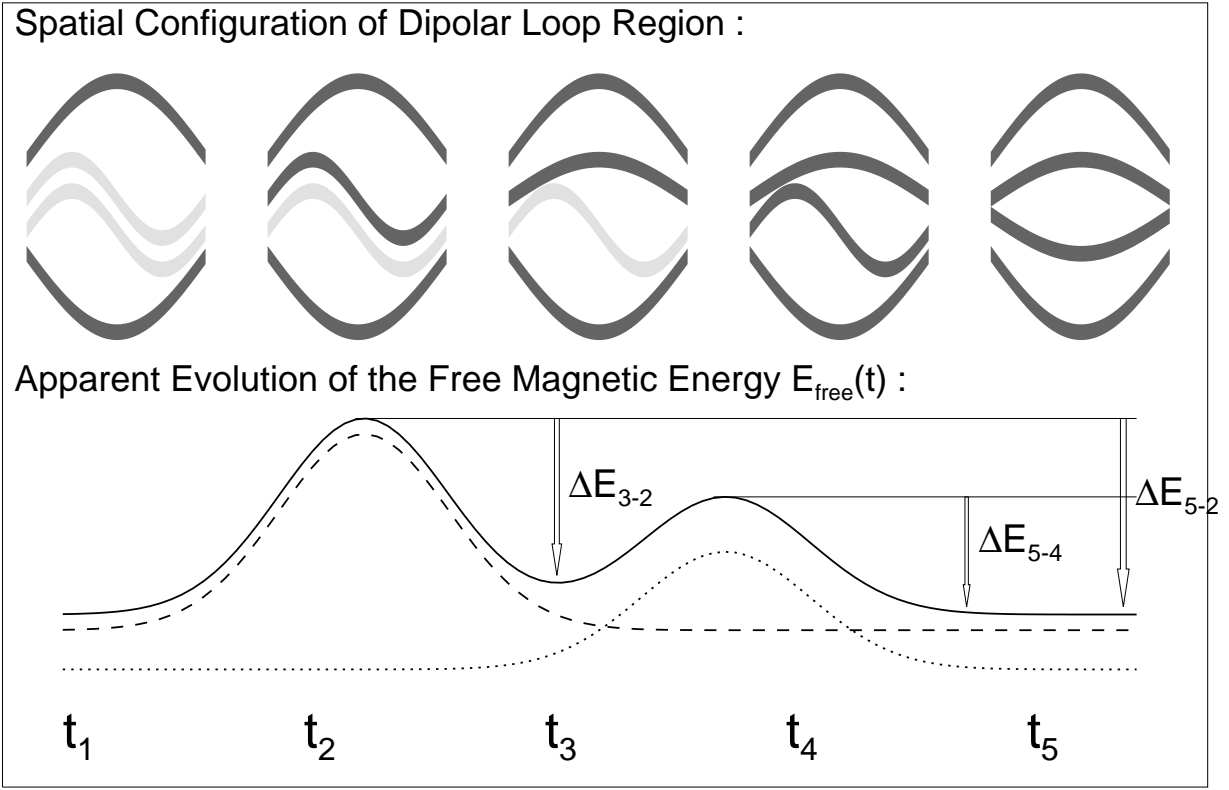}
\caption{Schematic diagram of the spatial loop configuration (top panel)
and evolution of free magnetic energy $E_{free}(t)$ during a flare.
Mostly potential loops are visible at the beginning of a flare ($t_1$),
while a first sigmoid is illuminated at $t_2$, which relaxes to a potential
loop at time $t_3$. A second sigmoid is illuminated at time $t_4$, which
relaxes to a potential loop at time $t_5$. The total energy difference
before and after the flare, $\Delta E_{5-2}$, is a lower limit to the
sum of all sequential energy releases $\Delta E_{3-2}$ and $\Delta E_{5-4}$,
and thus underestimates the total dissipated magnetic energy
(Aschwanden, Sun, \& Liu 2014).}
\end{figure}
\clearpage

\begin{figure}
\plotone{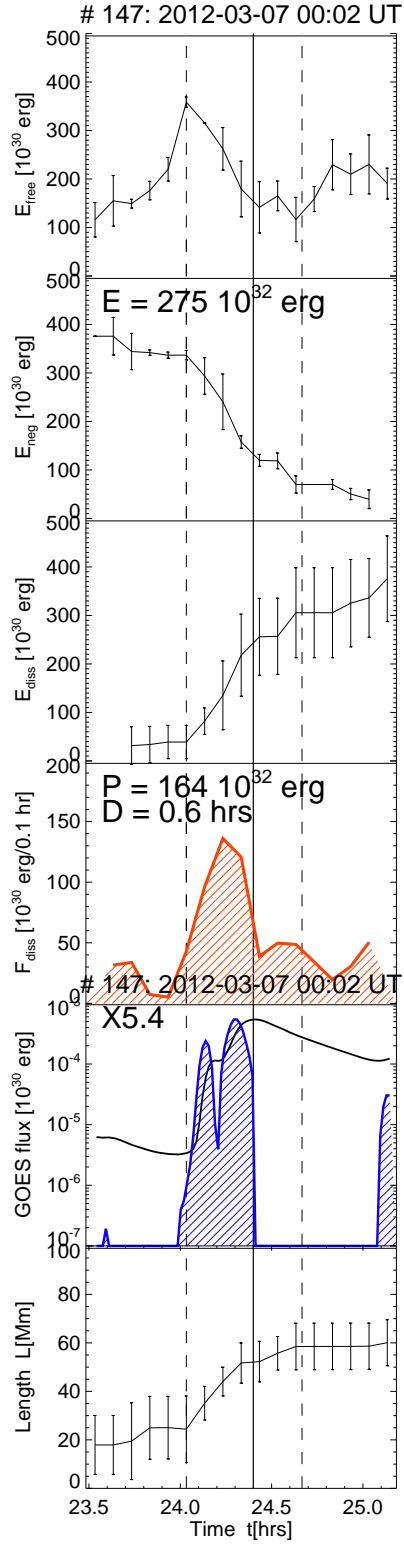}
\caption{An example of the evolutionary parameters as defined in text is
given here for the event \# 147, 2012-Mar-07 00:02 UT, an X5.4 GOES class 
flare: the free energy
$E_{free}(t)=E_n(t)-E_p(t)$ (top panel), the cumulative negative energy
$E_{neg}(t)$ (second panel), the dissipated energy $E_{diss}(t)$ (third
panel), the energy dissipation rate $F_{diss}=dE_{diss}(t)/dt$ (fourth 
panel; red), the GOES flux $F_{GOES}$ and its time derivative $d F_{GOES}(t)/dt$
(fifth panel; blue), and the length scale $L(t)$ of the cumulative flare
area (bottom panel). 
The error bars of the free energy (top panel) are derived from the
scatter of 3 runs with different loop selection parameters, while the other
error bars are propagated by Monte-Carlo simulations.}
\end{figure}

\begin{figure}
\plotone{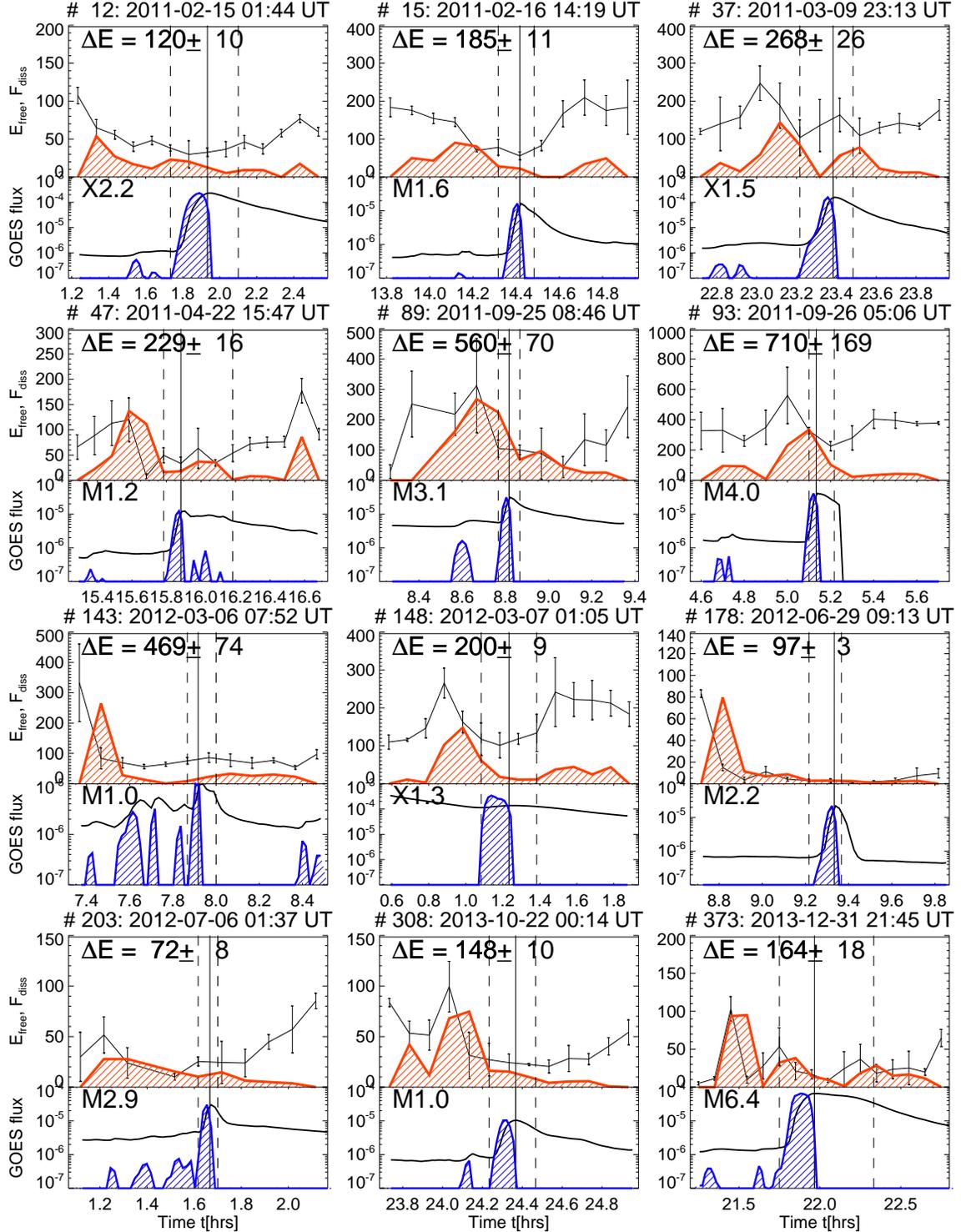}
\caption{Examples of 12 flares with magnetic energy dissipation starting
in the preflare phase. Each of the 12 panels shows the time evolution of
the free (magnetic) energy $E_{free}(t)$ in units of $10^{30}$ erg 
(black curve with error bars),
the energy dissipation rate $F_{diss}(t)$ in units of [$10^{30}$ erg / 0.2 hr]
(red curve with hatched area), the GOES 1-8 \ang\ light curve 
in units of W cm$^{-2}$ (solid black curves), and its time derivative 
(blue curve with hatched area).}
\end{figure}
\clearpage

\begin{figure}
\plotone{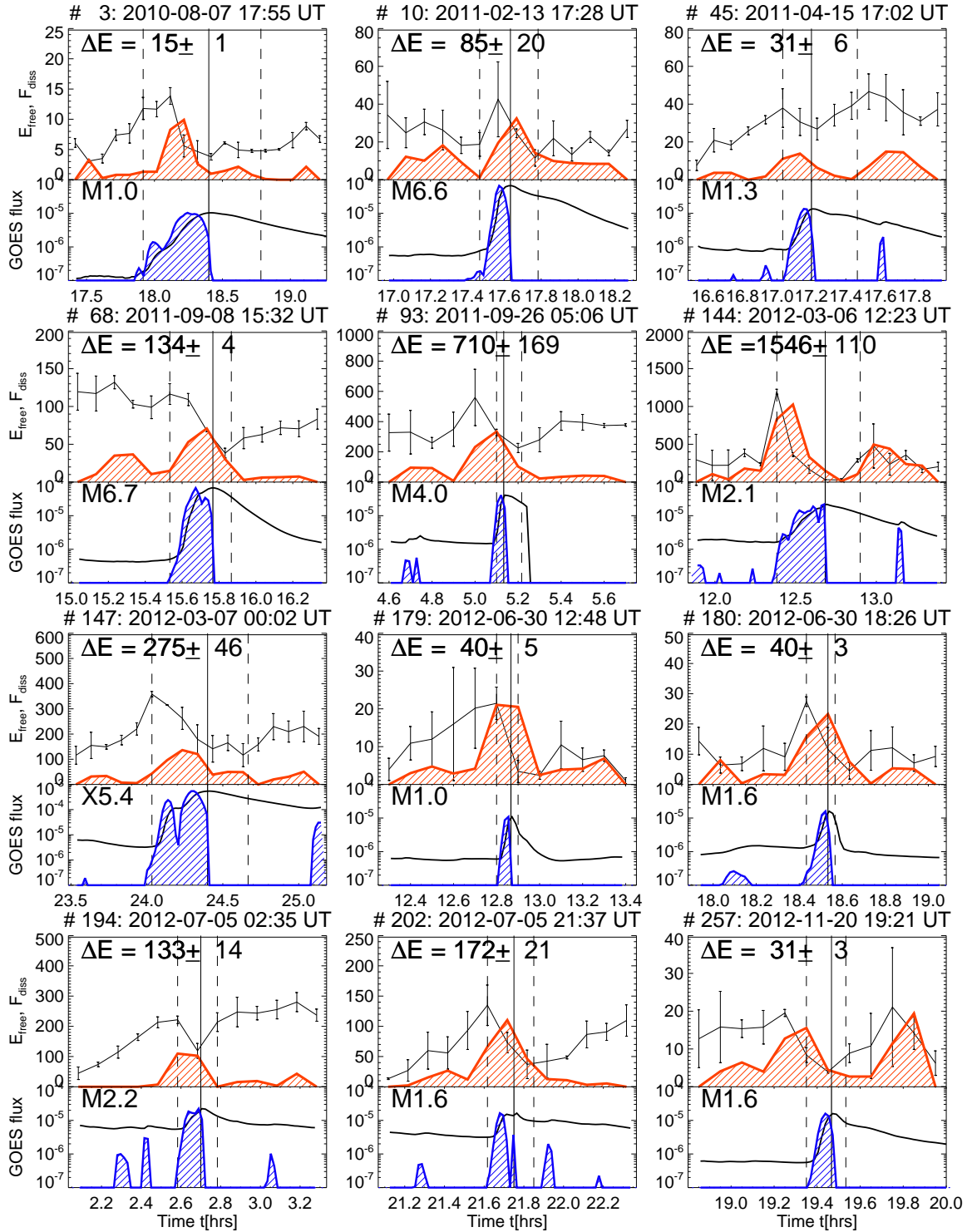}
\caption{Examples of magnetic energy dissipation coinciding with the
flare rise time. Representation otherwise similar to Fig.~5.}
\end{figure}
\clearpage

\begin{figure}
\plotone{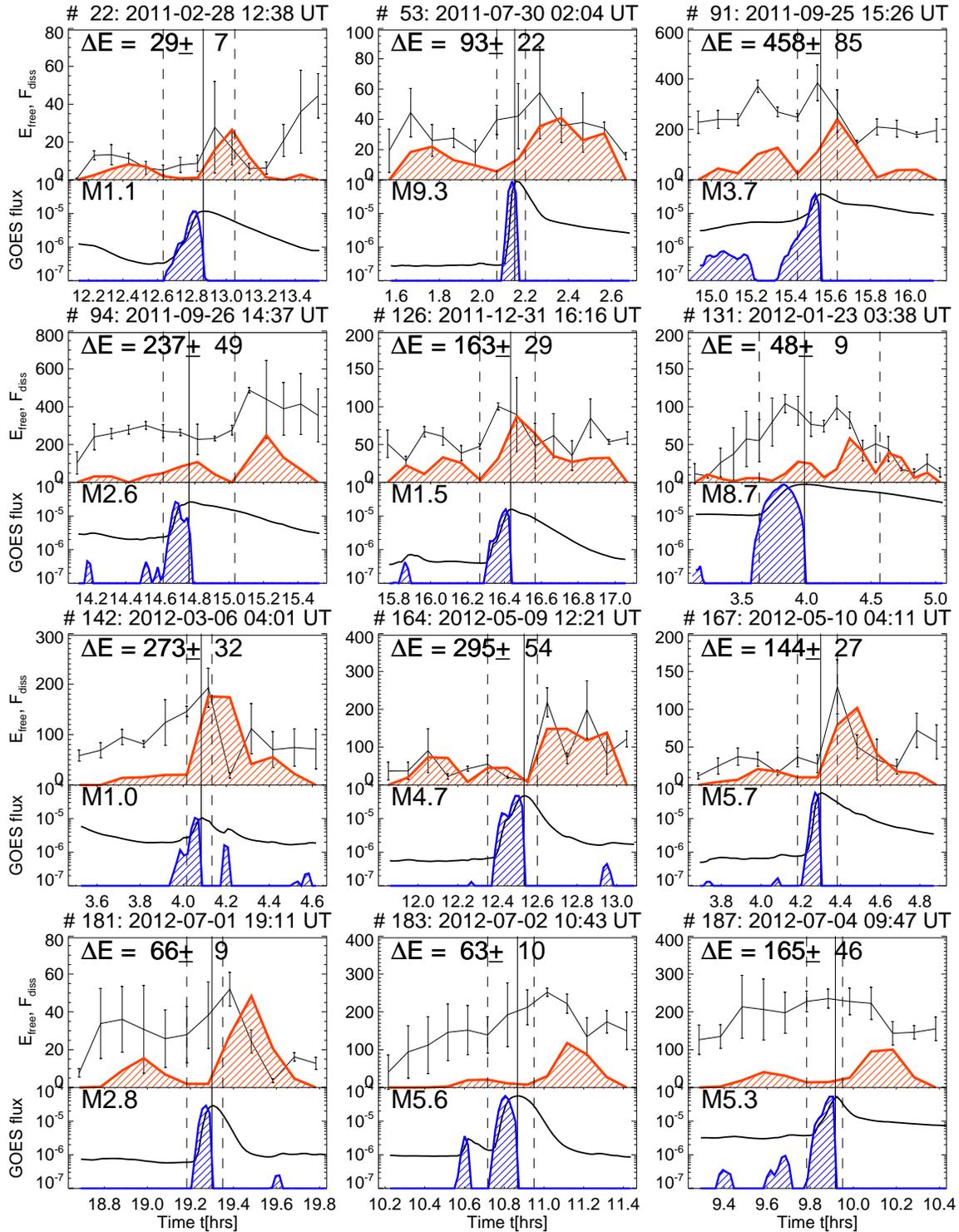}
\caption{Examples of events with magnetic energy dissipation occurring 
mostly during the flare decay phase. Representation otherwise 
similar to Fig.~5.}
\end{figure}
\clearpage

\begin{figure}
\plotone{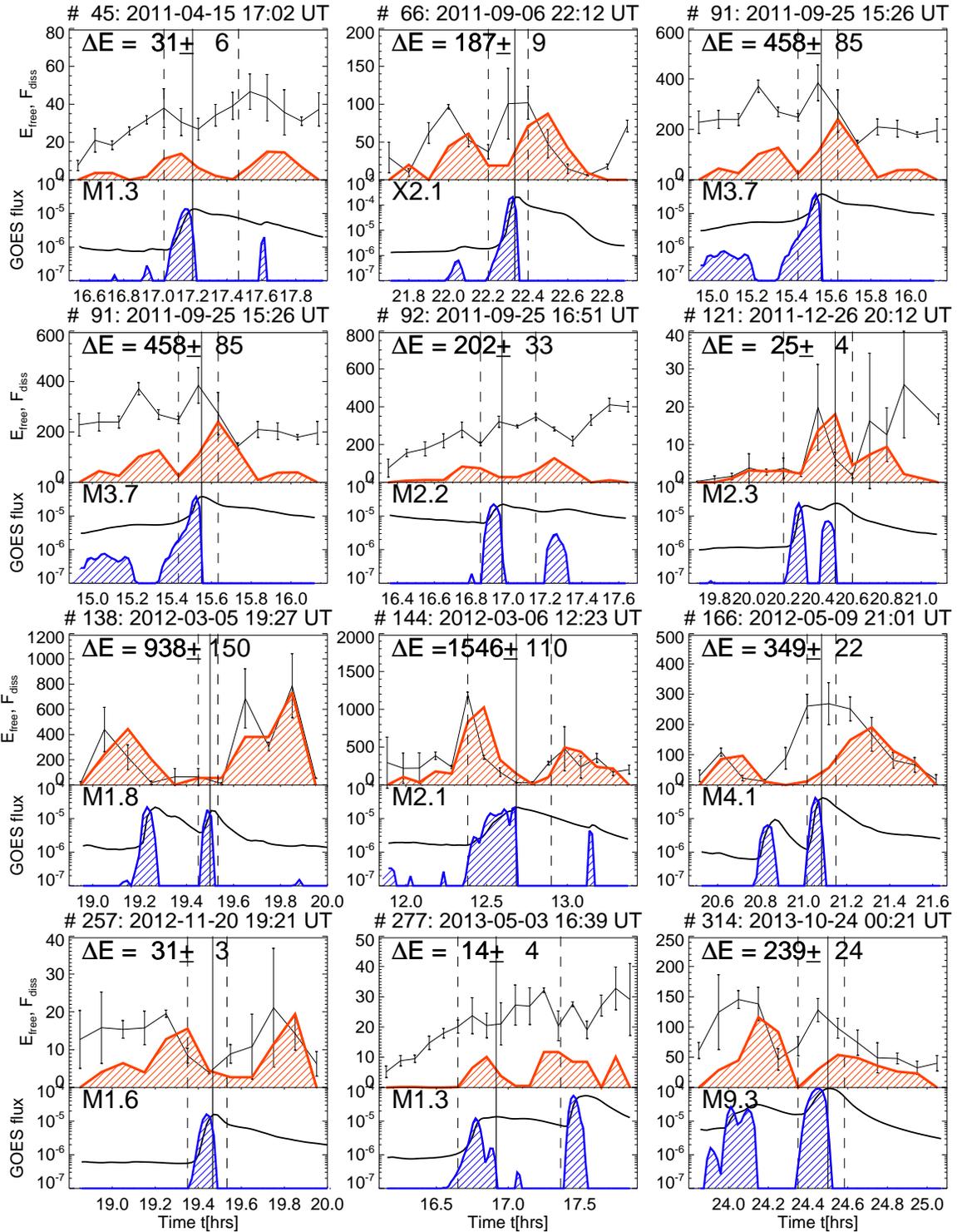}
\caption{Examples of dual flare events with magnetic energy dissipation
associated with the two flare peaks. Representation otherwise similar to
Fig.~5.}
\end{figure}
\clearpage

\begin{figure}
\plotone{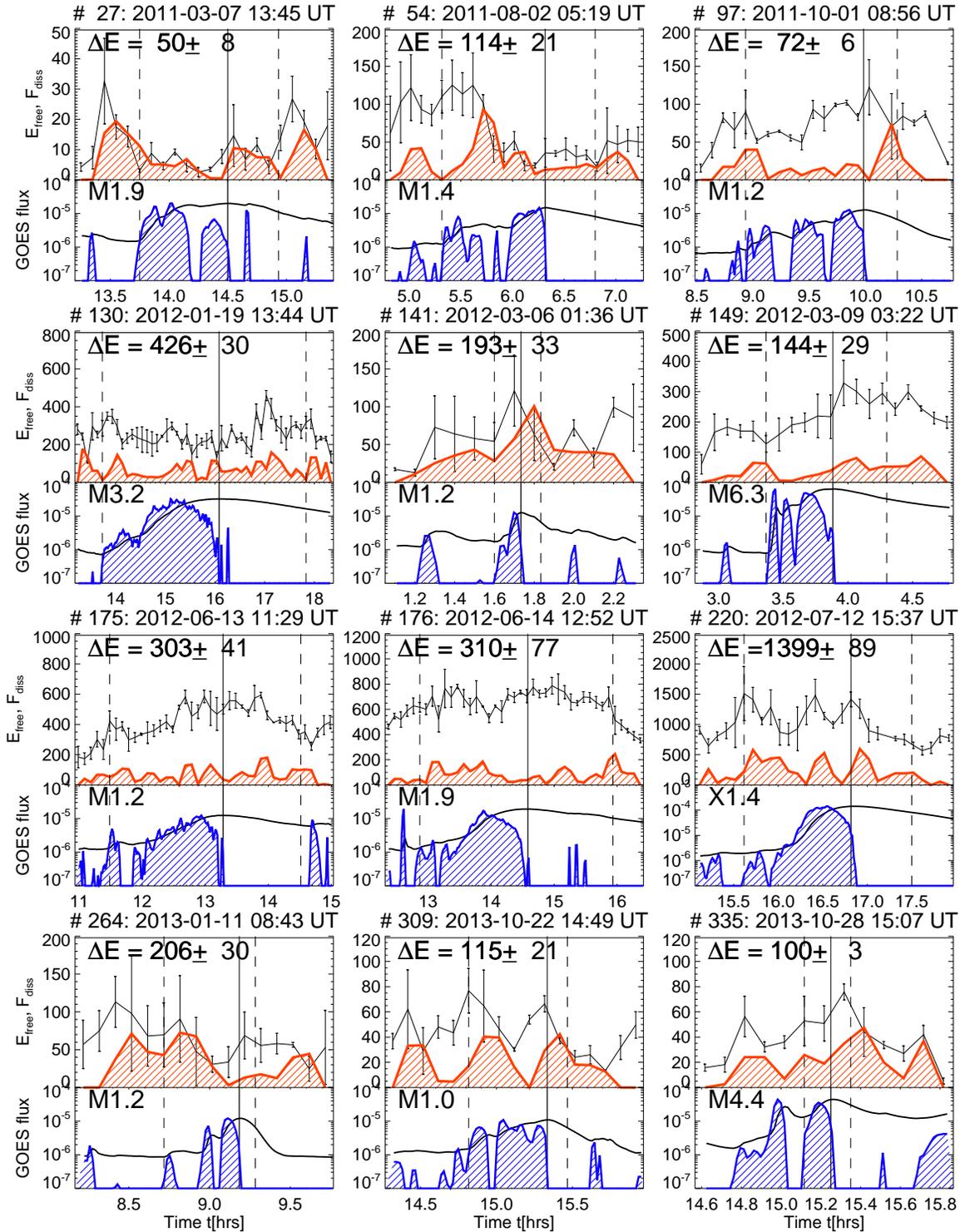}
\caption{Examples of complex flare events with multi-step magnetic 
energy dissipation associated with individual flare subpeaks. 
Representation otherwise similar to Fig.~5.}
\end{figure}
\clearpage

\begin{figure}
\plotone{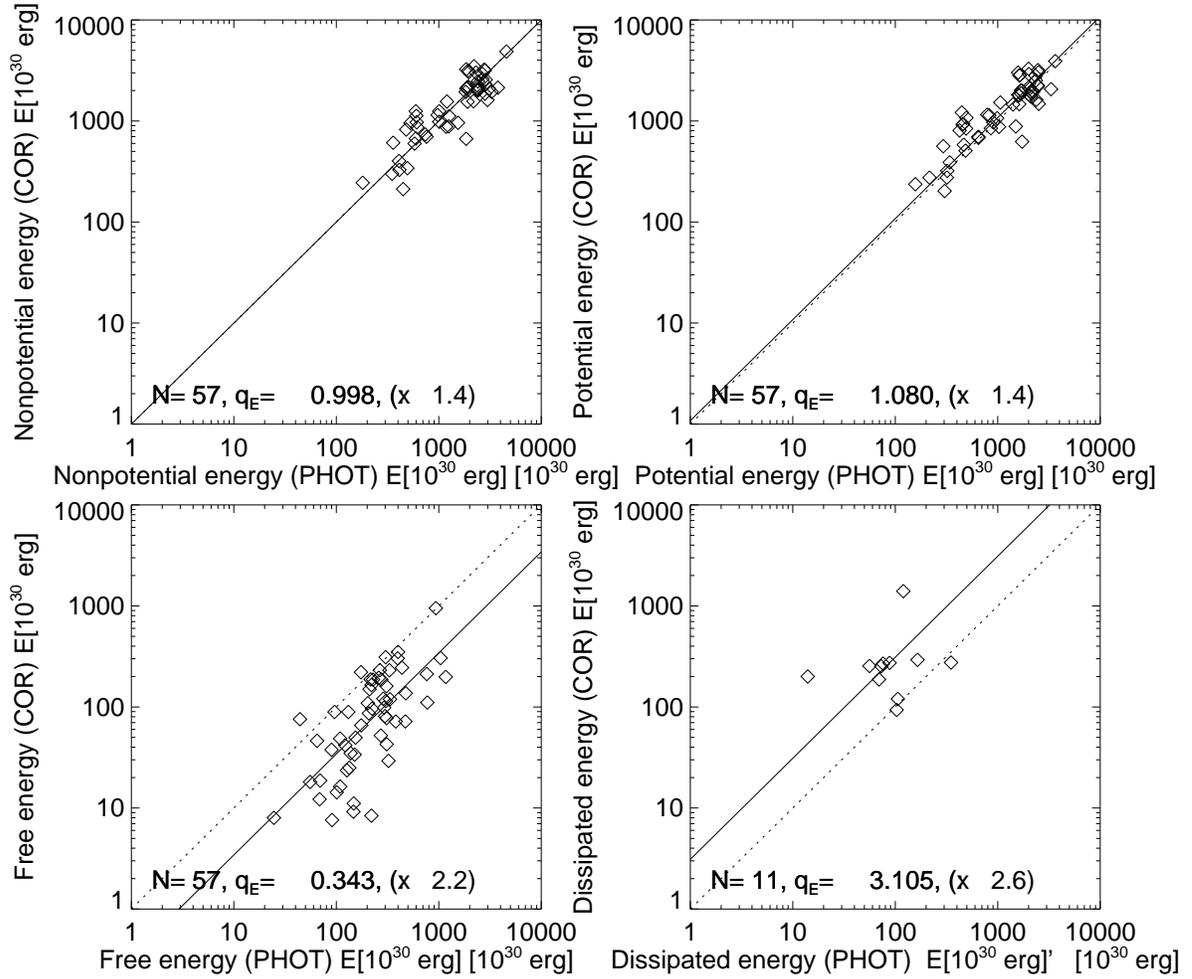}
\caption{Scatterplot of mean nonpotential energies (top left), potential
energies (top right), free energies (bottom left), and dissipated energies
(bottom right) between the PHOT-NLFFF (x-axis) and COR-NLFFF
codes (y-axis). The solid line indicates the mean proportionality ratio,
and the dottel line the unity ratio.}
\end{figure}
\clearpage

\begin{figure}
\plotone{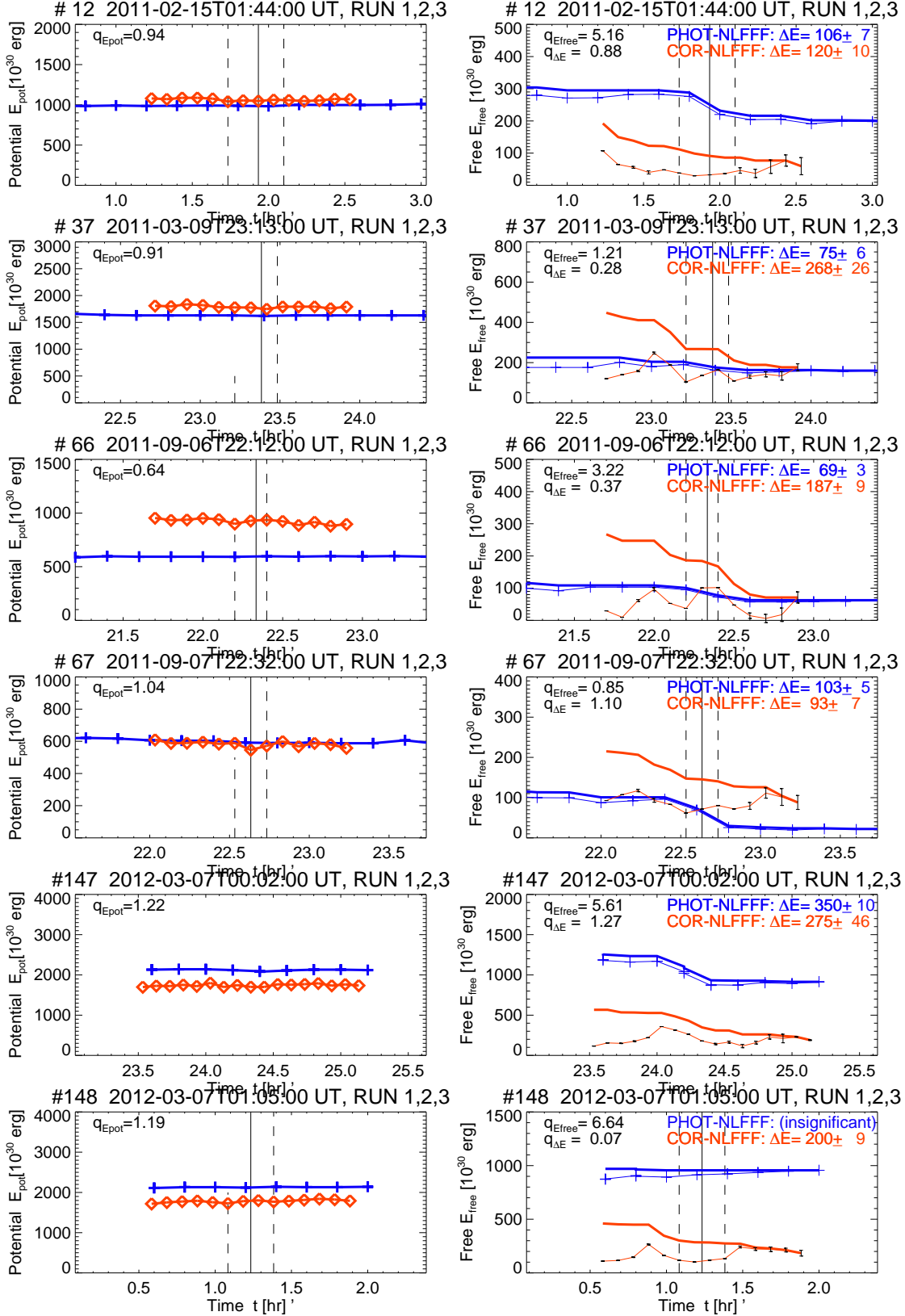}
\caption{Comparison of the evolution of the potential energy $E_p(t)$
(left panels), the free energy $E_{free}(t)=E_{np}(t)-E_p(t)$
(right panels, thin curves), and the cumulative negative energy
decreases $F_{neg}(t)$ (right panels, thick curves), for both the 
coronal COR-NLFFF code (red curves) and the photospheric PHOT-NLFFF code 
(red curves), for six X-class flares.}
\end{figure}
\clearpage

\begin{figure}
\plotone{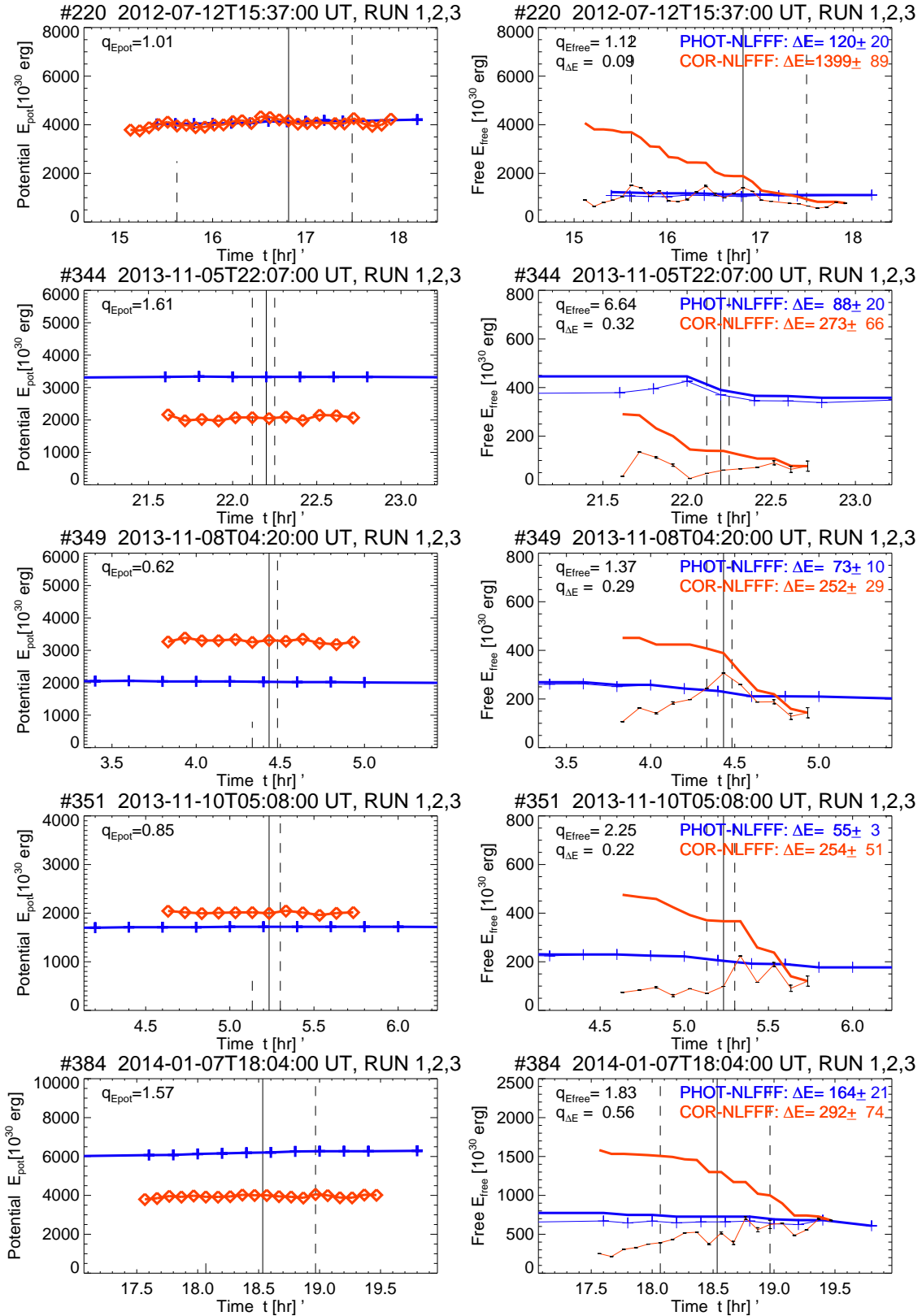}
\caption{Comparison of energy evolution for another five X-class flares,
with similar representation as in Fig.~11.}
\end{figure}
\clearpage

\begin{figure}
\plotone{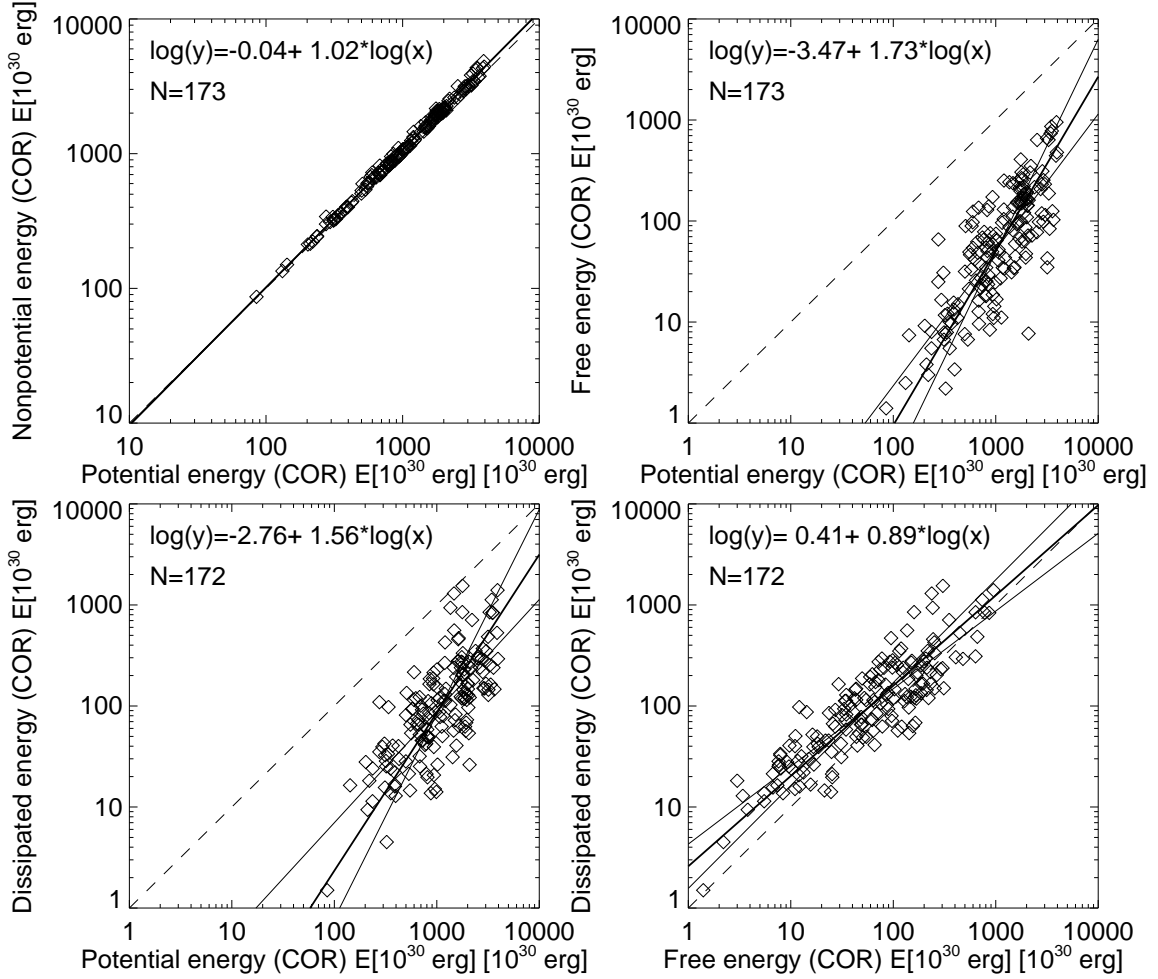}
\caption{Scaling laws between the nonpotential energy $E_{np}$, the potential
energy $E_p$, the free energy $E_{free}$, and the dissipated energy
$E_{diss}$, for energy values computed with the COR-NLFFF code.
Linear regression fits (thick solid lines) and uncertainties (thin lines)
are shown, while proportionality is indicated with a dashed line.}
\end{figure}
\clearpage

\begin{figure}
\plotone{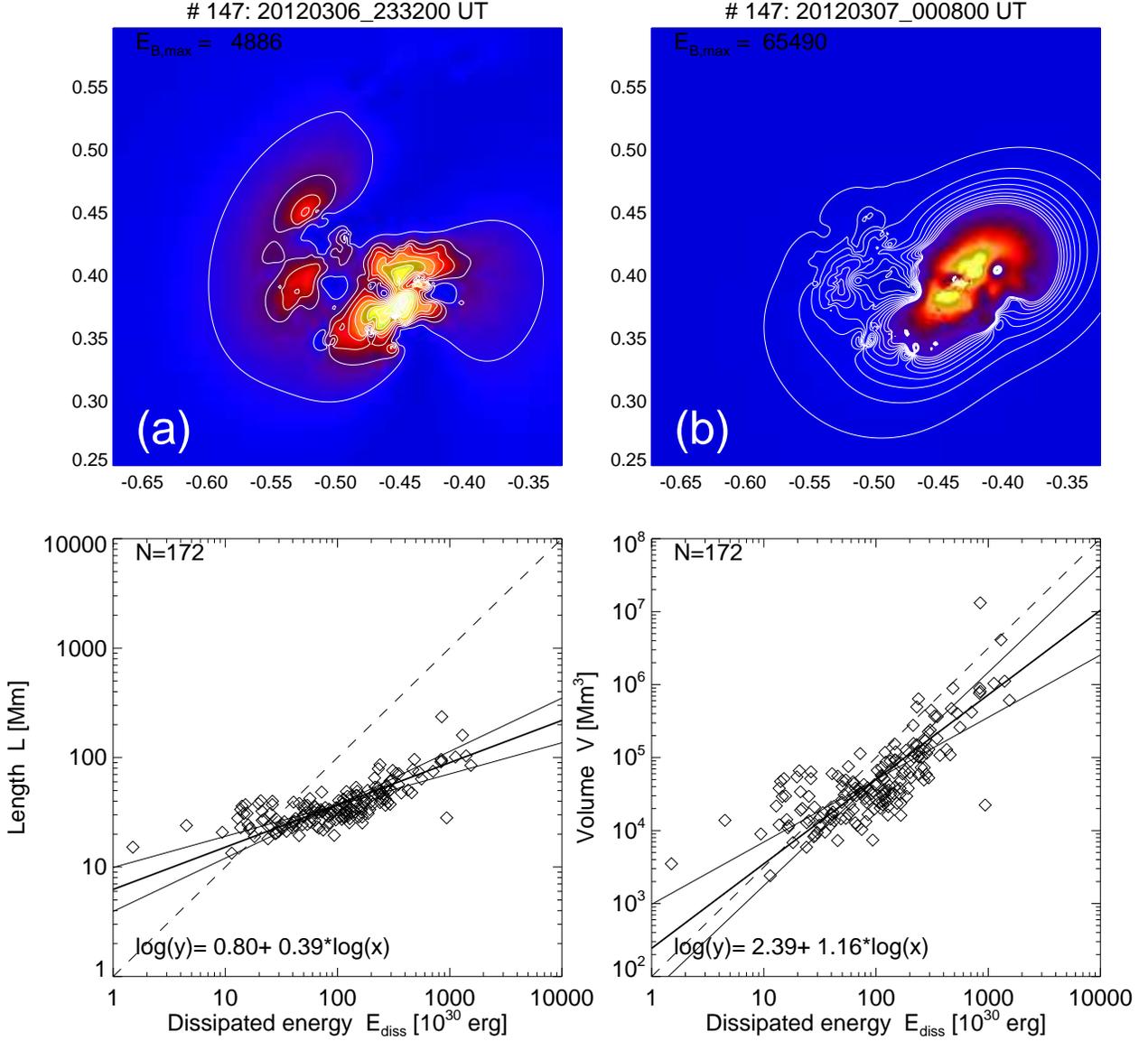}
\caption{(a) Contour plot of the free energy distribution $E_{free}(x,y)=
E_{np}(x,y)-E_p(x,y)=B_\varphi(x,y)^2/(8\pi)$, in contours of $B_i=5, 10, ...,
100$ G, before the flare (2011-Mar-06 23:32 UT), and (b) at the peak of the flare. 
Note the maximum of the dissipated energy distribution is a factor of 
13 higher at the flare peak. The flare area $A$ is measured from the number 
of pixels above some energy threshold, defining a length scale $L=A^{1/2}$
and a flare volume $V=A^{3/2}$. 
{\sl Bottom row:} Correlation plots of the dissipated flare energy $E_{diss}$ and
the flare length scale $L$ (bottom left panel) or the flare volume $V=L^3$
(bottom right panel) for all analyzed 172 flare events. 
Linear regression fits (thick solid lines), uncertainties (thin solid lines),
and proportionality (dashed line) are indicated, Note that the flare volume 
$V$ is almost proportional to the dissipated energies $E_{diss}$.}
\end{figure}
\clearpage

\begin{figure}
\plotone{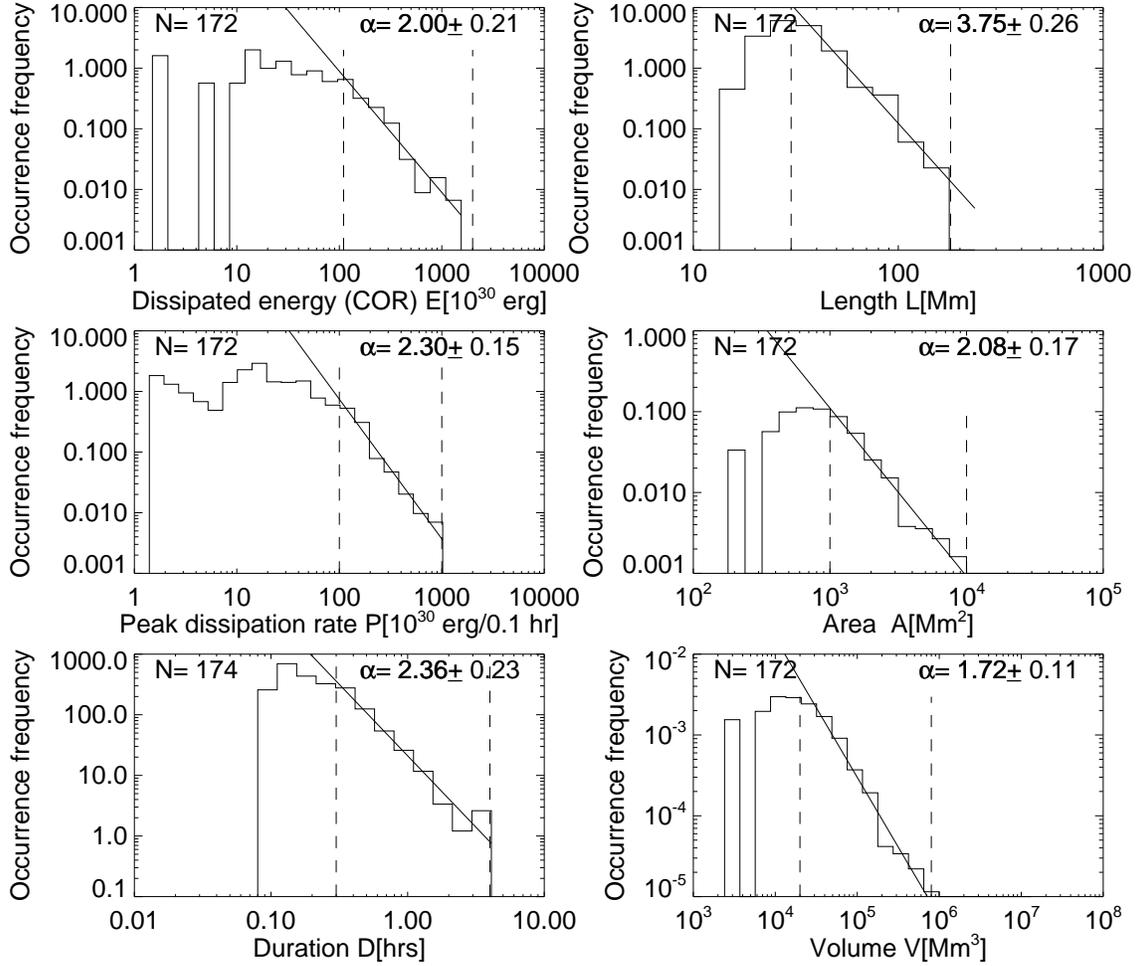}
\caption{Occurrence frequency distributions on a log-log scale for the
dissipated energy $E_{diss}$ (top left panel), the peak dissipation rate
$F_{diss}$ (second left panel), flare durations $T$ (bottom left panel),
the flare length scale $L$, the flare area $A$, and the flaring volume
$V$ for 172 analyzed M- and X-class flares.}
\end{figure}
\clearpage

\begin{figure}
\plotone{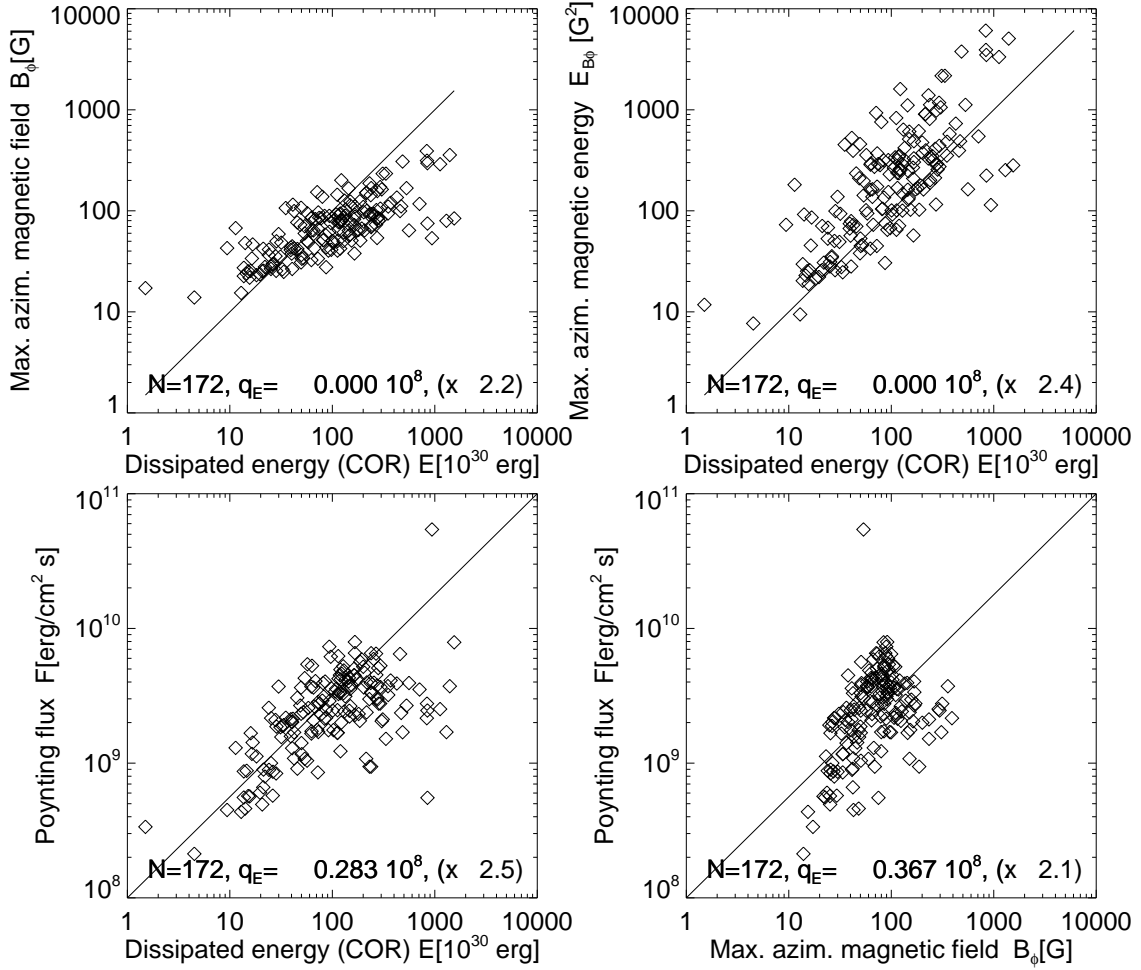}
\caption{Scatterplots between the dissipated flare energies $E$,
the maximum azimuthal magnetic field component $B_{\varphi}$
(top left panel), the maximum azimuthal magnetic energy per pixel
$E_{B\varphi}$ (top right panel), and the Poynting flux $F$
(bottom left and right panels). Note the proportionality between
$E$ and $E_{B\varphi}$ (top right panel).}
\end{figure}
\clearpage

\begin{figure}
\plotone{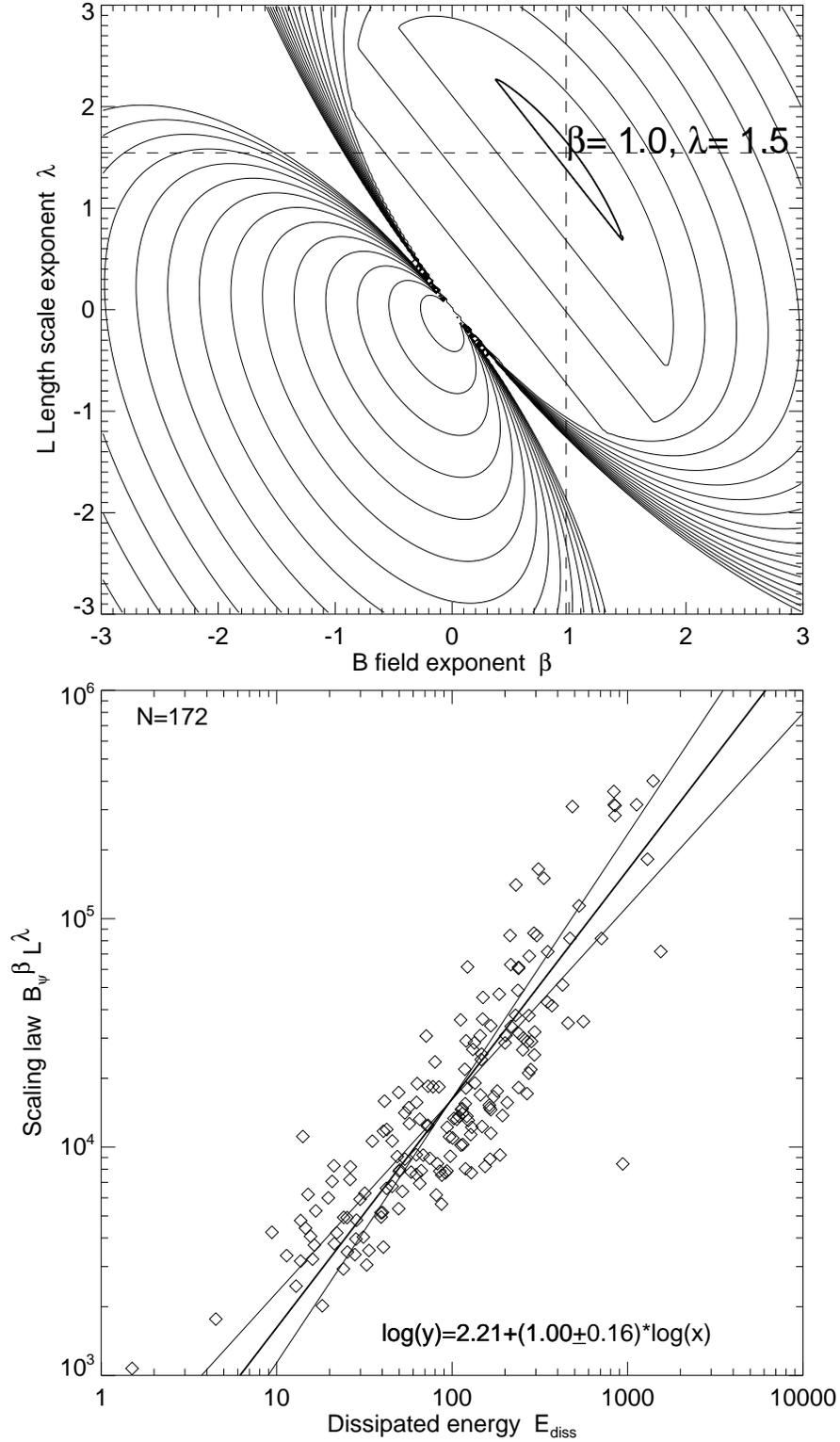}
\caption{Goodness-of-fit map in parameter space $\chi(\beta, \lambda)$
(top panel) for a scaling law of the free energy 
$E_{free} \propto B^{\beta} L^{\lambda}$. A linear regression fit
with the best-fit parameters $\beta=1.0$ and $\lambda=1.5$ is shown
in the bottom panel.}
\end{figure}
\clearpage

\begin{figure}
\plotone{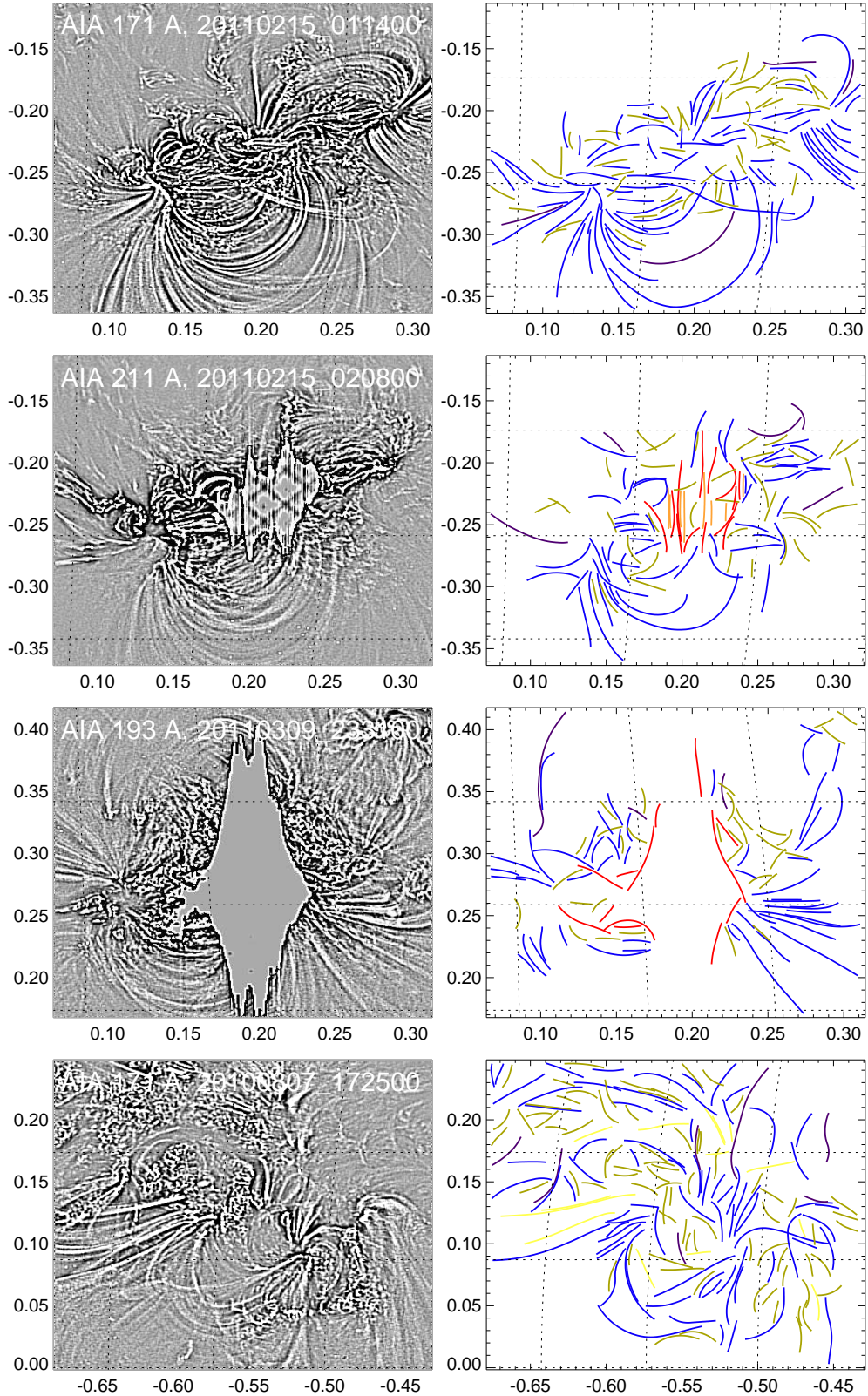}
\caption{Examples of automated pattern recognition of the COR-NLFFF code:
coronal loop structures (blue curves), boundaries of saturated image areas 
(red), vertical streaks from CCD pixel-bleeding (orange), and rippled (moss)
structures (green). The greyscale images (left) are highpass-filtered 
EUV images.}
\end{figure} 
\clearpage

\begin{figure}
\plotone{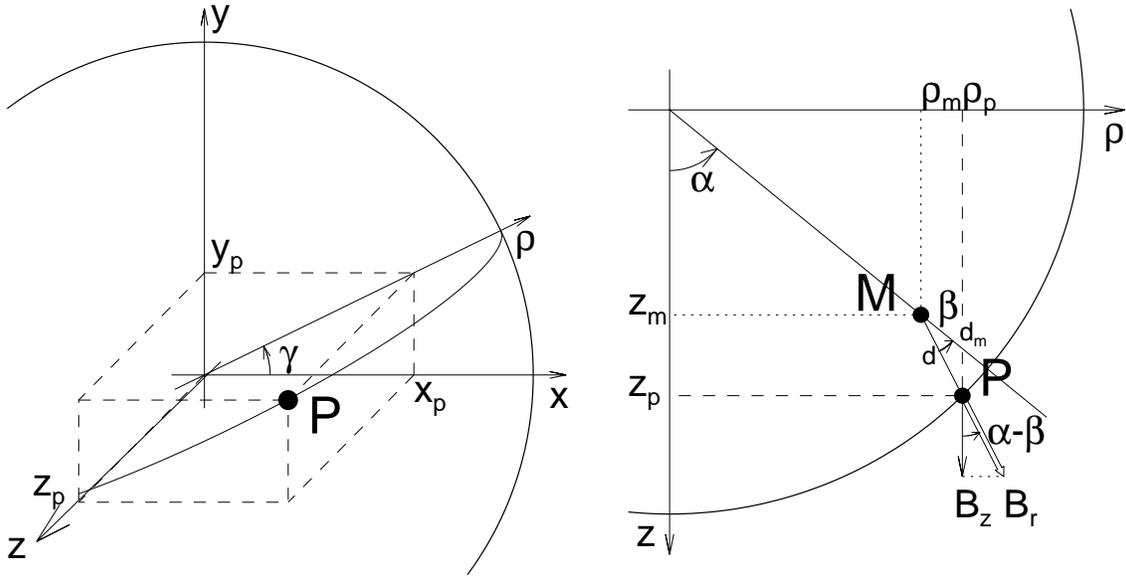}
\caption{For the decomposition of magnetograms into buried magnetic charges,
the 3D geometry of a point source $P=(x_p, y_p, z_p)$
in a cartesian coordinate system is shown (left), with the $z$-axis aligned
to the line-of-sight from Earth to Sun center. The plane through the
line-of-sight axis and the point source $P$ has a position angle $\gamma$
in the plane-of-sky with respect to the $x$-axis and defines the direction
of the axis $\rho$.
The geometry of a line-of-sight magnetic field component $B_z$ is shown
in the $(z,\rho)$-plane on the right hand side. A magnetic point charge $M$
is buried at position $(z_m, \rho_m)$ and has an aspect angle $\alpha$
to the line-of-sight. The radial component $B_r$ is observed on the solar
surface at location $P$ and has an inclination angle of $\beta$ to the local
vertical above the magnetic point charge $M$. The line-of-sight component
$B_z$ of the magnetic field has an angle $(\alpha-\beta)$ to the radial
magnetic field component $B_r$. (See details in Appendix A of
Aschwanden 2012a).}
\end{figure}
\clearpage

\begin{figure}
\plotone{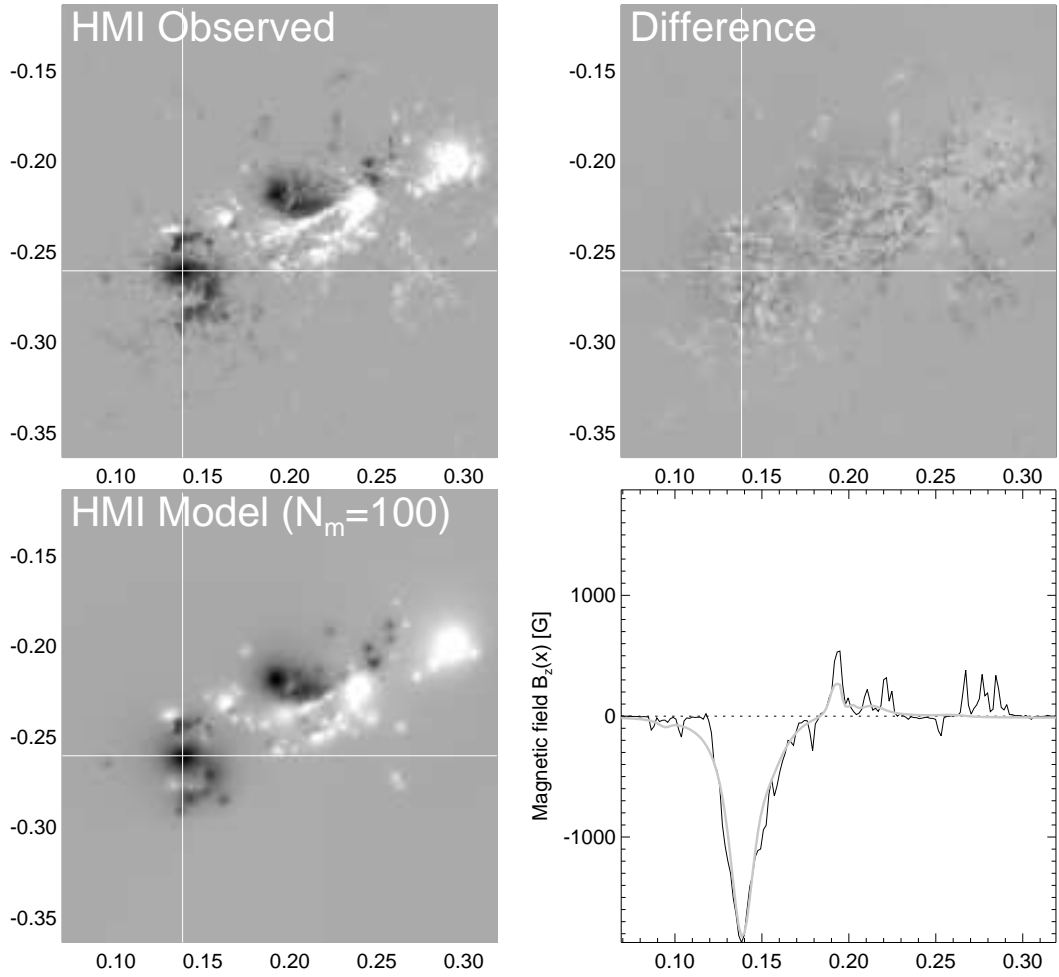}
\caption{Example of a line-of-sight magnetogram from 2011 February 15, 
01:56 UT, observed with HMI/SDO (top left), decomposed into $n_m=100$
unipolar magnetic charges (bottom left), with the difference shown
on the same grey scale (top right), along with a scan across the
sunspot with maximum field strength (bottom right: black profile is 
observed, grey profile is the model).}
\end{figure}
\clearpage

\begin{figure}
\plotone{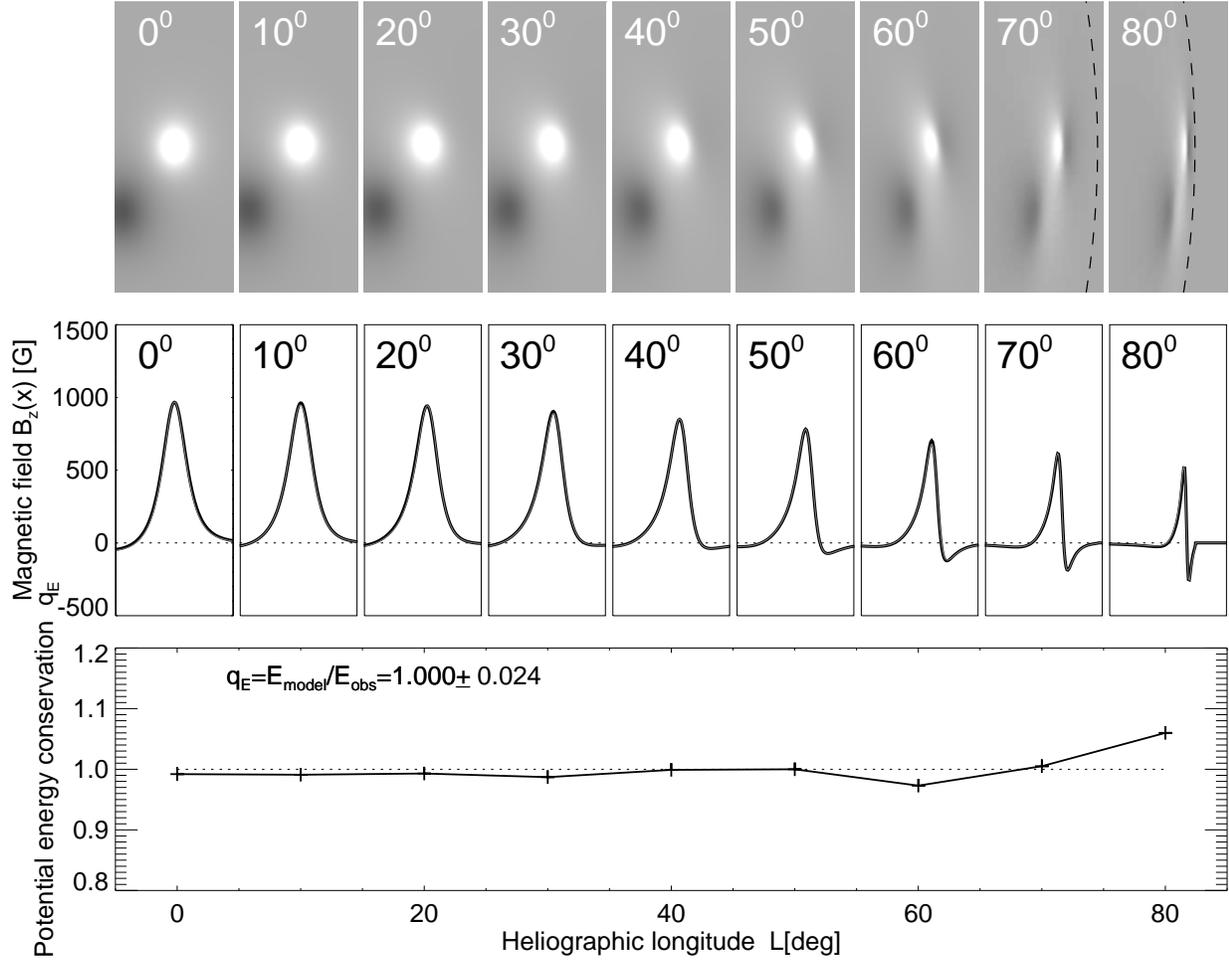}
\caption{Simulation of a magnetogram with a buried dipole as a function
of the longitude from Sun center in steps of $10^\circ$ to the west
limb at $80^\circ$ (top panels). The magnetic field profile of the 
LOS component $B_z(x)$ is shown from the numerically inverted
profile $B_z(x)$, which is indistinguishable from the simulated profile 
(black curves in middle panels). The conservation of the retrieved magnetic
energy $E_B=\int B^2(x,y)\ dx \ dy $ is shown in the bottom panel,
which demonstrates the invariance of the obtained magnetic energy
with respect to the solar rotation or the heliographic position.}
\end{figure}
\clearpage

\begin{figure}
\plotone{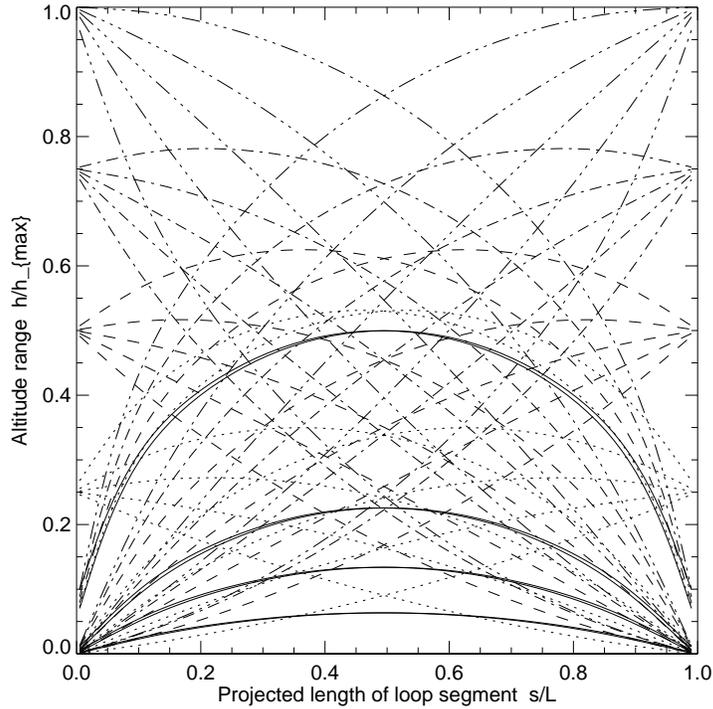}
\caption{Modeling of the altitude coordinate $h(s)$ of loop segments
observed in 2D, using a finite subset of ($2 \times n_h \times n_c$) 
circular geometries for the 3D reconstruction. The model loops have 
one footpoint at photospheric height and 
the other end of the loop segment at altitude $h=h_{max}(i/n_h), 
i=1,...,n_h, n_h=5$, and each case is fitted with $n_c=5$ different 
curvature radii.
The projected distance $s$ is scaled with the loop segment length $L$.}
\end{figure}
\clearpage

\end{document}